\documentclass{aa}  

\usepackage{txfonts}
\usepackage{hyperref}

\usepackage{graphicx}	
\usepackage{siunitx}         	
\usepackage{float}			
\usepackage{subfig}		
\usepackage{bm}
\usepackage{natbib}		
\usepackage{xcolor}
\usepackage{threeparttable}

\newcommand{\tCO}{$^{13}$CO}
\newcommand{\kms}{km~s$^{-1}$} 
\newcommand{\msun}{M$_{\sun}$}
\newcommand{\e}{$\times$10}
\newcommand{\tcm}{\e$^{23}$~cm$^{-2}$}
\newcommand{\nhp}{N$_{2}$H$^{+}$}
\newcommand{\nhhh}{NH$_3$}

\newcommand{\SFRtff}{SFR$_\text{ff}$}
\newcommand{\SFEtff}{$\epsilon_\text{ff}$}
\newcommand{\h}{H$_2$}
\newcommand{\Mline}{$M_\text{line}$}
\newcommand{\Nly}{$N_\text{ly}$}
\newcommand{\micron}{$\mu$m}

\begin{document}

   \title{Feedback from OB stars on their parent cloud:\\  Gas exhaustion rather than gas ejection}


   \author{E. J. Watkins
          \inst{1}
          \and
          N. Peretto\inst{1}
          \and
          K. Marsh\inst{2}
          \and
          G. A. Fuller\inst{3}
          }

   \institute{School of Physics and Astronomy, Cardiff University, The Parade, Cardiff CF24 3AA, UK \\
              \email{WatkinsEJ1@cardiff.ac.uk}
        \and
             	Infrared Processing and Analysis Center, California Institute of Technology 100-22, Pasadena, CA 91125, USA \\       	 
	\and
             	Jodrell Bank Centre for Astrophysics, School of Physics and Astronomy, The University of Manchester, Oxford Road, Manchester, M13 9PL\\
             }

   \date{Accepted June 13, 2019}

 
  \abstract
   {Stellar feedback from high-mass stars shapes the interstellar medium, and thereby impacts gas that will form future generations of stars. However, due to our inability to track the time evolution of individual molecular clouds, quantifying the exact role of stellar feedback on their star formation history is an observationally challenging task. }
   {In the present study, we take advantage of the unique properties of the G316.75-00.00 massive-star forming ridge to determine how stellar feedback from O-stars impacts the dynamical stability of massive filaments. The G316.75 ridge is 13.6~pc long ridge and contains 18,900~\msun\ of \h\ gas, half of which is infrared dark and half of which infrared bright.  The infrared bright part has already formed four O-type stars over the past 2~Myr, while the infrared dark part is still quiescent. Therefore, by assuming the star forming properties of the infrared dark part represent the earlier evolutionary stage of the infrared bright part, we can quantify how feedback impacts these properties by contrasting the two.}
   {We used publicly available {\it Herschel}/HiGAL and molecular line data to measure the ratio of kinetic to gravitational energy per-unit-length, $\alpha_\text{vir}^\text{line}$, across the entire ridge. By using both dense (i.e.\nhp\ and \nhhh) and more diffuse (i.e. \tCO) gas tracers, we were able to compute $\alpha_\text{vir}^\text{line}$ for a range of gas volume densities ($\sim$1\e$^2$--1\e$^5$ cm$^{-3}$).}
   {This study shows that despite the presence of four embedded O-stars, the ridge remains gravitationally bound (i.e. $\alpha_\text{vir}^\text{line} \le 2$) nearly everywhere, except for some small gas pockets near the high-mass stars. In fact, $\alpha_\text{vir}^\text{line}$ is almost indistinguishable for both parts of the ridge. These results are at odds with most hydrodynamical simulations in which O-star-forming clouds are completely dispersed by stellar feedback within a few cloud free-fall times. However, from simple theoretical calculations, we show that such feedback inefficiency is expected in the case of high-gas-density filamentary clouds. }
   {We conclude that the discrepancy between numerical simulations and the observations presented here originates from different cloud morphologies and average densities at the time when the first O-stars form. In the case of G316.75, we speculate that the ridge could arise from the aftermath of a cloud-cloud collision, and that such filamentary configuration promotes the inefficiency of stellar feedback. This does very little to the dense gas already present, but potentially prevents further gas accretion onto the ridge. These results have important implications regarding, for instance, how stellar feedback is implemented in cosmological and galaxy scale simulations.}


   \maketitle
%

\section{Introduction} \label{sec:into}

	Throughout their lifetime, high-mass stars ($>8$~\msun) affect their surroundings via gas ionisation, radiation pressure, stellar winds, protostellar outflows, and, at later evolutionary stages, supernova explosions.  These processes feed energy and momentum back into the surrounding gas and, by doing so, influence its ability to form new generations of stars.  The relative importance of each mechanism is a function of the scales (both space and time) that are taken into consideration \citep{krumholz_star_2014}.  At larger scales, feedback from OB-type stars play a central role in regulating the low star formation efficiency (SFE: $\epsilon_*=m_*/(m_*+m_\text{gas})\simeq 1\%$, where $m_\text{gas}$ is the cloud mass and $m_{*}$ is the stellar mass) observed in galaxies \citep{krumholz_slow_2007,hopkins_galaxies_2014}. However, how such feedback impacts the time evolution and structure of individual molecular clouds remains an open issue.  

\begin{figure*}
	\centering
		\includegraphics[width = \textwidth]{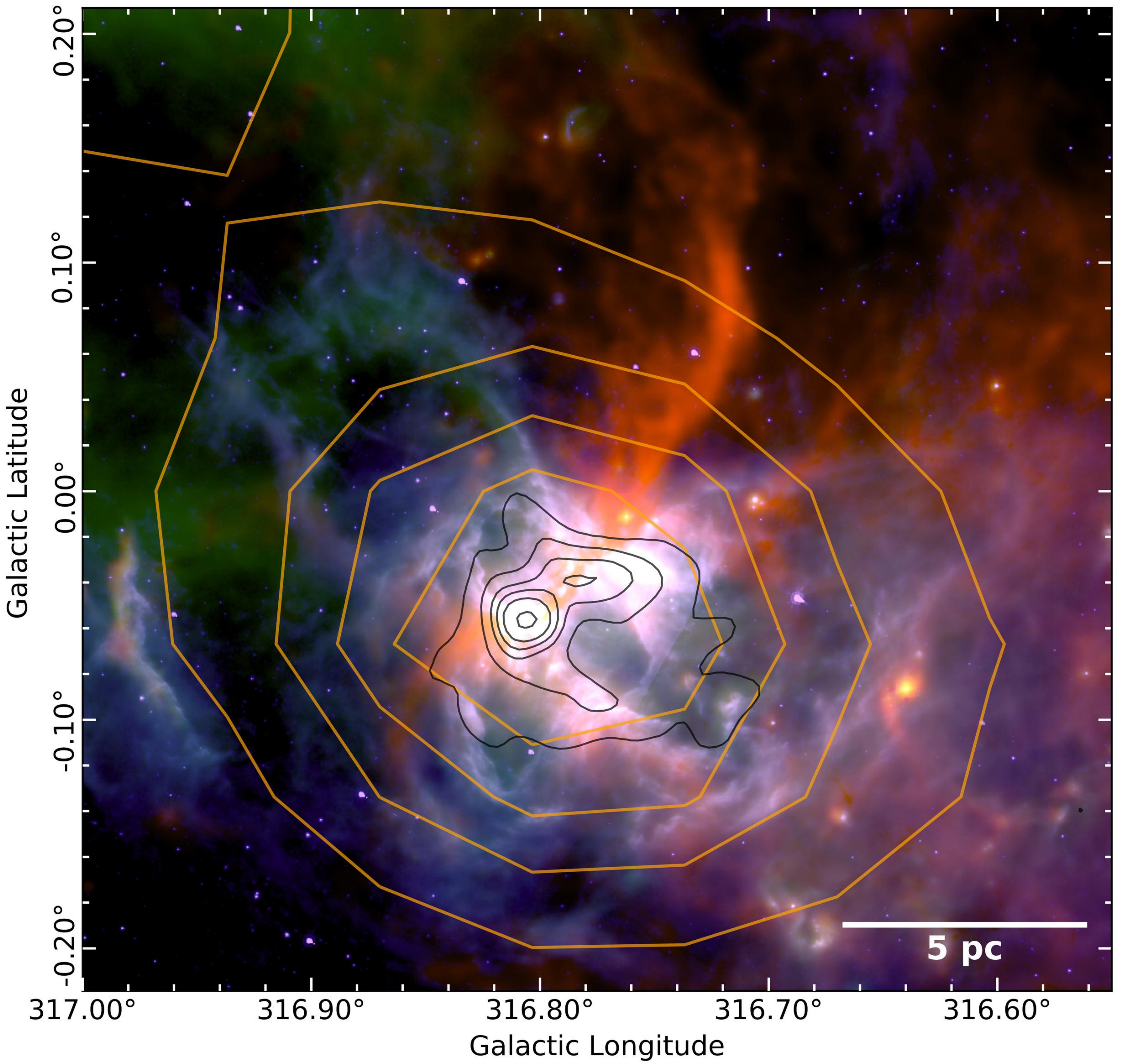}
	\caption{False-colour image of G316.75 made using {\it Spitzer} (\citealt{churchwell_spitzer/glimpse_2009}) and {\it Herschel} (\citealt{molinari_hi-gal:_2010}) observations.  Blue/pink:  {\it Spitzer} 8$\mu$m; Green: {\it Herschel} 70~\micron; Red: {\it Herschel} 250~\micron. The SUMSS  843 MHz radio emission is represented by black isocontours at 1.6 , 8.0, 17.7, 27.3, 40.1 and 64.2 MJy~sr$^{-1}$. The  CHIPASS 1.4~GHz emission is presented by orange isocontours at 1.6, 2.0, 2.3, 2.5 MJy~sr$^{-1}$.  The white line indicates a physical scale of 5~pc.}
	\label{fig:rgb_intro}
\end{figure*}

	In the past two decades, theories and numerical simulations have led to the development of two opposing views regarding the role of stellar feedback in star formation. In one view, stellar feedback provides a source of turbulent energy that helps to maintain giant molecular clouds (GMCs) in a state of quasi-static equilibrium \citep{krumholz_how_2005,krumholz_slow_2007,federrath_origin_2013}.  As a result, GMCs, and substructures within them, would live for tens of free-fall times with quasi-uniform star formation rates per free-fall time ($\epsilon_{\text{ff}}=\frac{m_*}{t_\text{SF}}\frac{t_\text{ff}}{m_\text{gas}}$, where $t_\text{ff}$ is the free-fall time and $t_\text{SF}$ the timescale over which the observed stellar mass $m_{*}$ has formed). Protostellar outflows on small scales  (e.g \citealt{wang_outflow_2010, krumholz_radiation-hydrodynamic_2012}) and supernova explosions on large-scales \citep{padoan_supernova_2016} would be the most likely sources of support against collapse, even though the exact amount of momentum and energy which is effectively deposited into the host clouds is still highly debated (e.g. \citealt{arce_complete_2010,duarte-cabral_molecular_2012,plunkett_assessing_2015,nakamura_observations_2015,maud_distance-limited_2015,drabek-maunder_jcmt_2016,yang_massive_2018,seifried_is_2018}).

	In the opposing view, feedback does not act as a stabilising agent against gravity, star formation is rapid, and clouds only live for a few $t_\text{ff}$ \citep{elmegreen_rapid_2007,dobbs_why_2011,vazquez-semadeni_high-_2009}.  In these models, the main impact of feedback is to disperse most of the cloud's mass via radiative and ionising feedback from recently formed OB stars (\citealt{matzner_role_2002,geen_feedback_2016,kim_modeling_2018,grudic_nature_2018,kuiper_first_2018,geen_indeterministic_2018}) and thereby limits the overall SFE of GMCs to a few percent. However, in the densest parts of the cloud \SFEtff\ could be as high as 30\% \citep{kim_modeling_2018}.
	
	Observational studies investigating stellar feedback have often focused on outflows \citep{arce_complete_2010,duarte-cabral_molecular_2012,maud_distance-limited_2015,yang_massive_2018} and triggered star formation \citep{zavagno_triggered_2007,deharveng_bipolar_2015,liu_alma_2017,lim_kinematic_2018,deb_case_2018}. However, from an observational perspective, measuring the impact of feedback within a given high-mass star-forming cloud is a challenging task since usually one cannot measure what the gas properties were before feedback occurred, nor track the time evolution of these properties. These observational hurdles have always limited the scope of any study on stellar feedback. Yet, in some specific situations, observations of star-forming regions under the influence of stellar feedback can still provide useful information regarding what feedback can or cannot do. For instance, by surveying the W51 high-mass star-forming region, \cite{ginsburg_toward_2016} showed the presence of dense cores in the direct vicinity of exposed young massive stars. They conclude that despite being exposed to large amounts of ionising radiation and powerful winds, these cores continue to form stars almost unperturbed. Other recent studies on stellar feedback such as \cite{rahner_forming_2018,rugel_feedback_2019} suggest that the gravitational potential of some of the most massive star-forming clouds in the local Universe (W49 and 30 Doradus) is deep enough to force initially expanding ionised HII regions to re-collapse and form second generation clusters. Here, we present the analysis of the G316.75 massive-star forming ridge whose morphology and current evolutionary stage allow us to circumvent the hurdle of time evolution.

	The G316.75 ridge (Fig.~\ref{fig:rgb_intro}) is located at a distance of 2.69$\pm$0.45~kpc from the Sun (using the \citealt{reid_trigonometric_2009} Galactic rotation model).  It consists of a 13.6 parsec-long ridge with an extended bipolar HII region emerging from the south end of the ridge.  The region is unique in that the north part of G316.75 is infrared dark and nearly free of young stellar objects, while the south part of the ridge is actively forming high-mass stars.  The infrared dark half of G316.75 has been classified as an infrared dark cloud (SDC316.786-0.044 \citealt{peretto_initial_2009}).  The high column densities and lack of significant 70~$\mu$m emission from this region made it a target to characterise the initial conditions that potentially lead to high-mass star-formation \citep{ragan_earliest_2012,vasyunina_chemistry_2011}. Its association with a bright IRAS source (IRAS14416$-$5937) has attracted a lot of attention to the infrared bright half of G316.75 in the past 20 years, often as part of large surveys investigating the early stages of high-mass star formation \citep{shaver_galactic_1970, shaver_galactic_1970-1,caswell_southern_1987,bronfman_cs2-1_1996,juvela_studies_1996,walsh_studies_1998,walsh_mid-infrared_2001,pirogov_n$mathsf_2$h$^+$10_2003,purcell_ch3cn_2006, pirogov_chemical_2007,longmore_multiwavelength_2007,beuther_ethynyl_2008,longmore_too_2009-1,longmore_ks-band_2009,purcell_h2o_2012,anderson_chasing_2014,caratti_o_garatti_near-infrared_2015,longmore_h2o_2017,samal_bipolar_2018}.  These studies have shown that G316.75 is a very active and young star-forming region, harbouring water, hydroxyl and methanol masers, HII and UCHII regions, a compact x-ray source, and very dynamic gas conditions.  So far, only three published studies focussed on the G316.75 ridge.  \cite{shaver_distance_1981} were the first to confirm that G316.75 is located at the near kinematic distance, and determined that the source of the HII region is likely to be a O6-type star. The second study, \cite{vig_infrared_2007}, estimated that not one, but two O-type stars of mass 45 M$_{\odot}$ and 25 M$_{\odot}$ are responsible for most of the infrared bright luminosity of G316.75 by performing radiative transfer modelling on the dust emission.  Based on 2MASS colour-magnitude and colour-colour diagrams they also conclude that a relative large number of B0 or earlier-type stars are present in the region.  They claim that as many as six of these stars are directly associated with the ridge (though, with no velocity information there is no certainty that they really are part of the ridge).  Finally, \cite{dalgleish_ionized_2018} have studied the kinematics of the ionised gas using radio recombination line emission, and conclude that the strong velocity gradient they observe could be the relics of the cloud's initial angular momentum.

	 The stark differences between the two halves of the G316.75 ridge provide us with the unique opportunity to quantify the exact impact of O-type stars on the gas properties of their host cloud. Indeed, it seems reasonable to assume that the gas properties, and in particular the gas velocity dispersion, within the active part of G316.75 before the formation of high-mass stars must have been very similar to that of the quiescent part. By comparing and contrasting the ridge properties in both halves, we are able to derive robust conclusions on the feedback's ability to destroy the cloud, and the star formation history of the ridge.  We note that this methodology makes the additional implicit assumption that the differences between the two parts of the ridge as observed today (i.e. one is actively forming stars the other one is quiescent) are not due to differences in the initial velocity dispersion of the gas or initial mass-per-unit-length of the two parts of the ridge, but rather a consequence of some asymmetries in the converging flows that led to the formation of the ridge in the first place.

		This paper is structured as followed.  Section \ref{sec:obs} introduces the data available on the G316.75 ridge.  Sections \ref{sec:stars}--\ref{sec:gas_temp} present the results, Section \ref{sec:analysis} analyses the results.  Section \ref{sec:discuss} discusses potential explanations and scenarios that best explain the data and Section \ref{sec:conclude} presents the concluding statements.  

\section{Observations} \label{sec:obs}

This paper makes use of a large set of publicly available data.  In the following we provide details on each of these datasets.  

\subsection{Hi-GAL data}

The Hi-GAL Galactic plane survey is a key project of the {\it Herschel} mission \citep{molinari_hi-gal:_2010}.  The survey mapped $\sim$1$^{\circ}$ above and below the Galactic plane over a 360$^{\circ}$ view at 70~\micron, 160~\micron, 250~\micron, 350~\micron, and 500~\micron\ with resolutions of 7$\arcsec$, 12$\arcsec$, 18$\arcsec$, 24$\arcsec$, and 35$\arcsec$, respectively.  The observations were split into $\sim$2.2 deg$^{2}$ fields. The G316.75 ridge was observed on the 21 August 2010 as part of the Hi-GAL survey of the galactic plane.  The archived data is already reduced with the ROMAGAL pipeline that optimise the {\it Herschel} observations \citep{molinari_hi-gal_2016}.  Hi-GAL observations are here used to determine the \h\ column density and dust temperature structure of G316.75.

\subsection{Mopra southern galactic plane CO survey} 

	 The {\it Mopra} Southern Galactic Plane CO Survey (MSGPCOS) is a millimetre molecular line survey of the Galactic plane \citep{burton_mopra_2013,braiding_mopra_2018}.  The data were taken with the 22~m Mopra radio telescope, located in Australia.  The MSGPCOS survey mapped the $^{12}$CO, \tCO, C$^{17}$O C$^{18}$O and rotational transitions from J=1$\to$0 between \emph{b} = $\pm$0.5$^{\circ}$ and \emph{l} = 270$^{\circ}$--360$^{\circ}$.  The data cubes have an angular resolution of 33$\arcsec$ and span a velocity range of $\sim$200~\kms\ with a spectral resolution of $\sim$0.1~\kms.  The $^{12}$CO and \tCO\ cubes are used in this paper to extract kinematic information about the low and medium density gas respectively.  The \tCO\ data has an rms of 1.3~K/channel and the $^{12}$CO data has an rms of 2.7~K/channel ($T_\text{mb}$).

\subsection{ThrUMMS}

	The THRee-mm Ultimate \emph{Mopra} Milky way Survey (ThrUMMS) is another millimetre molecular line survey of the Galactic plane \citep{barnes_three-mm_2015} taken with the 22~m Mopra radio telescope which mapped the $^{12}$CO, \tCO, C$^{18}$O and CN rotational transitions from J=1$\to$0. These data cover \emph{b} = $\pm$1$^{\circ}$ and \emph{l} = 300$^{\circ}$--360$^{\circ}$ with an angular resolution of 72$\arcsec$, a velocity range of $\sim$150~\kms, and a spectral resolution of $\sim$0.36~\kms.  As a result of its undersampling and lower integration time per pixel, ThrUMMS is superseded by MSGPCOS (the ThrUMMS $^{12}$CO data has an rms of 1.5~K/channel in $T_\text{mb}$ scale). Therefore, we only use the ThrUMMS $^{12}$CO data only when we need to view emission that extends beyond the latitude range covered by MSGPCOS.

\subsection{MALT90} \label{sec:malt90}

	The Millimetre Astronomy Legacy Team 90 GHz (MALT90) survey \citep{foster_millimeter_2011,foster_characterisation_2013,jackson_malt90:_2013} investigates the chemistry, evolutionary and physical properties of high-mass dense-cores at 3~mm.  This survey mapped 16 transitional lines including the cold dense tracer \nhp(1--0).  These data were taken with the Mopra radio telescope by individually observing each of the $\sim$2000 targeted clumps in a $3\arcmin\times3\arcmin$ data cube .  The data cubes have an angular resolution of 38$\arcsec$ and a spectral resolution of $\sim$0.11~\kms\ spanning from -200 to 200~\kms\ velocity range with an rms noise of 0.2~K/channel ($T_\text{A}^{*}$).  Four MALT90 observations were performed towards G316.75, covering most of the ridge but leaves a significant section of the infrared dark part unobserved.  The four corresponding data cubes were mosaicked together using Starlink \citep{currie_starlink_2014} in order to build a single \nhp(1--0) dataset of G316.75.  These data are used here to characterise the kinematics of the dense gas.  

\subsection{HOPS}

	The H2O Southern Galactic Plane Survey (HOPS) \citep{walsh_h2o_2011,purcell_h2o_2012,longmore_h2o_2017} observes 12~mm data including \nhhh\ (1,1) and \nhhh\ (2,2) using the Mopra telescope.  This blind survey was undertaken from \emph{l} = 30-290$^{\circ}$, 0.5$^{\circ}$ above and below the galactic plane, covering G316.75.  The beam size of this survey is $\sim$2.2$\arcmin$ with a velocity resolution of 0.52 ~\kms\ between 19.5-27.5 GHz and a velocity resolution of 0.37~\kms\ between 27.5-35.5~GHz.  The median rms is 0.20$\pm$0.05~K ($T_\text{mb}$). The HOPS data are used to analyse the kinematics of the dense gas.  Given its lower critical density compared to \nhp\ (1--0), the \nhhh\ (1,1) emission is more extended an so probes gas down to lower densities, which are intermediate between \tCO\ (1--0) and \nhp\ (1--0).

\subsection{Radio continuum observations}

\subsubsection{SUMSS survey}

	 The Sydney University Molonglo Sky Survey (SUMSS) \citep{mauch_sumss:_2003} is a southern sky radio continuum survey at 843 MHz measured using the Molonglo Observatory Synthesis Telescope (MOST).  The MOST telescope comprises two cylindrical paraboloids each 778$\times$11.6~m in size.  The telescope has a FWHM resolution of $\sim45\arcsec\times45\arcsec$ cosec|$\delta$| (where $\delta$ is the source declination and a rms flux sensitivity of 6 mJy beam$^{-1}$ at G316.75 declination \citep{mills_molonglo_1981-1}.  The high resolution reconstructed by this interferometer allows us to locate the brightest ionising sources.

\subsubsection{CHIPASS survey}
Continuum HI {\it Parkes} All-Sky Survey (CHIPASS) is a single dish radio continuum and HI survey at frequency of 1.4~GHz observed with the 64~m Parkes telescope \citep{calabretta_new_2014}.  The survey covers declinations of +25$^{\circ}$ with a 14.4\arcmin\ beam and reaches an rms sensitivity of 40~mK ($\sim$6~mJy).  Even though these single dish observations are at low resolution, they recover the extended emission that is lost in the SUMSS interferometric data. We therefore make use of the CHIPASS radio continuum data to estimate the total mass of the stellar cluster.

\begin{figure*}
	\centering
		\includegraphics[width = \textwidth]{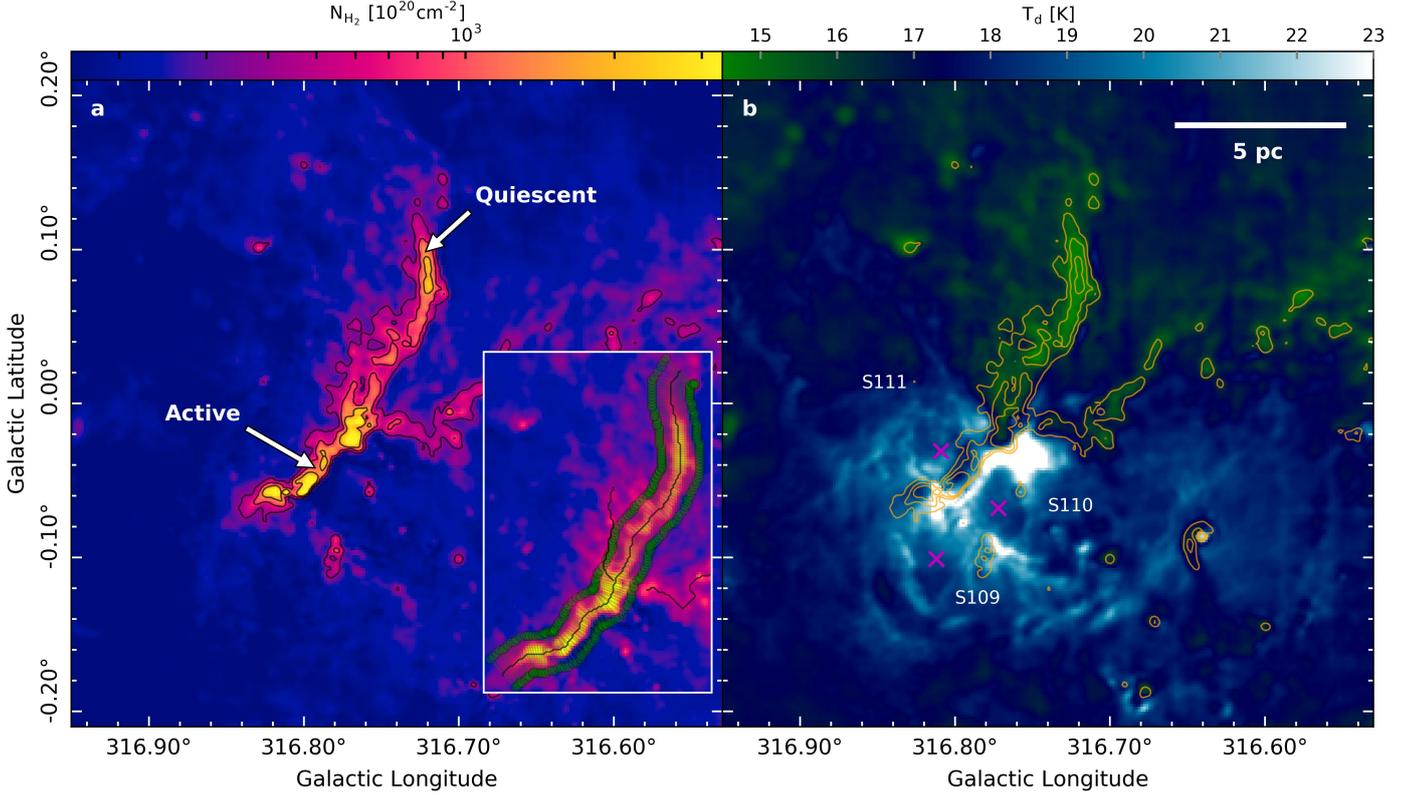}
	\caption{\h\ Column density map {\bf(a)} and dust temperature map {\bf(b)} derived using PPMAP. The black contours on {\bf(a)} and the orange contours on {\bf b} show \h\ column density contours at 400, 800 and 1600 \e$^{20}$~cm$^{-2}$.  Inaxes figure shows the three spines traced by Hessian methods as solid black lines.  Green-dashed lines show the perpendicular cuts along the ridge offset from the spine centre by 0.63~pc.  Theses are used for further analysis.  Labeled arrows point toward the active and quiescent parts of the filament and labeled magenta crosses show the location of three bubbles, S109, S110, and S111, cataloged in \citet{churchwell_bubbling_2006}.  The white solid line indicates a physical scale of 5~pc.}
	\label{fig:temp_col}
\end{figure*} 

\section{Stellar masses and HII region dynamical age} \label{sec:stars}

In order to quantify the impact of the embedded G316.75 stellar cluster on its parent molecular cloud, we first need to compute the cluster mass, and provide constraints on its age.  The stellar mass can be estimated by comparing the amount of ionising photons \Nly\ required to form the G316.75 HII region to stellar model predictions.  We calculate \Nly\ using the following equation \citep{martin-hernandez_high_2005,panagia_radio_1978}:

\begin{equation} 
	N_{ly} = 7.603\times10^{46}\left(\frac{S_\nu}{1 \text{Jy}}\right)\left(\frac{T_e}{1.0\times10^4 \text{K}}\right)^{-0.33}\left(\frac{d}{1 \text{kpc}}\right)^{2} 
	/ b(T_\text{e},\nu)
	\label{eq:Nly}
\end{equation}

\noindent where $S_{\nu}$ is the radio continuum flux density at frequency $\nu$, $T_\text{e}$ is the electron temperature, $d$ is the heliocentric distance to the ridge, and $b(T_\text{e},\nu)$ is given by:

\begin{equation} 
	b(T_\text{e},\nu) = 1 + 0.3195\log_{10}\left(\frac{T_\text{e}}{10^4 \text{K}}\right) - 0.213\log_{10}\left(\frac{\nu}{1 \text{GHz}}\right)
	\label{eq:b_T,nu}
\end{equation}

\noindent This method assumes that the HII region is spherical, optically thin, and homogeneous.  It is also important to note that this method provides only a lower limit to the Lyman-alpha photons therefore a lower limit on the stellar masses as this method assumes that all the Lyman-alpha photons emanating from the stars are used up to form the HII region, while some might either escape the HII region or be absorbed by dust grains \citep{binder_multiwavelength_2018}.  In the following calculations we use T$_\text{e}$ = 6600~K for the electron temperature of the HII region, as estimated in \cite{shaver_galactic_1983}.

	First, we use the CHIPASS single dish data.  As we can see in Fig.~\ref{fig:rgb_intro}, the CHIPASS observations of G316.75 detect extended emission that matches very well the morphology of the bipolar HIII region seen at 8 \micron.  By integrating the intensity over the entire HII region, we estimate a total flux density of 61.3$\pm$4.0~Jy, which, according to Eq.~(\ref{eq:Nly}), corresponds to $\log_{10}$(\Nly $[s^{-1}]) =49.63\pm0.11$.  Rather than attributing this flux density to one dominant ionising object, we calculated the cluster mass needed to reproduce this emission using the following equation from \citet{lee_observational_2016}:

\begin{equation} 
	M_* = 1.37N_\text{ly}\times1.6\times10^{-47} \ M_\odot
	\label{eq:Nly_mass}
\end{equation}
 
\begin{table*}
	\centering
	\begin{threeparttable}
	\caption{Table containing properties describing the clumps within G316.75.}
	\begin{tabular}{cccccc c c c c} 
		\hline
		ID \#  &  $l$               &  $b$             &  Area\tnote{\emph{a}}        & $R_\text{eff}$\tnote{\emph{b}} & $\lambda_\text{sep}$\tnote{\emph{b}} & $\overline{N}_{\text{H}_\text{2}}$  & Mass\tnote{\emph{a}}      & $\overline{T}_\text{Dust}$ & Notes \\
		          & ($^{\circ}$)  & ($^{\circ}$)  & (pc$^2$)  & (pc)                & (pc)     		       &  (\e$^{23}$~cm$^{-2}$)  	           & (\msun) & (K)                                    &           \\ 
		\hline
		1   & 316.81785  & -0.05777 & 0.21 & 0.26 & 0.31 & 2.41 & 820   & 19.6$\pm$1.3 & In active half     \\
		2   & 316.81119  & -0.05777 & 0.03 & 0.01 & 0.31 & 2.03   & 87       & 28.9$\pm$0.7 & In active half       \\
		3   & 316.79952  & -0.05611 & 0.35 & 0.33 & 0.55 & 2.51 & 1507& 19.8$\pm$1.6 & In active half       \\
		4   & 316.78785  & -0.03944& 0.19 & 0.25 & 0.96 & 1.44& 357   & 19.8$\pm$0.5 & In active half       \\
		5   & 316.76784  & -0.02443 & 0.17 & 0.23 & 0.65 & 3.18 & 748   & 16.8$\pm$0.8 & In active half       \\
		6   & 316.75784  & -0.01443 & 0.04 & 0.12 & 0.35 & 1.36  & 71      & 18.4$\pm$0.2 & In active half       \\
		7   & 316.76451  & -0.01110 & 0.09 & 0.17 & 0.35 & 3.13& 404   & 17.0$\pm$1.2 & In active half       \\
		8   & 316.74451  & 0.02891 & 0.29 & 0.31 & 0.72 & 0.75 & 311  & 15.0$\pm$0.3 & In quiescent half     \\
		9   & 316.74117  & 0.04391 & 0.09 & 0.17 & 0.72 & 0.65  & 72      & 14.9$\pm$0.1 & In quiescent half     \\
		10 & 316.72450  & 0.05391  & 0.06 & 0.14 & 0.91 & 0.93  & 71      & 14.6$\pm$0.1 & In quiescent half     \\
		11 & 316.71950 & 0.07725  & 0.10 & 0.18 & 0.47 & 1.78  & 225   & 13.5$\pm$0.2 & In quiescent half     \\
		12 & 316.71950 & 0.08725  & 0.08 & 0.16 & 0.47 & 1.69  & 166   & 13.5$\pm$0.1 & In quiescent half     \\	
	
		\hline
	\end{tabular} \label{tab:core_mass}
	\begin{tablenotes}
      \footnotesize
      \item[\emph{a}]{Error propagated from the distance error and is equal to a fractional error of 1/3}
      \item[\emph{b}]{Error propagated from the distance error and is equal to a fractional error of 1/6}
      \item[]{ ID \# corresponds the the clump identification number from the most southernly position (see Fig.~\ref{fig:col_temp_str} and \ref{fig:velo}a-c). $l$ and $b$ are the galactic coordinates of the peak intensity of the clump. Area is calculated from the number of pixels contained within the isocontour extracted using \texttt{astrodendro}. $R_\text{eff}$ is the effective radius, $R_\text{eff}=\sqrt{\text{Area}/\pi}$. $\lambda_\text{sep}$ is the minimum separation to the neighbouring clump. $\overline{N}_{\text{H}_\text{2}}$ is the mean column density contained within the \texttt{astrodendro} isocontour for each clump. Mass is the average of the bijective and clipped mass contained within the \texttt{astrodendro} isocontour for each clump.  $\overline{T}_\text{Dust}$ is the mean {\it Herschel} dust temperature within the clumps isocontour and Notes specifies whether the clump is located in the active or quiescent region.}
   \end{tablenotes}
   \end{threeparttable}
	 
\end{table*}

\noindent where 1.6\e$^{-47}$ is the normalisation factor for the amount of ionising photons needed to power the HII region for a star cluster distributed according to a modified Muench initial mass function (IMF) \citep{muench_luminosity_2002,murray_star_2010} and 1.37 accounts for dust absorption.  Equation (\ref{eq:Nly_mass}) assumes that the IMF is fully sampled.  For $\log_{10}$(\Nly$[s^{-1}]) = 49.63\pm0.11$, we obtain a stellar cluster mass $M_\text{cl}=930\pm230$~\msun, that includes four O-type stars one of which has a mass larger than 48~\msun.  This number of high-mass stars is comparable to the six B0 stars, or earlier, identified by \cite{vig_infrared_2007} using near-infrared colour-colour diagrams.  We used the SUMSS observations of G316.75 to identify the location of the strongest ionising sources.  These observations resolve the HII region into two separate radio continuum peaks (see Fig.~\ref{fig:rgb_intro}).  Assuming that this emission is associated with two dominant ionising sources, we measure radio flux densities of 23$\pm$2 and 12$\pm$1~Jy, corresponding to $\log_{10}$(\Nly $[s^{-1}]) =49.1\pm0.1$ and $\log_{10}$(\Nly $[s^{-1}]) =48.8\pm0.1$, respectively.  Depending on the stellar model used \citep{panagia_physical_1973,vacca_lyman-continuum_1996,sternberg_ionizing_2003}, we estimate that the corresponding two high-mass stars have spectral types between O8.5 and O7 V with a stellar mass of 28-38~\msun, and O6.5 and O6 V with stellar mass 34-55~\msun\ for the small and large intensity peaks respectively.  These stellar masses are consistent with the cluster mass estimated using the CHIPASS data.

	Finally, the dynamical age of the HII region can be estimated by calculating how long it takes for an HII region of internal pressure $P_\text{I}$ to expand into a turbulent molecular cloud of pressure $P_\text{turb}$ up to the observed HII region radius. Using numerical simulations,
\cite{tremblin_age_2014} constructed isochrones of expanding HII regions as a function of their radius, pressure, and ionisation rate.  According to these models, a G316.75-like HII region with a radius of 6.5~pc and \Nly\ = $10^{49.63} s^{-1}$ is $\sim2$~Myr old. The lifetime for O6 stars is $\sim4$~Myr \citep{weidner_masses_2010}, consistent with the estimated dynamical age of the HII region.  It also indicates that in a couple of Myr a supernova explosion should occur in G316.75, which might drastically change the star formation history of the ridge.

\begin{figure}
	\centering
		\includegraphics[width = \columnwidth]{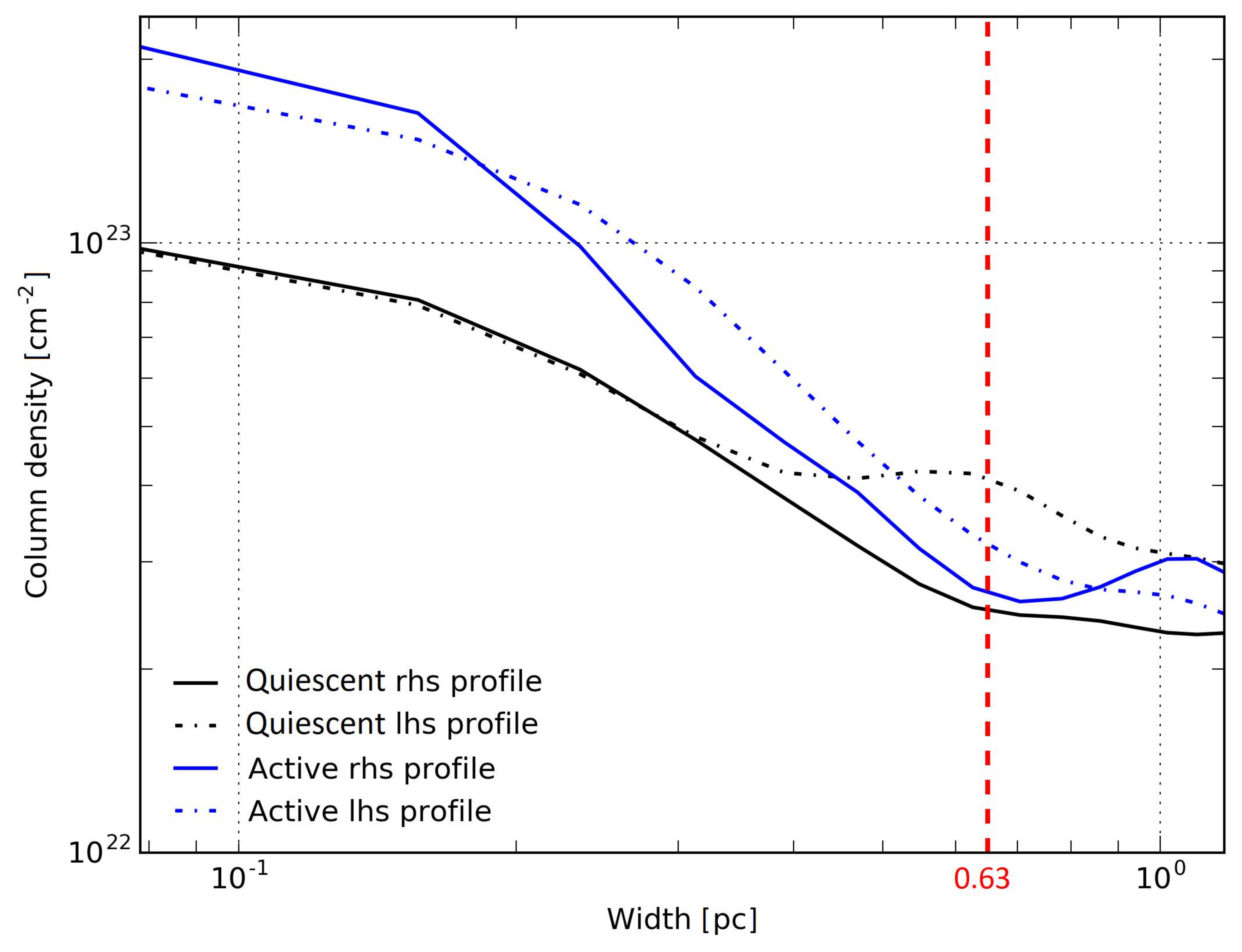}
		\caption{Mean transverse \h\ column density profiles averaged in the longitudinal direction from  0 and 6~pc for the active region, and from 6 to 13.6 ~pc for the quiescent region. The left-hand-side (lhs) and right-hand-side (rhs) profiles denotes the eastern (i.e. from 0 to -0.63~pc) and western ( from 0 to +0.63~pc) offsets from the spine centre. The red dashed line marks the position of a power-break that fits best the four profiles. We interpret this as the transition between the compact ridge and the more diffuse material surrounding the ridge. }
	\label{fig:1d_col_log}
\end{figure} 

\begin{figure*}
	\centering
		\includegraphics[width = \textwidth]{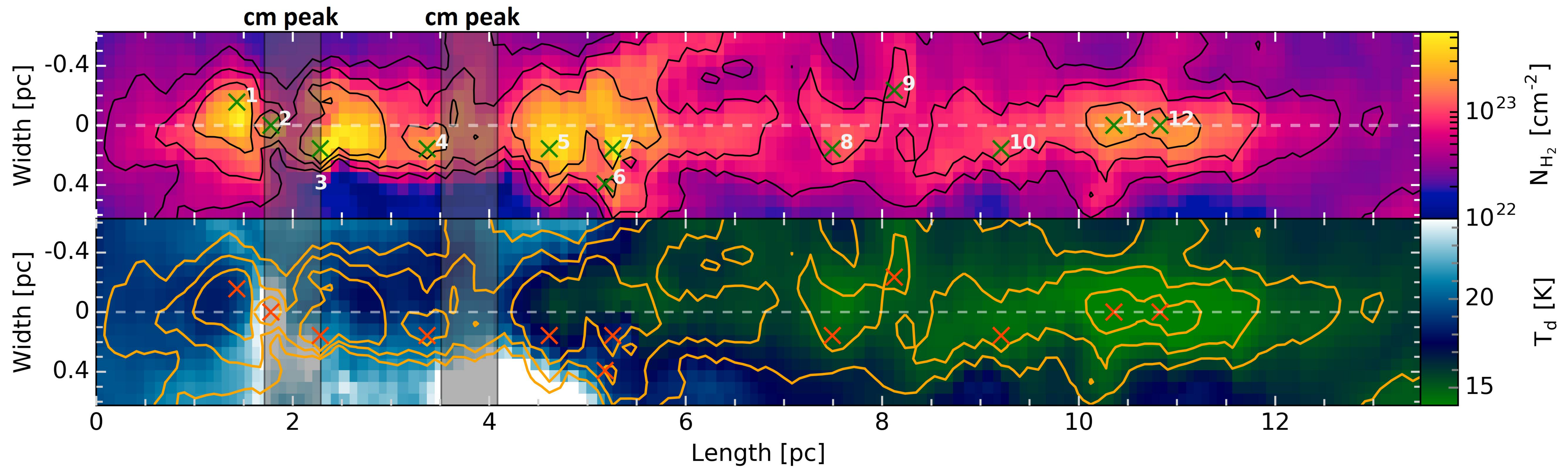}
		\caption{Straightened \h\ column density map {\bf top} and dust temperature map {\bf bottom} with a width of $2\times0.63$~pc and length of 13.6~pc. Negative values along the y axis corresponds to the eastern offset from the spine centre, while along the x axis 0~pc corresponds to the most southern position of the spine. The black contours on {\bf(a)} and the orange contours on {\bf b} show \h\ column density contours at 0.4, 0.8 and 1.6 \e$^{23}$~cm$^{-2}$. Green and orange crosses mark the locations of the peak column density for each clump along with their ID \#.  White-dashed line shows where the centre of the spine is.  The grey-shaded regions labeled 'cm peak' correspond to the radio continuum peaks seen in SUMSS (see Fig.~\ref{fig:rgb_intro}.}
	\label{fig:col_temp_str}
\end{figure*} 

\section{\h\ column density and dust temperature maps} \label{sec:ppmap}

	The dust temperature and \h\ column density  maps (Fig.~\ref{fig:temp_col}) were derived using PPMAP \citep{marsh_temperature_2015} on all five {\it Herschel} images, resulting in an angular resolution of 12$\arcsec$ with a pixel size of 6$\arcsec$, which corresponds to 2 pixels along the beam.  PPMAP is a multi-wavelength Bayesian method that derives the differential \h\ column density for a set of temperature bands (in the case of G316.75 we used 12 temperature bands from 8 to 50~K) from dust emission, by assuming optically thin emission and using a constant dust emissivity law $\kappa_{\lambda}=0.1\times\left(\frac{\lambda}{300\mu\text{m}}\right)^{-2}$~cm$^2$/g \citep{hildebrand_determination_1983}.  For the purpose of calculating \h\ column densities we used a mean molecular weight $\mu$=2.8. On simulated data, PPMAP reproduced the column densities a factor of two times better than traditional SED fitting methods. Here, the uncertainty in the model selection is only few percent meaning the uncertainty in column density is still dominated by the dust emissivity law \citep{roy_reconstructing_2014}.  For a full account of this method see \cite{marsh_temperature_2015}.  Consequently we have adopted this method over traditional single-temperature SED techniques not only for its potential to better characterise column densities, but because PPMAP offers the higher resolutions that are crucial for investigating the impact feedback has at smaller scales.  

\begin{figure*}
	\centering
		\includegraphics[width = \textwidth]{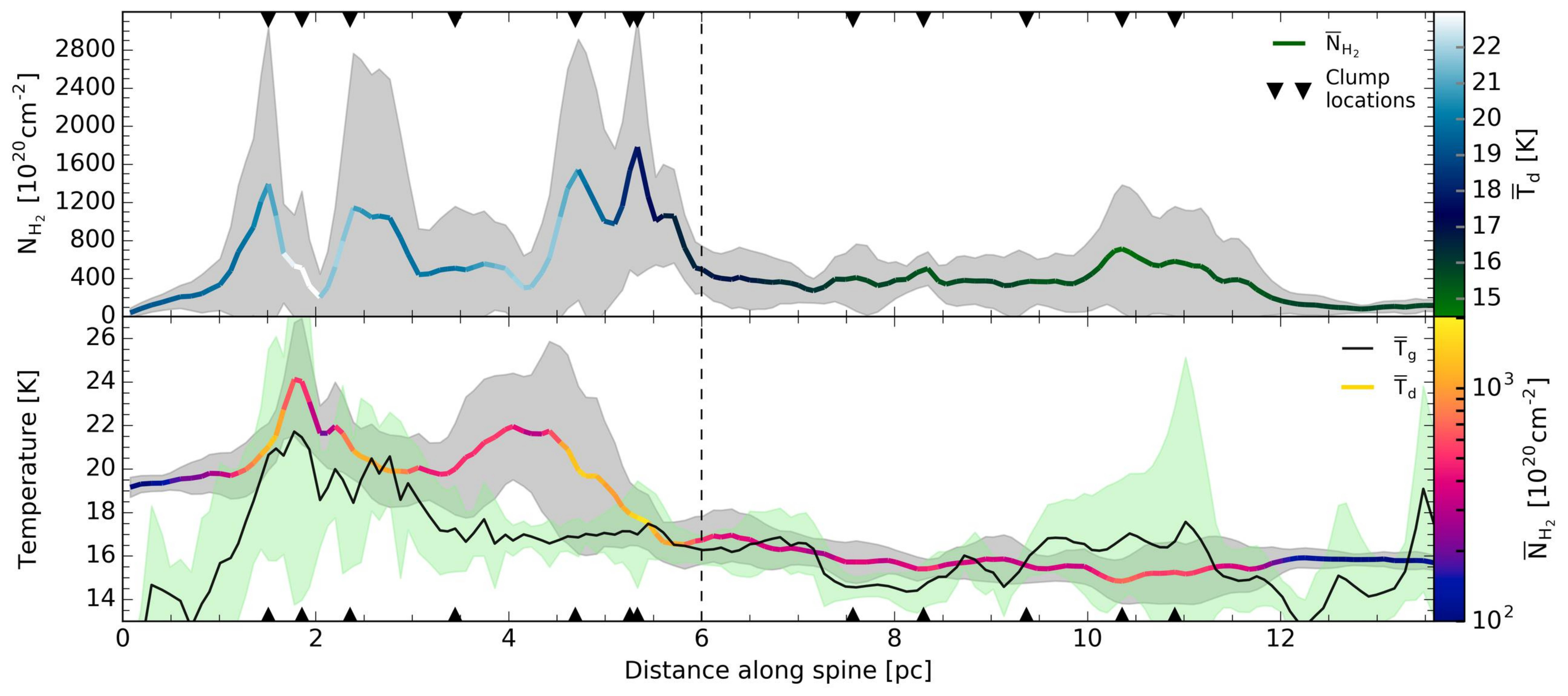} 
	\caption{Background subtracted longitudinal column density {\bf(top)}, and longitudinal dust temperature (solid coloured line) and the gas temperature (solid black line) {\bf(bottom)} between 0 and 13.6~pc and averaged over $2\times0.63$~pc (i.e. 0.63~pc either side of the spine).  The thin black vertical dashed line marks the separation between the active half of the ridge (left) and the quiescent half (right). The colour of the solid line in the {\bf top} panel corresponds to the dust temperature longitudinal profile of the {\bf bottom} panel and the colour of the dust temperature line in the {\bf bottom} panel corresponds to the \h\ column density longitudinal profile in the {\bf top} panel.  Black triangles mark the clump positions and the translucent regions shows the 1-sigma uncertainty range for the values.}
	\label{fig:temp_col_height}
\end{figure*} 

	The differential column density cube created by PPMAP (see Appendix~\ref{sec:PPMAP}) was collapsed to output the total \h\ column density (Fig.~\ref{fig:temp_col}a) and the mean \h\ column density weighted dust temperature (Fig.~\ref{fig:temp_col}b).  These figures reveal that the ridge is 13.6~pc long (assuming zero inclination towards our line of sight for the entire region).  Focusing on the ridge, we can already see some basic differences between the quiescent and active regions as the latter exhibits higher \h\ column density and dust temperatures compared to the quiescent region.  Typically, the \h\ column density and dust temperature are anti-correlated in both the active and quiescent regions.  The main exception to this anti-correlation is near the SUMSS radio peak, a location that contains one of the young ionising high-mass stars (see Sec.~\ref{sec:stars}).  Here, both the dust temperature and \h\ column density increase together.  Additional notable features are three warm bubbles of gas swelling from the active region in the dust temperature map, resembling the same features in 8~\micron\ Spitzer (see Fig.~\ref{fig:rgb_intro}).  We can also see that some diffuse dust emission around the active part reaches temperature of 50~K (see Appendix ~\ref{sec:PPMAP}).

	The \h\ column density map also shows that the ridge is surrounded by diffuse emission from the galactic plane. This background material is composed of a large number of unrelated diffuse clouds that are accumulated along the line of sight across the entire Galaxy, and consequently, can reach large column density values \citep[e.g.][]{peretto_mapping_2010,peretto_initial_2016}. At $\sim0.23\pm0.03$~\tcm, this {\it Herschel} background becomes comparable to the extended emission from the ridge itself (see Sec~\ref{subsec:hessian}). Below this value, it is impossible to disentangle what fraction of the diffuse dust emission is associated to the ridge itself. We therefore  used this background column density value of 0.23$\pm0.03$~\tcm\ to define the borders of the dense inner part of the ridge and removed it from any mass calculations. As a result, we estimate that G316.75 has a total gas mass of 18900$\pm$6500~\msun\ of which 11200$\pm$3800~\msun\ (59\%) is located in the active region and 7700$\pm$2700~\msun\ (41\%) in the quiescent.  These masses were calculated by averaging the clipped and bijective masses \citep{rosolowsky_structural_2008} within the first isocontour located three standard deviations above 0.23~\tcm.  The relatively high background column density underestimates the ridge mass but ensures that measured properties are those of the dense gas. The mass uncertainties cover the spread resulting from the clipped and bijective schemes propagated with the distance error.  The same isocontour was used to calculate a mean dust temperature of 19.6$\pm$2.3~K in the active region and 15.6$\pm$0.8~K in the quiescent region.  The quoted uncertainties measure the spread of dust temperatures for the active and quiescent regions.

\subsection{Clump identification} \label{sec:dendro}

\begin{figure}
	\centering
		\includegraphics[width = \columnwidth]{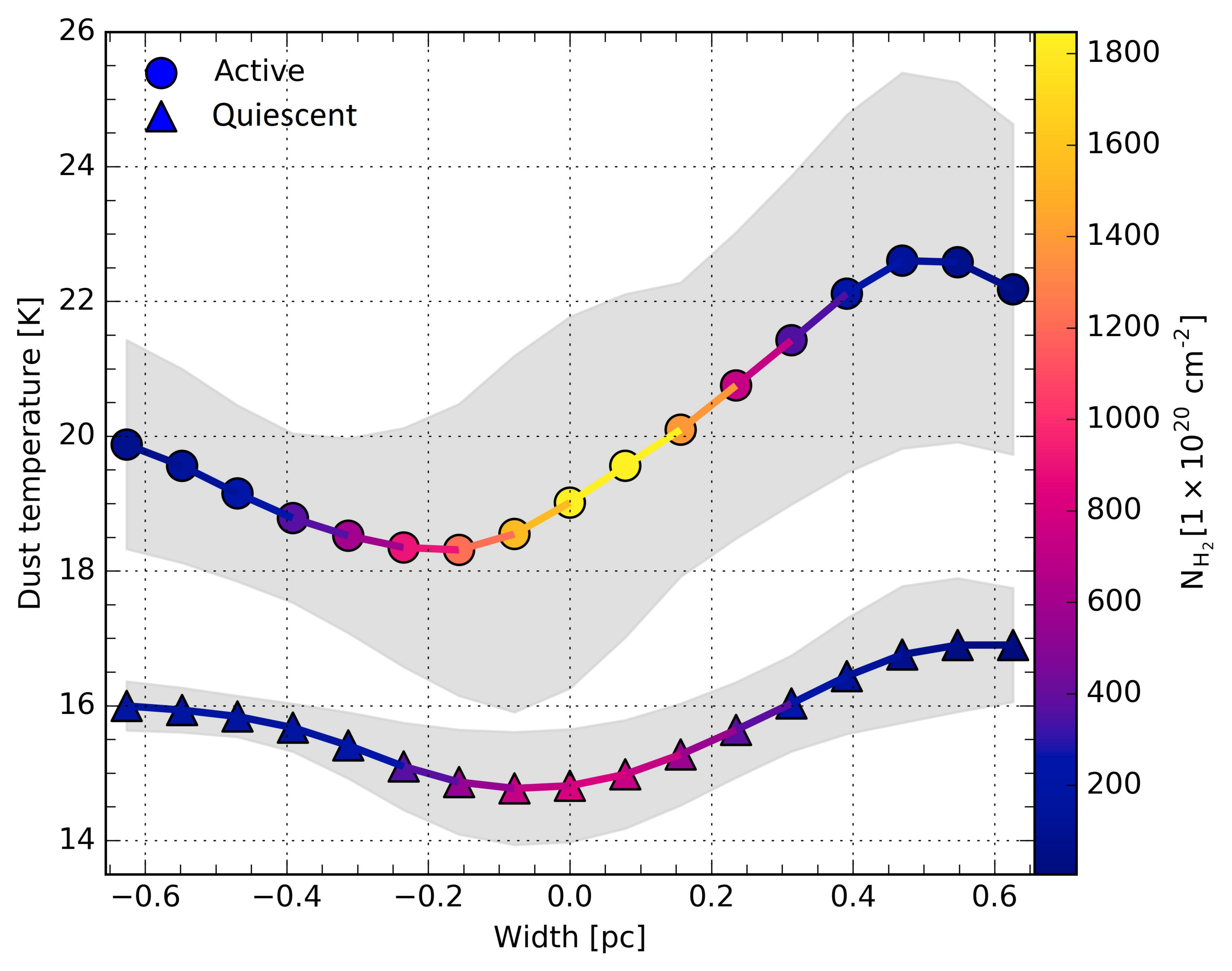}
	\caption{Mean transverse temperature profile averaged over the same limits as Fig.~\ref{fig:1d_col_log}. The colour of solid line denotes the mean {\it Herschel} \h\ column density profile where the circle and triangle markers show the dust temperature profile for the active and quiescent regions respectively.  The translucent region shows the 1-sigma uncertainty range for the dust temperature.}
	\label{fig:1d_col_with_temp}
\end{figure}

\begin{figure*}
	\centering
		\includegraphics[width = \textwidth]{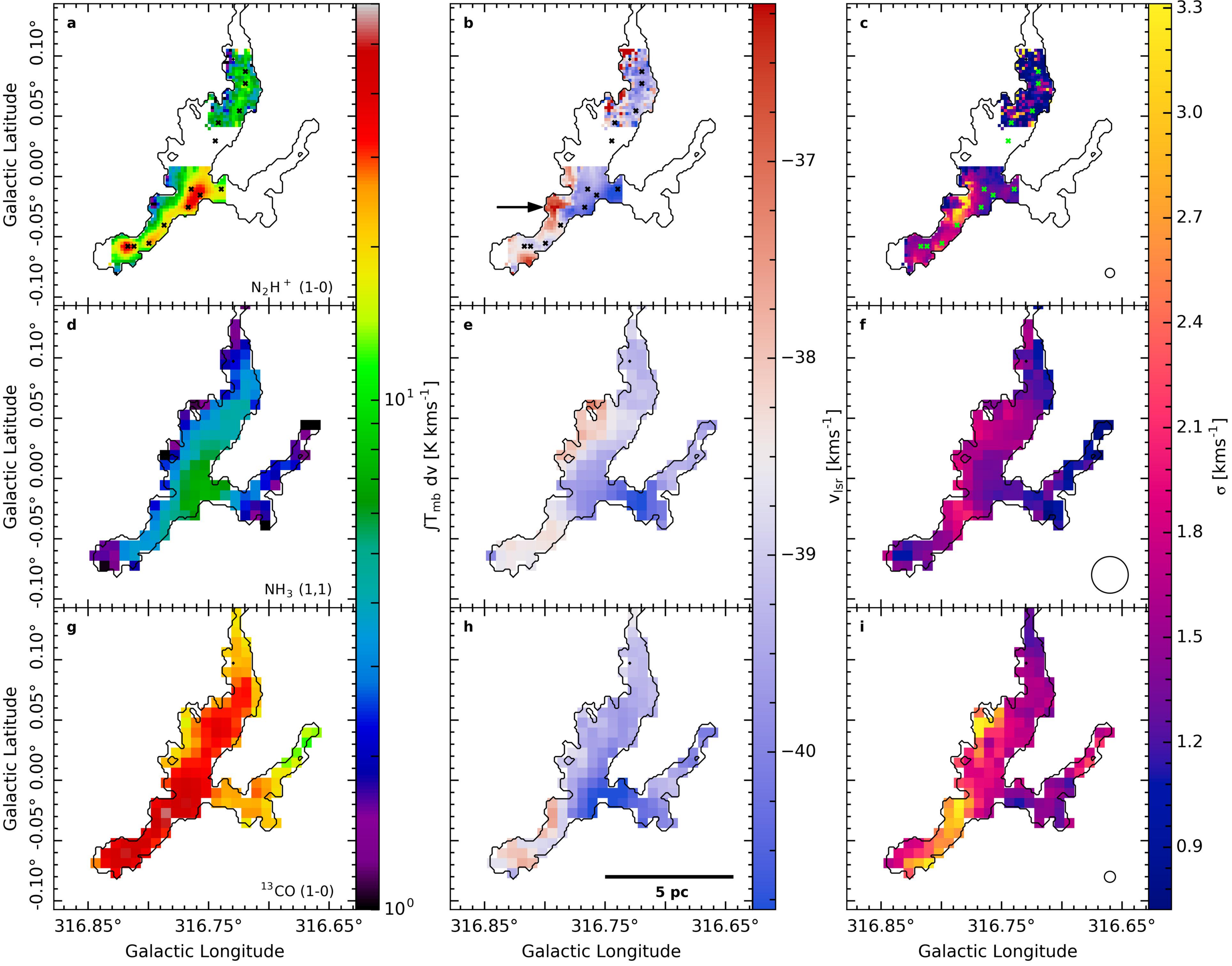}
		\caption{Molecular transition analysis for \nhp\ (J=1--0) {\bf(top)}, \nhhh\ (1,1) {\bf(middle)} and \tCO\ (J=1--0) {\bf(bottom)}.  First column shows the integrated intensity between -42.5 to -33~\kms.  The second column shows the radial velocity and the third column shows the velocity dispersion. The over-plotted contour corresponds to an \h\ column density of 0.312~\tcm\ (see Sec.~\ref{sec:dendro}).  Each dataset has been masked according to this contour and is used to estimate the properties of G316.75 in Table \ref{tab:velo split}. The molecular transition used has been labeled in the bottom right of of the first column. The open black circle in the last column shows the FWHM beam size of the observations.  The crosses in the first row show the positions of the clumps and the black arrow in {\bf b} show the location and the direction along which the \nhp\ spectra presented in Fig.~\ref{fig:nhp_2_comp} were taken.}

	\label{fig:velo}
\end{figure*}

	The \h\ column density map reveals a number of local peaks clumps.  To identify these clumps and estimate their masses, we used the python package \texttt{astrodendro} \citep{rosolowsky_structural_2008}.  Dendrograms trace the morphological hierarchy of an image based on isocontours.  In this case these are \h\ column density isocontours.  The three basic parameters defining a dendrogram are the value of the lowest isocontour, the lowest amplitude of a significant structure between its local maximum and minimum, and the minimum size of a significant structure.  The dendrogram we used has a lowest isocontour of 0.23~\tcm, with a minimum difference of 0.2~\tcm.  The smallest size considered significant was based of the PPMAP resolution of 12\arcsec, resulting in a minimum of five pixels.  This dendrogram extracts seven clumps in the active region, five in the quiescent region and three in the eastern filament.  The clump properties are given in Table~\ref{tab:core_mass} and are numbered from 1 to 12 from south to north.  We over-plotted their identified location in Fig.~\ref{fig:col_temp_str} and \ref{fig:velo}a-c.  The three clumps identified in the western filament are mentioned for completeness but their properties are not presented since we do not analyse them further.  As for the ridge mass, the clump masses provided in Table~\ref{tab:core_mass} are calculated by averaging the clipped and the bijective masses.  We also calculate the minimum separation distance, $\lambda_\text{sep}$ for each clump and provide these distances in Table~\ref{tab:core_mass}.

\subsection{Ridge tracing} \label{subsec:hessian}
\begin{table*} 
	\centering
	\small
	\begin{threeparttable}
	\caption{Average properties the active and quiescent parts of the ridge within the 3.12~\tcm\ \h\ column density contour (see Sec.~\ref{sec:dendro}).}
	\begin{tabular}{l r c c c c c c c}
		\hline
		Region  &  Area\tnote{\emph{a}}  &  Length\tnote{\emph{b}}  & Width & $\overline{N}_{\text{H}_\text{2}}$  & $M_\text{ridge}$   & $\overline{T}_\text{d}$ & $\overline{\lambda}_\text{sep}$    & $\overline{M}_\text{clump}$   \\
			     &  (pc$^2$)   & (pc)        & (pc)    & (\e$^{23}$~cm$^{-2}$)                     &(\msun)           & (K)                      & (pc)                                        &  (\msun) \\
		\hline
		Active   & 5.86   & 6.0 & 0.63$\pm$0.08 & 0.85 & 11200$\pm$3800 & 19.6$\pm$2.3 & 0.50$\pm$0.22 & 570$\pm$470 \\
		Quiescent & 10.08 & 7.6 & 0.63$\pm$0.08 & 0.34 &   7700$\pm$2700 & 15.6$\pm$0.8 & 0.66$\pm$0.17 & 170$\pm$90\\
		\hline
	
	\end{tabular} \label{tab:split}
	\begin{tablenotes}
      \footnotesize
      \item[\emph{a}]{Error propagated from the distance error and is equal to a fractional error of 1/3}
      \item[\emph{b}]{Error propagated from the distance error and is equal to a fractional error of 1/6}
   \end{tablenotes}
   \end{threeparttable}

\end{table*}

\begin{figure*}
	\centering
		\includegraphics[width = \textwidth]{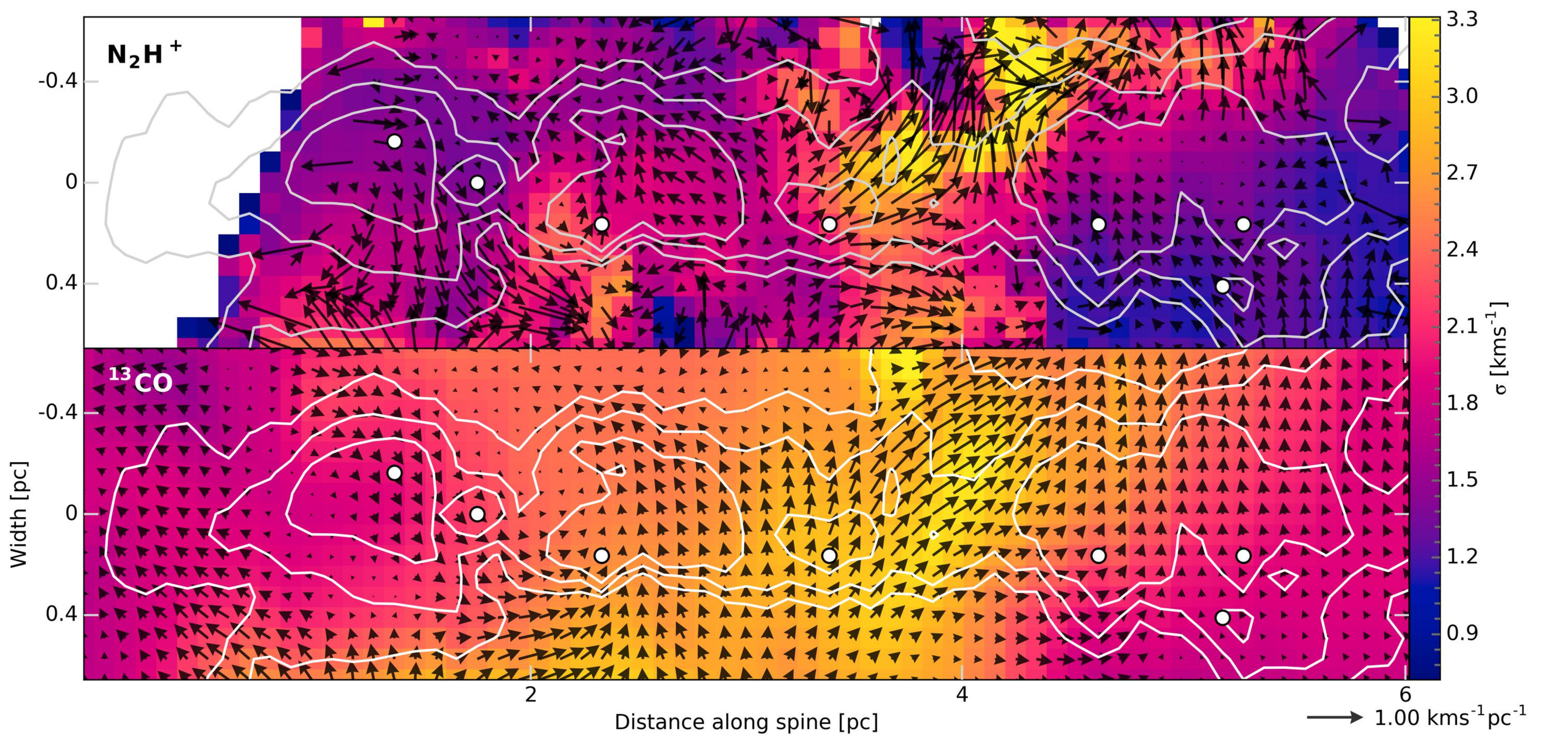}
	\caption{Two panels showing the straightened velocity dispersion \nhp (J=1--0) {\bf(top)} and \tCO (J=1--0) {\bf(bottom)}. The length and width and contours used are identical to Fig.~\ref{fig:col_temp_str}. Over-plotted on each panel are the velocity gradient vectors calculated across a beam size for each corresponding data set. The magnitude scale of the gradient vectors is indicated at the bottom right of the plot. The white dots denote the positions of the clumps.}
	\label{fig:straight kin}
\end{figure*} 

	A Hessian method was used to extract the filament crests and the perpendicular angles of the crests.  This method traces ridges by diagonalising the Hessian (i.e second derivative) matrix of the \h\ column density map (see \citealt{schisano_identification_2014,williams_gravity_2018,orkisz_dynamically_2019}) to determine the two eigenvalues, $\varepsilon_1$ and $\varepsilon_2$, and two eigenvectors, $\bm{u_1}$ and $\bm{u_2}$, for each pixel of the map.  Filaments are regions of the map where at least one of the eigenvalues, for instance $\varepsilon_1$, is negative with the condition that $|\varepsilon_1| > |\varepsilon_2| $ (i.e. the curvature along the $\bm{u_1}$ direction is larger).  Since the goal here is to obtain a reliable skeleton of the G316.75 ridge, and not identify the full filament network of the region, we first convolved the \h\ column density image to reduce noise and then normalised the eigenvalues between -1 to 1 to identify regions where $\varepsilon_1 <-0.15$.  This ensures only the most prevalent structures are selected.  This map was then binarised and then skeletonised using a medial axis transform and the remaining map identified three filaments in the vicinity of G316.75: the G316.75 ridge; a small filament to the east of the main ridge; and a third spine west of the main region (Fig.~ \ref{fig:temp_col}).  The third filament does not connect to the main ridge due to an absence of \h\ column density on the right side of the ridge.

	The advantage of using Hessian methods for ridge tracing is that not only does it trace the spine from the eigenvalues, but the eigenvectors trace the direction parallel and perpendicular to the spine (see Fig.~\ref{fig:temp_col}).  One can therefore easily compute the transverse dust temperature and \h\ column density profiles for any position along the ridge skeleton by interpolating the maps across the perpendicular direction to the spine. Using these individual transverse profiles, we first computed the mean \h\ column density profile for the active and quiescent regions (see Fig~\ref{fig:1d_col_log}). We find that these profiles have a power-break at $\sim$0.23~\tcm at a distance of $\sim$0.63~pc from the spine. This is the point where emission from the galactic plane becomes comparable to the extended emission from the ridge (see Sec.~\ref{sec:ppmap}). As a result, we adopt the latter as the radius of the dense part of the ridge. In Fig.~\ref{fig:col_temp_str}, we plot the interpolated dust temperature and \h\ column density profiles of the ridge as straightened projections. These provide a new perspective of the ridge and show a clear view of the correlation/anti-correlation of column density and temperature.  One particular feature to notice is the decrease of column density in combination with strong temperature peaks around clump \#3 and between clump \#4 and \#5.  These locations are coincident with the two radio continuum peaks observed in SUMSS observations, (see Fig.~\ref{fig:rgb_intro}) which strongly suggests that here feedback from embedded O-stars are impacting the ridge.  This feature is also nicely visible in the longitudinal profile of the ridge (see Fig.~\ref{fig:temp_col_height}).  On the other hand, the temperature and column density profiles of the quiescent part of the ridge do not show any sign of embedded star formation activity.  Another feature to note is that the temperature minimum of the active part of the ridge is offset with respect to the ridge spine, unlike the \h\ column density maximum (see Fig.~\ref{fig:1d_col_with_temp}).  This shows that the temperature offset is a genuine physical effect and not the result of the skeletonisation process, which would offset the \h\ column density maximum and temperature minimum equally.  

\begin{figure}
	\centering
		\includegraphics[width = \columnwidth]{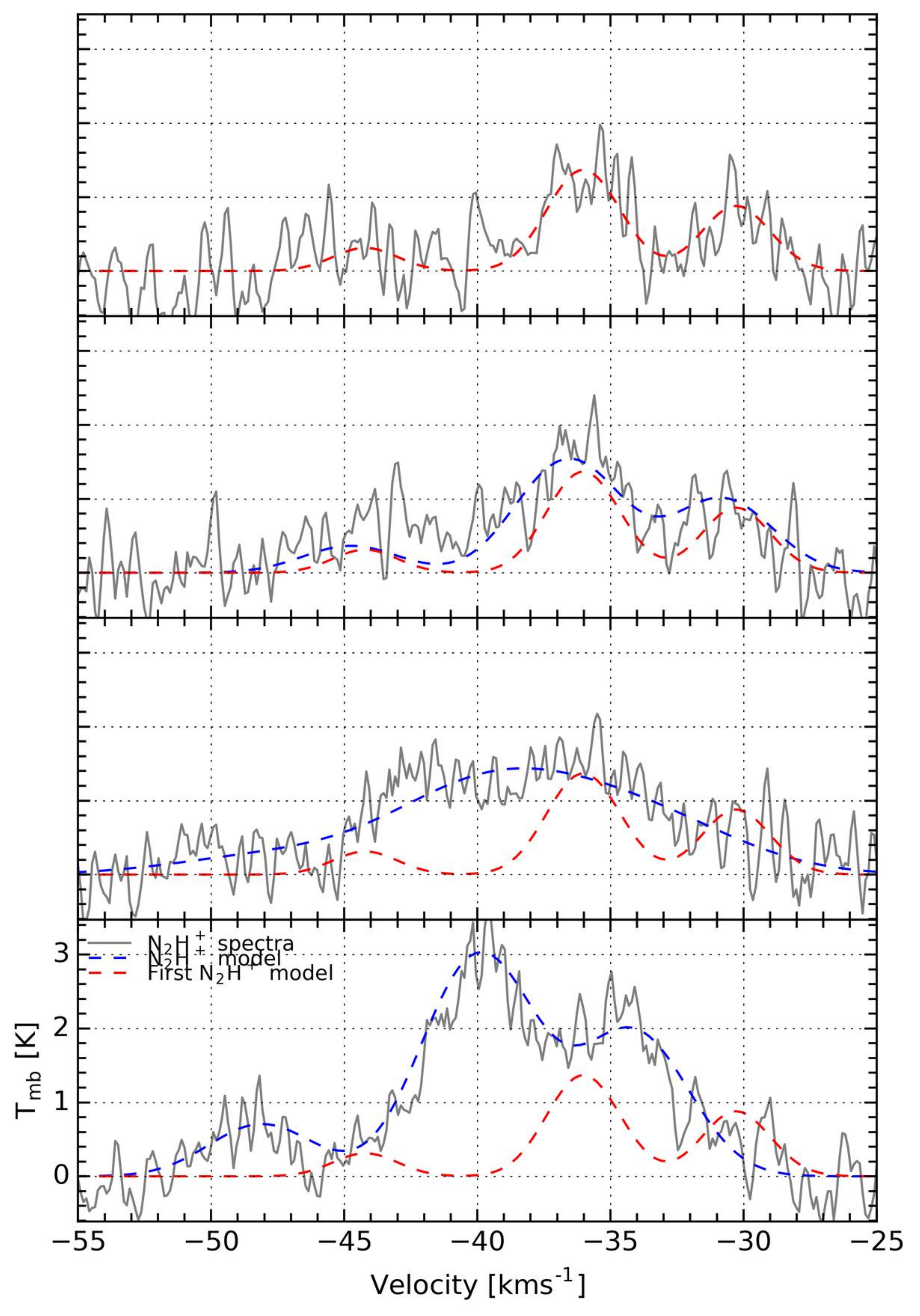}
	\caption{Four vertical panels showing \nhp\ (J=1--0) spectra and corresponding fitted models at four different positions from left to right as indicated by the arrow in Fig.~\ref{fig:velo}b.  The solid grey line are the observed spectra at each location.  The red dashed line corresponds to the best fit model of the top panel spectrum, which has then been over-plotted on the bottom three panels for reference. The dashed blue lines correspond to the best fit models to the bottom three spectra. All fits have been calculated using GILDAS.} 
	\label{fig:nhp_2_comp}
\end{figure} 

\section{Kinematics and gas temperature} \label{sec:gas_temp}

	The main goal of the present study is to evaluate the impact of OB stars on their parent cloud.  However, the effect that feedback has on the surrounding gas is most likely a function of its density \citep{thompson_sub-eddington_2016}.  Here, we analysed line data from four different molecules, each tracing relatively different gas density regimes: $^{12}$CO (1--0) and \tCO\ (1--0) trace gas at $\sim 1\times10^2$~cm$^{-3}$; \nhhh\ (1,1) traces gas at $\sim 1\times10^3-10^4$~cm$^{-3}$; and \nhp\ (1--0) traces gas at $\sim 1\times10^4-10^5$~cm$^{-3}$ (see \citealt{shirley_critical_2015}).  Overall, these data probe gas that span more than 2-3 orders of magnitude in density.  

\begin{table*} 
	\centering
	\begin{threeparttable}
	\caption{Kinematic properties and gas temperature of the ridge calculated within using the masked datasets from Fig.~\ref{fig:velo}}  
	\begin{tabular}{llllllll} 
		\hline
		Region & $\overline{v}_\text{lsr}$ \nhp & $\overline{v}_\text{lsr}$ \nhhh & $\overline{v}_\text{lsr}$ \tCO & $\overline{\sigma}$ \nhp & $\overline{\sigma}$ \nhhh & $\overline{\sigma}$ \tCO & $\overline{T}_\text{gas}$ \\
			 &  (\kms)		   		 &  (\kms)		     		   &  	(\kms)				   &  (\kms)			    &  (\kms)			        &  	(\kms)		        	& (K) 			       \\
		\hline
		Active   & -38.7$\pm$1.1 & -39.0$\pm$0.6 & -39.1$\pm$1.0 & 1.7 $\pm$0.6 & 1.5 $\pm$0.3 & 2.3$\pm$0.5 & 17.4$\pm$2.6 \\
		Quiescent & -38.7$^{+1.1}_{-1.0}$ \tnote{\emph{a}}& -38.8$\pm$0.5 & -39.4$\pm$0.3 & 1.0$^{+0.3}_{-0.4}$ \tnote{\emph{a}} & 1.5$\pm$0.2  & 1.8$\pm$0.5 & 14.4$\pm$2.4 \\
		\hline
	\label{tab:velo split}
	\end{tabular}
	\begin{tablenotes}
        \footnotesize
        \item[\emph{a}]{Median and interquartile ranges are used to estimate average values}
   \end{tablenotes}
   \end{threeparttable}
\end{table*}

\subsection{\nhp\ (J=1--0)}

	The kinematics of the dense gas is investigated using the MALT90 \nhp\ (J=1--0) data (see Sec.~\ref{sec:malt90}).  The velocity channels were first smoothed to 0.22~\kms\ to improve the signal-to-noise ratio, after which the seven hyperfine components of the  \nhp\ (J=1--0) were fitted using the HFS routine within GILDAS. At every position, one velocity component has been fitted. The resulting integrated intensity, centroid velocity, and velocity dispersion maps are shown in Fig.~\ref{fig:velo}a-c. Here, it becomes immediately obvious that the full ridge has not been observed by the MALT90 survey predominately in the quiescent region.  Yet from the coverage we have, we see that the integrated intensity has a similar morphology to the \h\ column density and has a number of peaks that coincide with {\it Herschel} clumps (see Sec.~\ref{sec:dendro}).  However, the \nhp\ (J=1--0) emission in the active region is asymmetrically concentrated to the west side of the \h\ column density, which might be a sign that the relative abundance of \nhp\ is affected by the different physical conditions of the ridge at this location.  This is somewhat reminiscent of what has been observed in the SDC335 massive star-forming infrared dark cloud \citep{peretto_global_2013}.  
	
	The centroid velocity map in Fig.~\ref{fig:velo}b reveals a very dynamic environment, in particular towards the active part of the ridge where large velocity gradients and dispersions' coincide.  More specifically, on small spatial scales, we see large velocity gradients perpendicular and parallel to the ridge in the active region.  For a more quantitative comparison between the velocity gradients and velocity dispersion, we constructed a vector plot of these gradients using the centroid velocity map for each tracer on Fig.~\ref{fig:straight kin}.  The velocity gradient was calculated using the central difference over the beam size for each tracer. We straightened the plots in the same way as Fig.~\ref{fig:col_temp_str} to emphasise the spatial correlation between these two quantities. On Fig.~\ref{fig:straight kin}, one can see that the velocity dispersion in the active region is large, with a non-weighted average of 1.7~\kms, reaching peaks of 3.9~\kms.  These velocity dispersion peaks correlate spatially with velocity gradient peaks of $\sim 10$~\kms~pc$^{-1}$.  This strongly suggests that unresolved gas flows are present the ridge.  In fact, a closer inspection of the \nhp\ (J=1--0) spectra around where the largest velocity gradients are located, near
clumps \#2 and \#4 (see Fig.~\ref{fig:nhp_2_comp}), reveals that they could be the result of two overlapping velocity components (one at $\sim$-36.5~\kms\ and one at $\sim$-40~\kms).  More unresolved velocity structures might be present in the ridge, potentially contributing to the measured velocity dispersion along the ridge. As far as the ridge stability analysis is concerned, such gas flows will, in the case where gravity is responsible for their development, lead to an overestimate of the kinetic pressure term. Therefore, their presence can only strengthen the results presented in Sec.~\ref{sec:stability}.

	The centroid velocity within the quiescent region is relatively uniform, despite being noisier.  Visual inspection of the spectra indicates that they have low signal-to-noise ratios (S/N).  These low S/N results in poor fits and so when discussing the quiescent region we restrict ourselves to using the median centroid velocity and median velocity dispersion, along with their corresponding interquartile ranges (see Table~\ref{tab:velo split}).  Median values are less impacted by erroneous fits.

\begin{figure}
	\centering
		\includegraphics[width = \columnwidth]{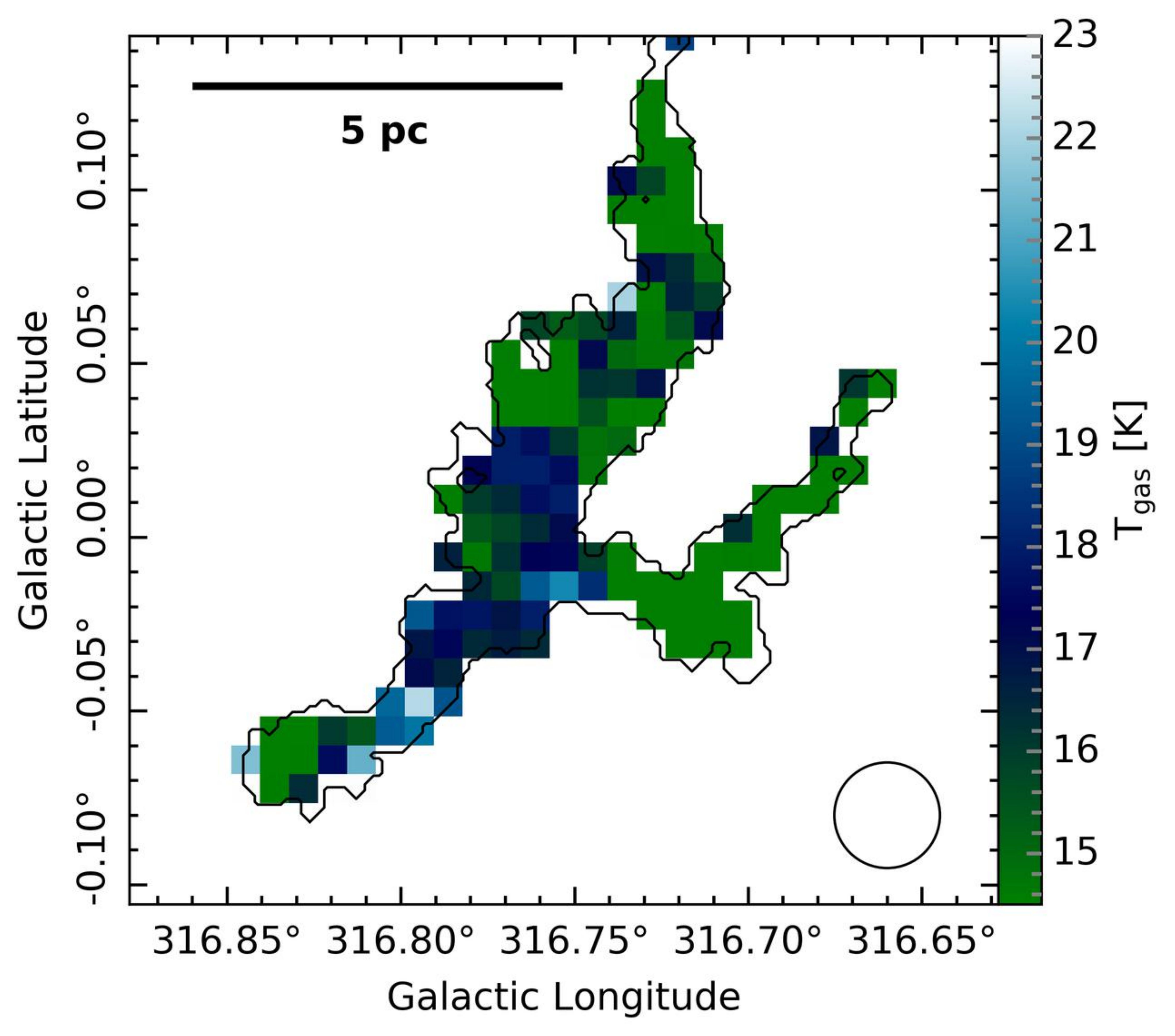}
	\caption{Rotation temperature derived from \nhhh\ (1,1) and (2,2) observations. Limits of the colour map are equal to the limits of the dust temperature in Fig~\ref{fig:temp_col}.  The black contour and mask used are identical to Fig~\ref{fig:velo}. The open black circle shows the beam size of the observations.}
	 \label{fig:rot_temp}
\end{figure} 

\begin{figure}
	\centering
		\includegraphics[width = \columnwidth]{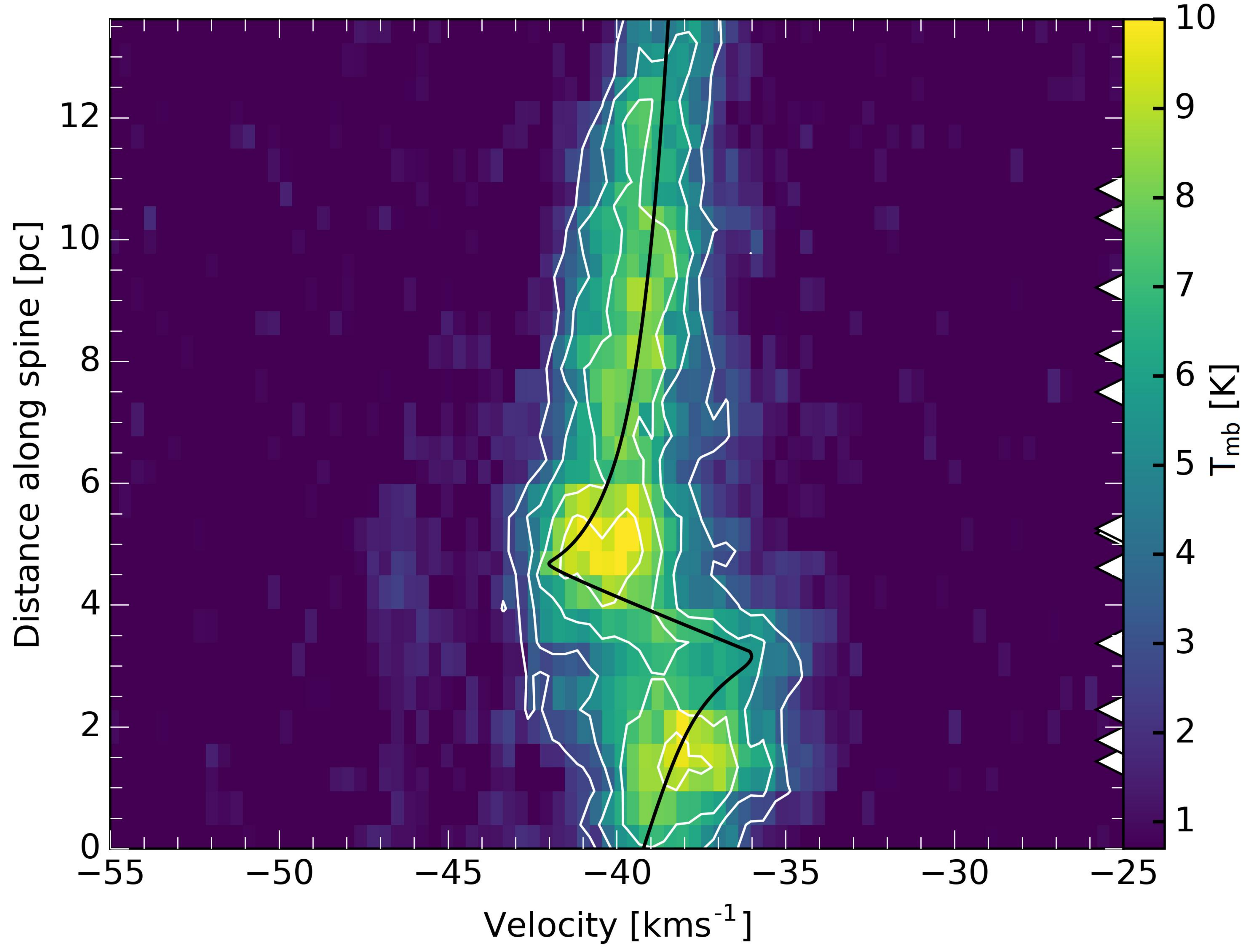}
	\caption{Position-velocity diagram of \tCO\ along the spine centre from 0 and 13.6~pc.  The solid black line traces the centre of the emission and the solid white contours mark the intensity of the emission at 3, 5, 7 and 9 K.  White triangles mark the clump positions.}
	 \label{fig:vel_dist}
\end{figure}

\begin{figure}
	\centering
		\includegraphics[width = \columnwidth]{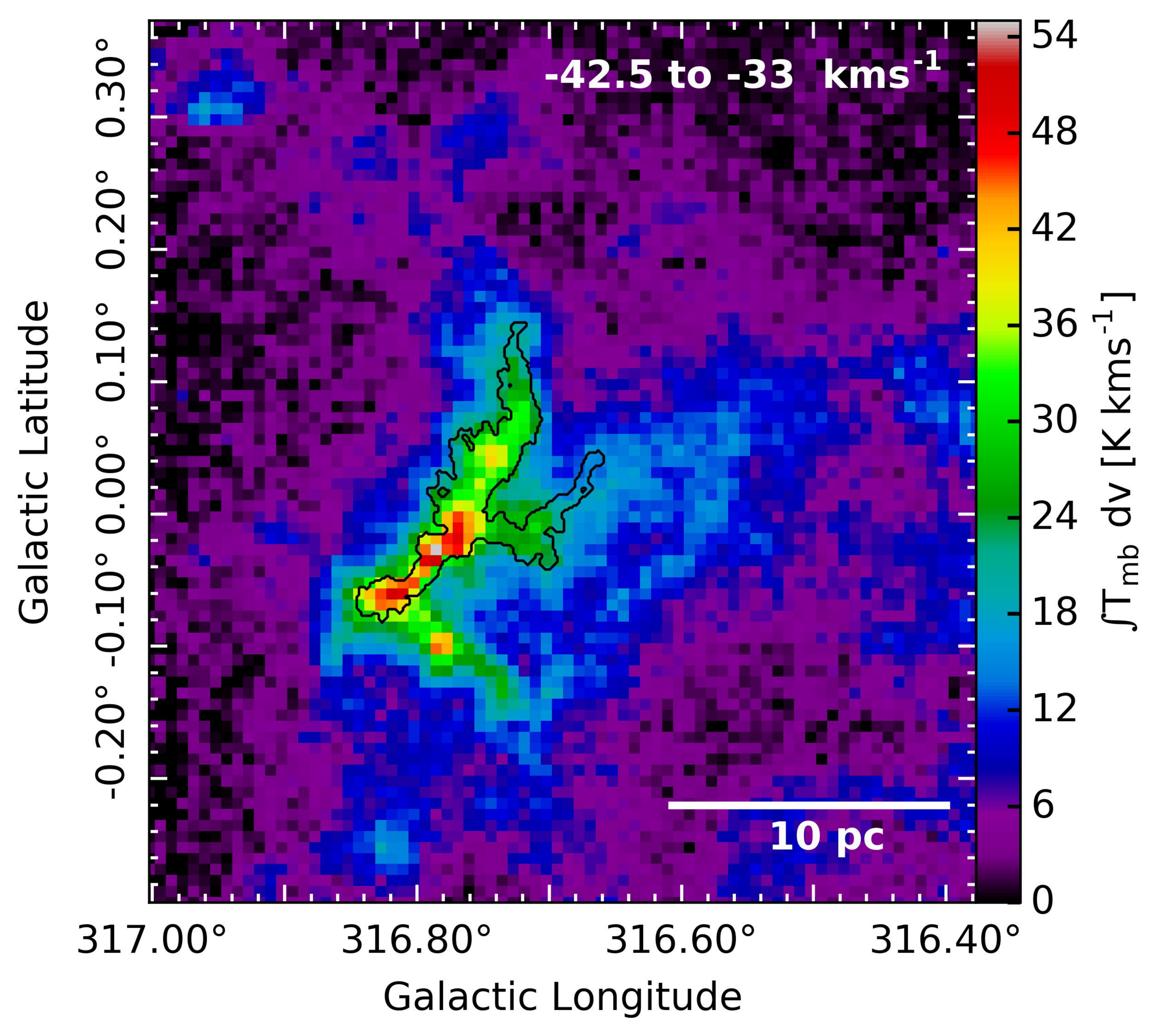}
	\caption{\tCO(1--0) integrated intensity  of G316.75 between -42.5 and -33~\kms.  The black contour is identical to Fig~\ref{fig:velo}.}
	 \label{fig:13co_int}
\end{figure}

\begin{figure*}
	\centering
		\includegraphics[width = \textwidth]{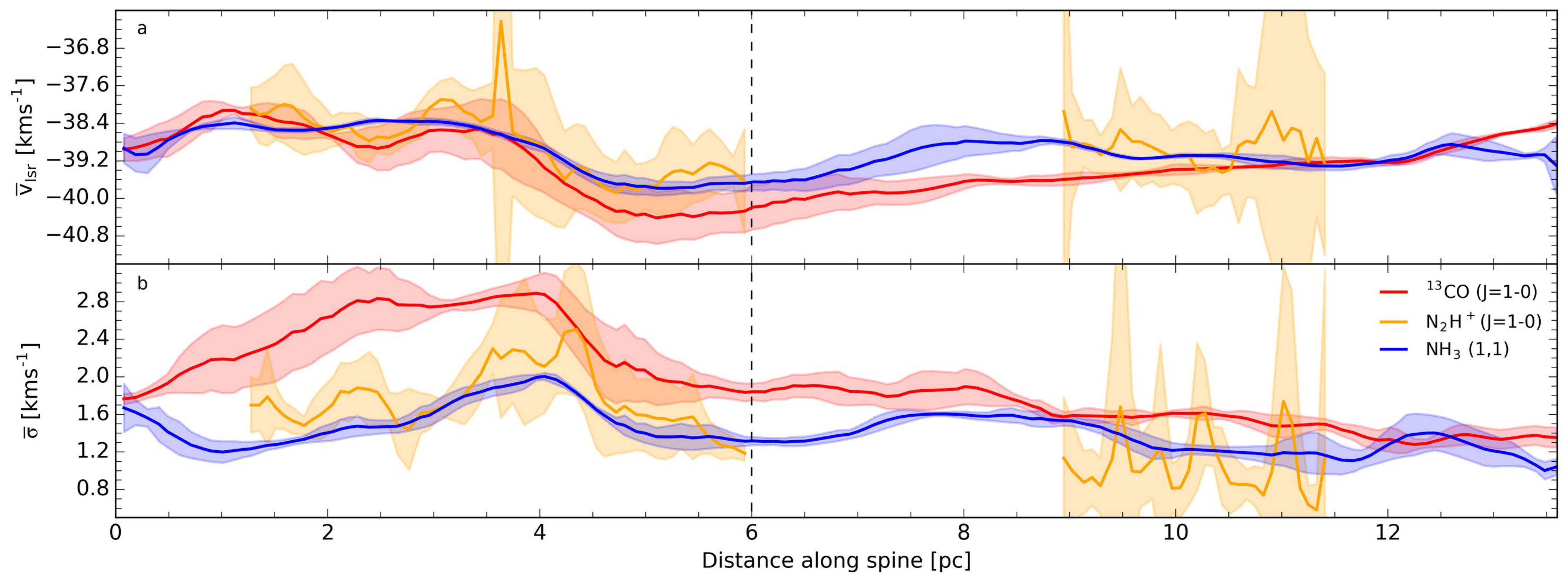}
	\caption{Longitudinal centroid velocity {\bf(a)}, and velocity dispersion {\bf(b)} in \kms\ for the molecular transitions shown in Fig.~\ref{fig:velo} averaged over the same width and length as Fig.~\ref{fig:temp_col_height}. The solid red line is \tCO\ (1--0), the solid yellow is \nhp\ (1--0) and the solid blue is \nhhh.  The thin black vertical dashed line marks the separation between the active half of the ridge (left) and the quiescent half (right).  Translucent regions shows the 1-sigma uncertainty values.}
	\label{fig:mean kin long}
\end{figure*} 

\subsection{\nhhh\ (1,1) and (2,2)} \label{sec:nh3}

	The \nhhh\ (1,1) and \nhhh\ (2,2) HOPS observations cover all of the G316.75 ridge, allowing us to investigate the dense gas that is not mapped in \nhp.  The velocity channels were smoothed to 1~\kms.  As for \nhp, both datasets were fitted using the HFS routine of GILDAS, using one velocity component.  The integrated intensity, centroid velocity and dispersion maps resulting from the fit of the \nhhh\ (1,1) transition are shown in Fig.~\ref{fig:velo}d-f.  We do not show the (2,2) maps in this study since they trace the same structures as to the (1,1) maps (but at lower S/N).

Overall, all these maps resemble the \nhp\ maps but at a lower resolution.  It is interesting to see that the quiescent region has a significant velocity gradient across the ridge where the \nhp observations were not mapped (1.5--2.5 \kms~pc.  See Fig.~\ref{fig:velo}).  Visual inspection of the \nhhh\ spectra do not show multiple velocity components.  

	The excitation of the ammonia inversion lines is dominated by collisions at low temperatures.  We can therefore use ammonia to derive the temperature of the gas \citep{ho_interstellar_1983}.  Following the same method as that presented in, for example, \cite{ho_interstellar_1983,williams_gravity_2018}, we compute the ammonia rotational temperature map (see Fig.~\ref{fig:rot_temp}).  We can see that the general morphology of the gas temperature matches that seen in dust temperature; the active region is warmer than the quiescent with mean gas temperatures of 17.4$\pm$2.6~K and 14.4$\pm$2.4~K respectively.  In Fig.~\ref{fig:temp_col_height} we plot the longitudinal gas temperature to better illustrate the matching features in both the gas and temperature maps.  One can see that both temperatures are consistent with each other along most of the ridge, and differences can be explained by the order of magnitude difference in angular resolution between HOPS maps and PPMAP datasets.

\begin{figure}
	\centering
		\includegraphics[width = \columnwidth]{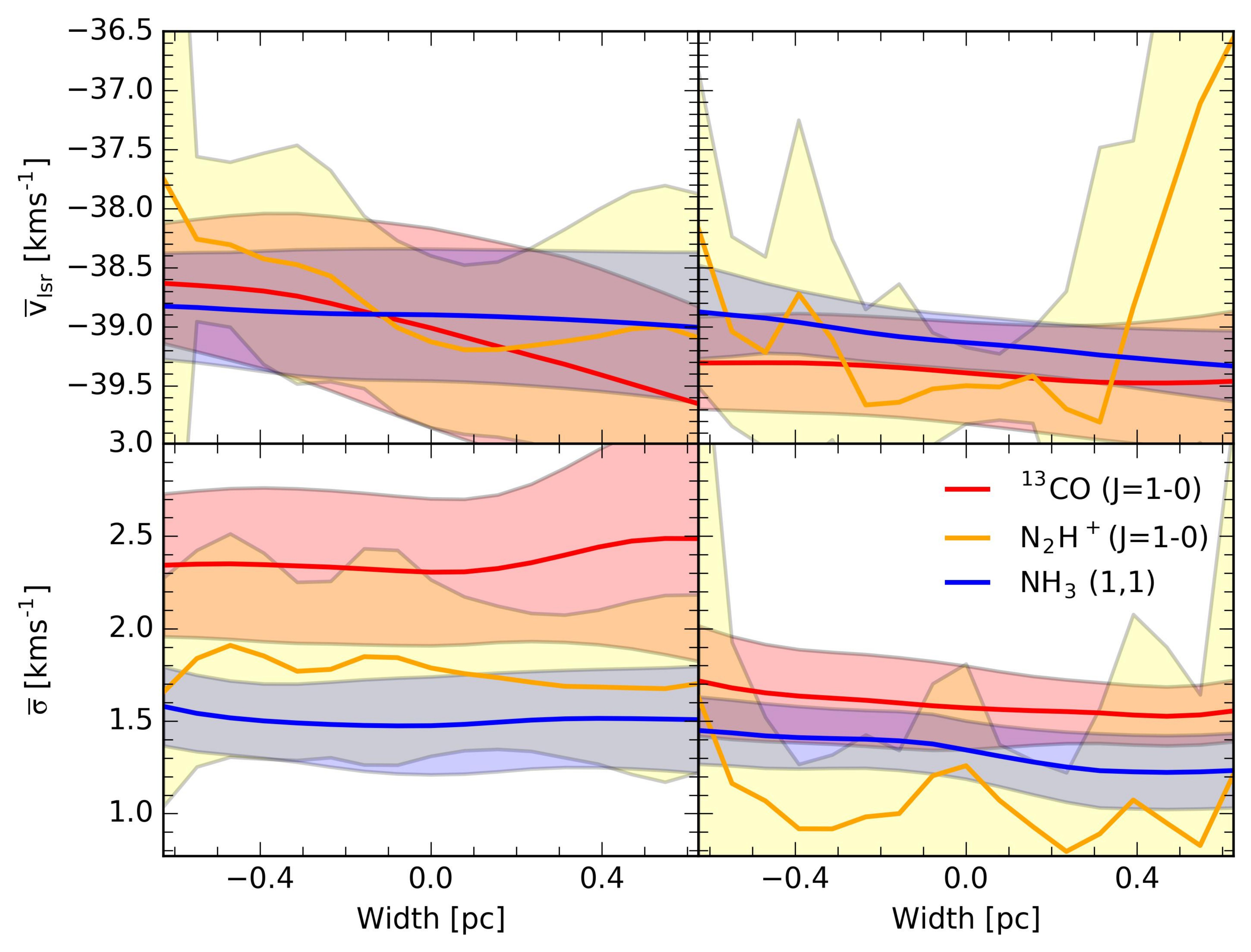}
	\caption{Transverse centroid velocity ({\bf top}), and velocity dispersion ({\bf bottom}) profiles are shown using the same the molecular transitions and line colours shown in Fig.~\ref{fig:mean kin long}. Profiles are averaged over the same range as Fig.~\ref{fig:1d_col_with_temp} for the active region ({\bf left} column) and quiescent regions ({\bf right} column).  Translucent regions describes the 1-sigma uncertainty values.}
	\label{fig:mean kin profiles}
\end{figure}

\begin{figure}
	\centering
		\includegraphics[width = \columnwidth]{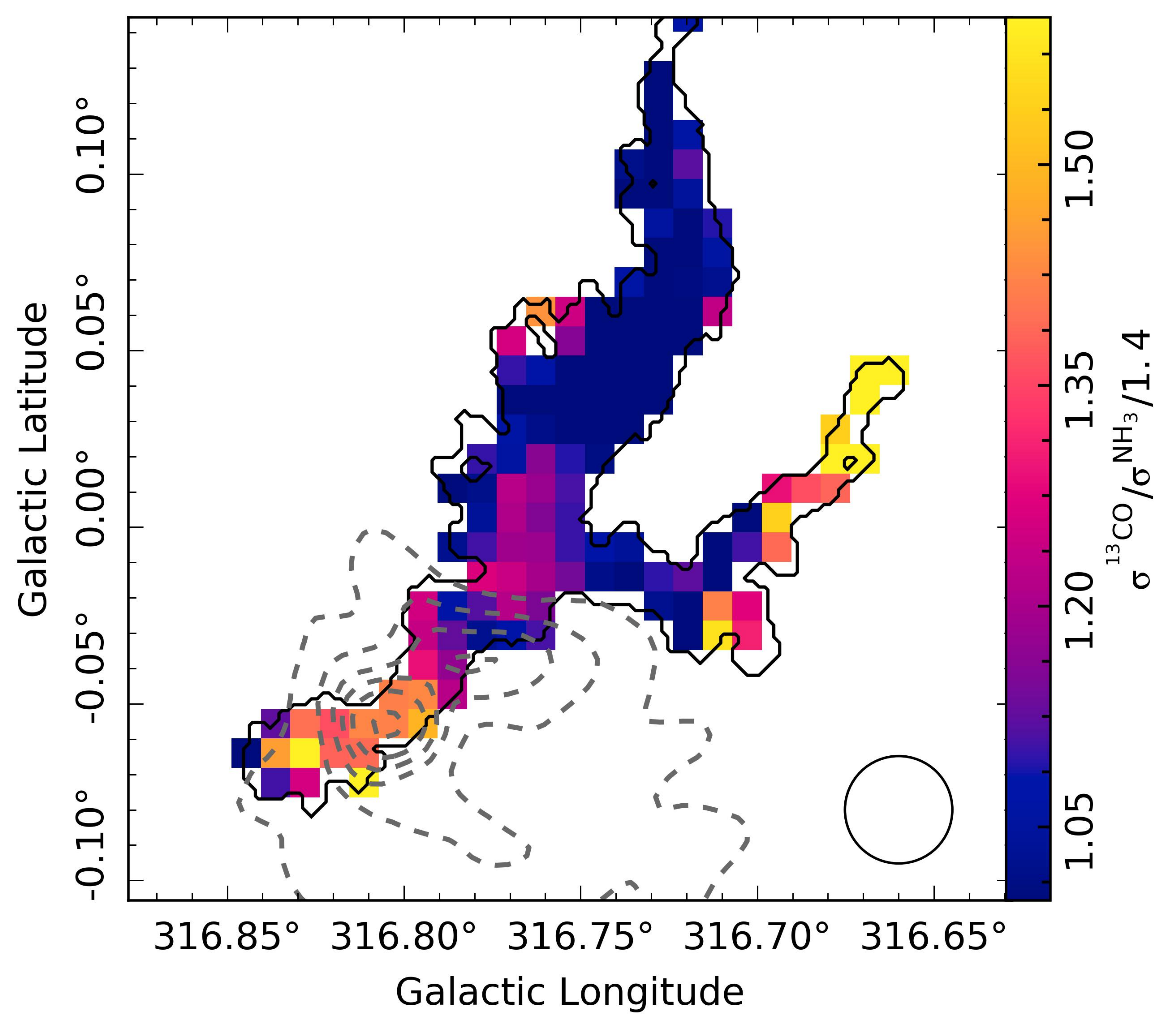}
	\caption{Ratio map of \tCO\ against \nhhh\ velocity dispersion divided by the mean value of the ratio in the quiescent region, (i.e 1.4). The black contour and mask used are identical to Fig~\ref{fig:velo}. Dashed grey contours show SUMSS radio emission using the same contour levels presented on Fig.~\ref{fig:rgb_intro}.  The open black circle shows the beam size of the ratio map}.
	 \label{fig:13co/nh3}
\end{figure}

\subsection{\tCO\ (J=1--0)} \label{sec:13co}

\begin{figure}
	\centering
		\includegraphics[width = \columnwidth]{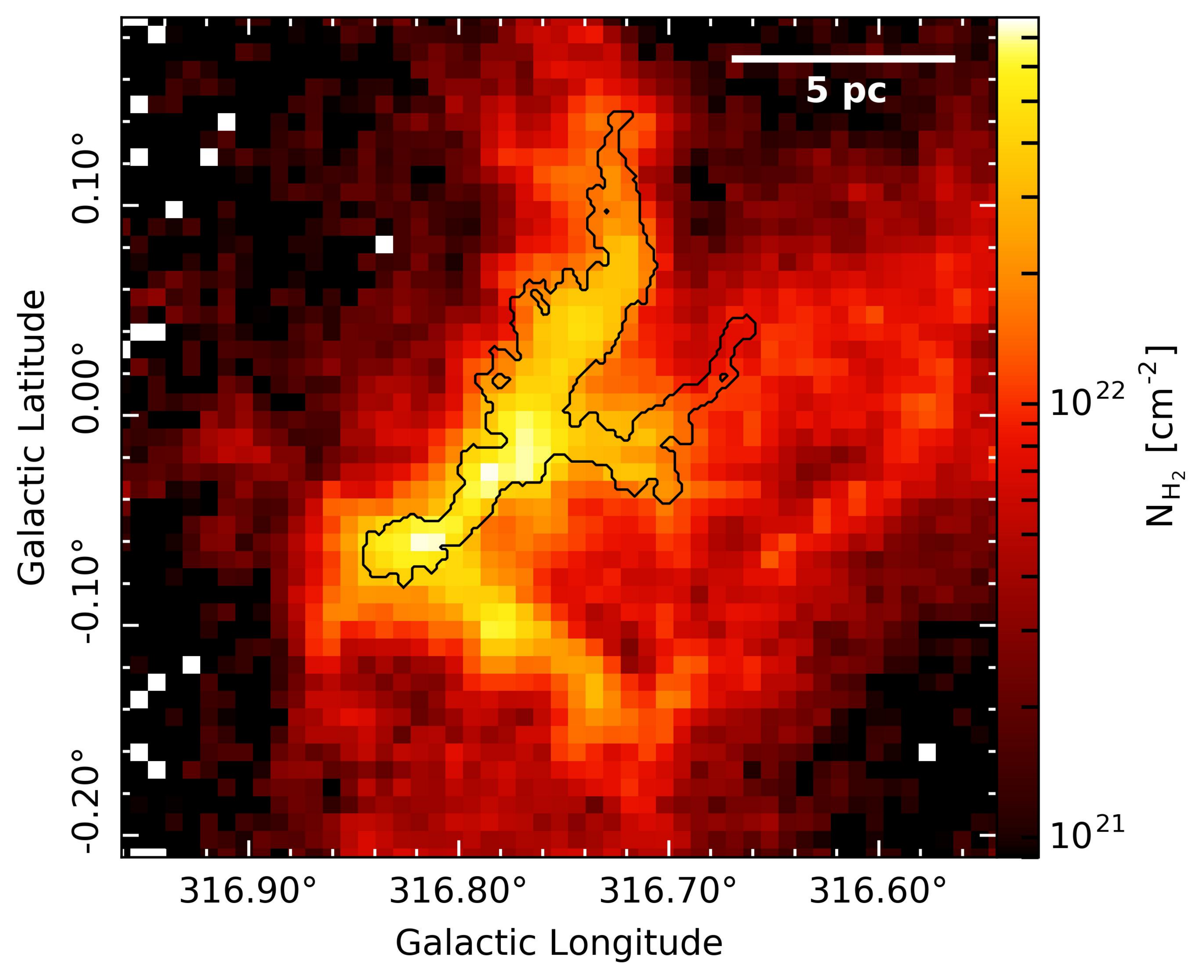}
	\caption{\tCO-based \h\ column density map derived using LTE approximation (see Appendix \ref{sec:LTE}). The black contour and mask used are identical to Fig~\ref{fig:velo}.}
	\label{fig:LTE_col}
\end{figure}

\begin{figure*}
	\centering
		\includegraphics[width = \textwidth]{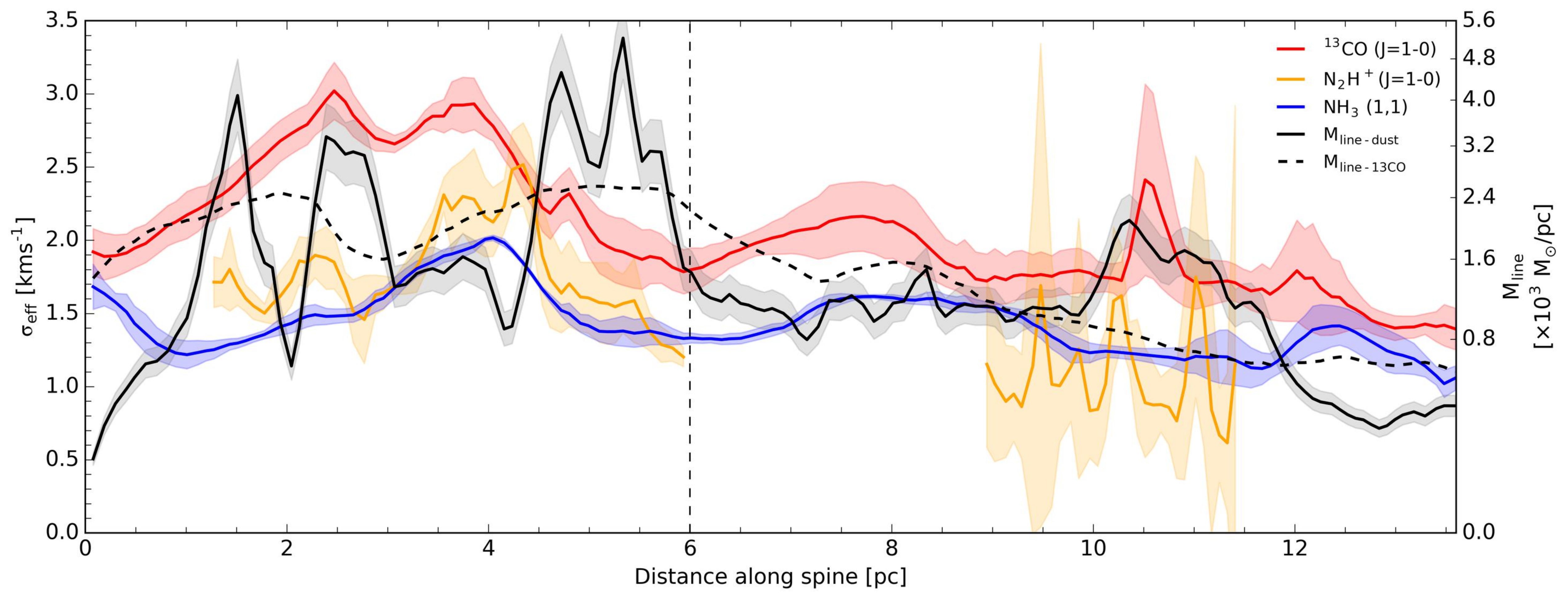}
	\caption{Longitudinal effective velocity dispersion and corresponding $M_\text{line}^\text{crit}$ are shown using the same molecular transitions and line colours as shown in Fig.~\ref{fig:mean kin long}.  Also plotted are $M_\text{line-dust}$ and $M_\text{line-13CO}$ (and corresponding $\sigma_\text{eff}^\text{crit}$). Except for \tCO-based measurements, the values are averaged over the same width and length as in Fig.~\ref{fig:temp_col_height}, whereas \tCO-based values are averaged over a 2.82~pc scale (i.e. 1.41~pc either side of the ridge).  The y axis on the left portrays the velocity dispersion and the y axis on the right is the \Mline.  The axis appears stretched by a square root scale since, in the critical case, $\sigma_\text{eff}\propto$\Mline$^{0.5}$.  Translucent regions describes the 1-sigma uncertainty values each molecular line tracer.  1-sigma uncertainty values are not shown for $M_\text{line-13CO}$.}
	 \label{fig:sigma_eff_mline}
\end{figure*}

	The MSGPCOS \tCO\ (J=1--0) integrated intensity, centroid velocity, and velocity dispersion maps were calculated by fitting a single Gaussian at each position after the velocity channels were smoothed to 0.37~\kms (see Fig.~\ref{fig:velo}g--i).  The integrated intensity of \tCO\ (J=1--0) resembles the \h\ column density and unlike \nhhh\ and \nhp, the integrated intensity peaks along the spine of the ridge.  The centroid velocity map looks relatively similar to that of the \nhhh\ centroid velocity map, even though the perpendicular gradient observed in the quiescent region is not as apparent in the \tCO\ (J=1--0) data.  To check this gradient we have produce a position-velocity diagram along the spine of the ridge in Fig.~\ref{fig:vel_dist}. Indeed, this shows that there is a velocity gradient present. The black line within the figure indicates that the gradient steepens toward the centre.  This feature could indicate gas infall towards the centre of the ridge \citep{hacar_gravitational_2017,inoue_formation_2018} but the complexity of the region stops us from making robust conclusions about the origin of this velocity gradient.  The \tCO\ (J=1--0) velocity dispersion map exhibits, with the exception of a couple of pixels with unresolved N2H+(1-0) velocity components, larger values than any of the other two tracers. We observe with values as high as $\sigma = 3.5$~\kms\ in the active region.  As for the other two tracers, large velocity dispersion peaks are also matched by large velocity gradients (Fig.~\ref{fig:straight kin}).

	In Fig.~\ref{fig:13co_int} we show the integrated intensity of the \tCO\ (J=1--0) line over a larger field than that showed in Fig.~\ref{fig:velo}d.  On this image, one can see that there is a significant amount of diffuse emission surrounding the ridge, with a clear asymmetric morphology of the emission with respect to ridge's spine, most of it being located south of the ridge.  This map also reveals an additional filamentary structure extending perpendicularly to the southern end of the active part of the ridge.  This is not seen in the dense gas tracers, but does match a warm and low column density structure present in the {\it Herschel} column density map (see Fig.~\ref{fig:temp_col} and A1). The \tCO\ (J=1--0) spectra observed along this structure have low intensities but spread over a large velocity range $>10$~\kms.

\subsection{Comparison of the gas kinematics} \label{sec:comp_gas_kin}

	In order to get a more concise view of the ridge kinematics, we computed the centroid velocity and velocity dispersion profiles of \tCO(1--0), \nhhh(1,1) and \nhp\ (1--0) in both longitudinal and transverse directions. The longitudinal profiles (see Fig.~\ref{fig:mean kin long}) for the centroid velocity show a mostly smooth velocity gradient along the ridge that steepens at around 4-5~pc (see also Fig.~\ref{fig:vel_dist}).  Owing to its higher angular resolution, the \nhp\ profile shows more small-scale structures that are not recovered in the other two profiles. Regarding the velocity dispersion, it is quite clear that the velocity dispersion traced by \tCO\ (1--0) is larger than the other two dense gas tracers everywhere in the ridge. \nhhh\ (1,1) and \nhp\ (1--0) also show larger velocity dispersions in the active part of the ridge, however not to the same extent as \tCO. The transverse profiles (see Fig.~ \ref{fig:mean kin profiles}) are less structured than the longitudinal ones.  The active region displays a small transverse velocity gradient, mostly evident in \tCO, while its quiescent counterpart seems to be flat.  Regarding the velocity dispersion, the transverse profiles show similar trends as the longitudinal ones for both parts of the ridge.

	As demonstrated above, the measured \tCO\ velocity dispersion is larger than the dense gas tracers. This is generally the case in any star-forming cloud as \tCO\ tends to tracer more diffuse gas. However, we would like to evaluate to what extent feedback from the embedded O-stars are contributing to this increase. For that purpose, we convolved the \tCO\ (1--0) data to the \nhhh\ (1,1) resolution, and regridded the data to the same grid.  We then recalculated the velocity dispersion map for \tCO\ and computed the ratio of \tCO\ (1--0) to \nhhh\ (1,1) velocity dispersion (see Fig~\ref{fig:13co/nh3}). From this ratio map, we find that, in the quiescent part of the ridge where feedback from O-stars is minimal, the \tCO\ (1--0) velocity dispersion is on average 1.4 times larger than the \nhhh\ (1,1) velocity dispersion. This confirms that \tCO\ includes large velocity dispersion gas that is not probed with dense gas tracers. Taking this ratio of 1.4 as the non-feedback-contaminated velocity dispersion ratio between the two tracers, we can then evaluate what is the contribution of feedback on the  \tCO\ (1--0) velocity dispersion. To do this, we simply divided the velocity dispersion ratio map by this natural value of 1.4.  The resulting map is shown in Fig.~\ref{fig:13co/nh3}. Here, we notice that the active part of the ridge exhibits up to 65\% larger velocity dispersion ratios than the average value of 1.4. The spatial correlation of this increase with respect to the location of the embedded O-stars strongly suggests that the increase is due to stellar feedback.

 \begin{figure*}
	\centering
		\includegraphics[width = \textwidth]{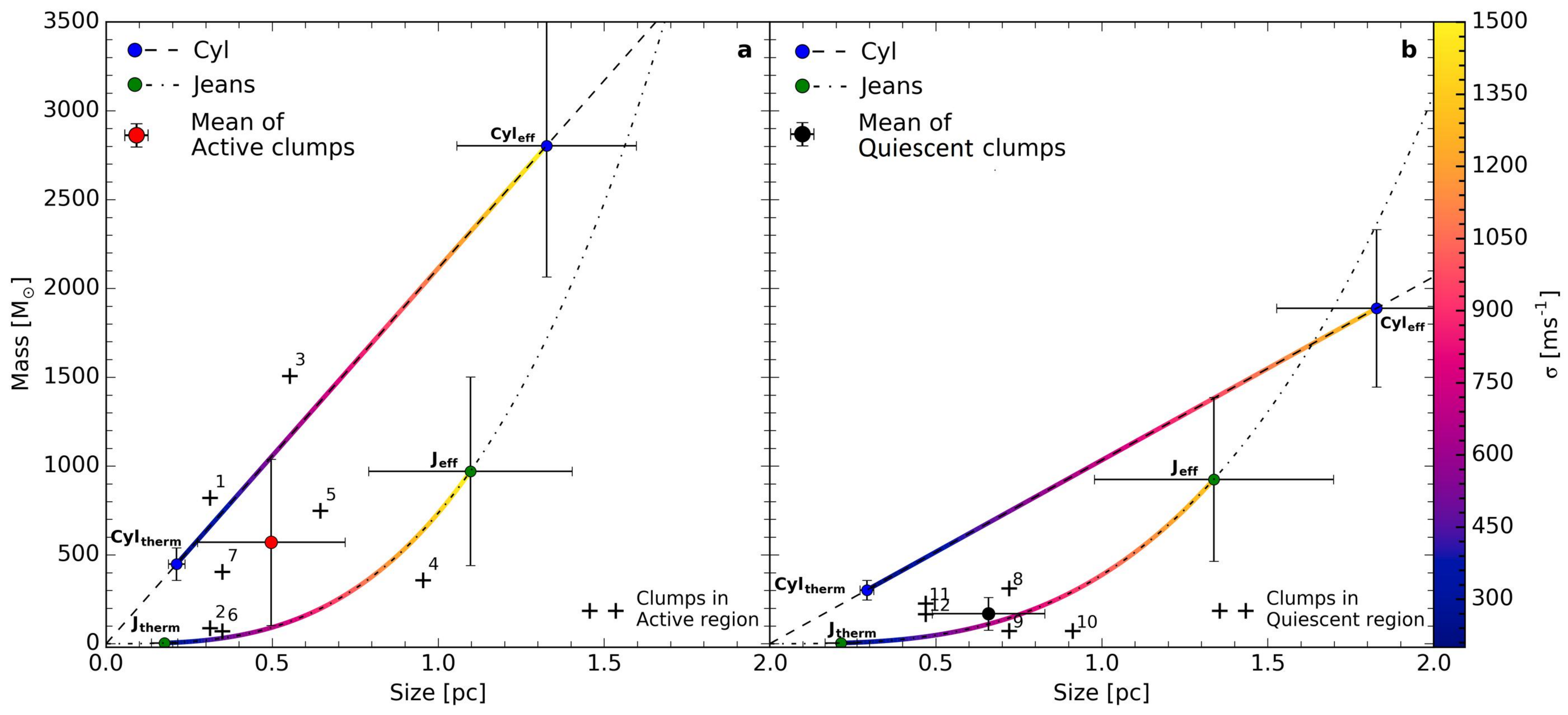}
	\caption{Fragmentation scales (both mass and length) for increasing values of velocity dispersion for the active {\bf (a)} and quiescent {\bf (b)} regions calculated using {\it Herschel} column density.  The dashed-black line shows the cylindrical fragmentation case, while the dashed-dotted-black line shows the Jeans fragmentation case.  The mass and length calculated for thermal and effective fragmentation modes are shown as blue and green circles. The coloured lines show how the mass and length fragmentation scales change with increasing velocity dispersion (from thermal to effective velocity dispersion). The black pluses show the mass and minimum separation of the clumps tabulated in Table~\ref{tab:core_mass}. The numbers ascribed to the clumps match the numbering given in Table~\ref{tab:core_mass}.  The red circle marks the mean mass and minimum separation of the clumps for the active region, and the black circle marks the mean mass and minimum separation of the clumps for the quiescent region. The errorbars for the mean show the 1-sigma spread in values.}
	\label{fig:frag}
\end{figure*}

\begin{table*} 
   \centering
   
	\caption{Table with derived average values for both parts of the ridge. Except for \tCO-based measurements, the averages tabulated use the straightened values over the same width and length as in Fig.~\ref{fig:temp_col_height}, whereas \tCO-based values are averaged over a 2.82~pc scale (i.e. 1.41~pc either side of the ridge). All mean values use 1-simga errors.  The remaining errors are calculated using error propagation unless otherwise stated. }
	\resizebox{\textwidth}{!}{
\begin{threeparttable}
	\begin{tabular}{cccccccccccccc} 
		\hline
		Region  & $\overline{\sigma}_\text{eff}$ \tCO & $\overline{\sigma}_\text{eff}$ \nhhh & $\overline{\sigma}_\text{eff}$ \nhp & $n_\text{c}$\tnote{\emph{a}} & $\overline{n}$ & $\lambda_\text{J}$  & $\lambda_\text{cyl}$ & $M_\text{J}$ &  $M_\text{cyl}$ & $\lambda_\text{J}^\text{eff}$  & $\lambda_\text{cyl}^\text{eff}$ & $M_\text{J}^\text{eff}$ &  $M_\text{cyl}^\text{eff}$\\
			  & (\kms)	& (\kms) & (\kms) & (10$^{4}~$cm$^{-3}$)  & (10$^{4}~$cm$^{-3}$) & (pc) & (pc) & (\msun) &  (\msun)  & (pc) & (pc)	 & (\msun)	&  (\msun)	\\
		\hline
		Active   & 2.4$\pm$0.4 & 1.5$\pm$0.2 & 1.5$\pm$0.4 & 17$\pm$4 & 2.0$\pm$0.9 & 0.18$\pm$0.04 & 0.21$\pm$0.02 & 4$\pm$1 & 450$\pm$90 & 1.10$\pm$0.31 & 1.33$\pm$0.27 & 971$\pm$531 & 2803$\pm$739\\
		Quiescent & 1.8$\pm$0.2 & 1.3$\pm$0.2 & 0.6$\pm$0.5 &   7$\pm$2 & 1.0$\pm$0.4 & 0.21$\pm$0.05  & 0.29$\pm$0.02 & 4$\pm$1 & 302$\pm$55 & 1.38$\pm$0.36 & 1.83$\pm$0.30 & 925$\pm$462 & 1888$\pm$443\\
		\hline
		\end{tabular}
	\begin{tablenotes}
        \footnotesize
        \item[\emph{a}]{Uncertainty estimated from 1-sigma confidence in $\chi^2$ residuals in Appendix \ref{sec:plummer}.}
	\item[]{$\overline{\sigma}_\text{eff}$ is the mean effective dispersion for the transitions listed, $n_\text{c}$ is the central number density and $\overline{n}$ is the mean number density. $\lambda$ and M refers to fragmentation length and mass, where the subscript J refers to Jeans values and cyl refers to cylindrical fragmentation values. Any superscripts with eff show that the fragmentation scale was calculated using $\overline{\sigma}_\text{eff}$ rather than the thermal dispersion.}
        \end{tablenotes}
   \end{threeparttable}
		}

   \label{tab:analysis}
\end{table*}

\section{Analysis} \label{sec:analysis}

In the following section we evaluate the ability of the G316.75 ridge to collapse and fragment.  We also explore if current feedback has the capacity to disrupt the ridge.

\subsection{Density, mass-per-unit-length, and effective velocity dispersion} \label{sec:analysis_param}

In order to determine the global stability and fragmentation ability of the ridge, we need to compute a number of key physical quantities, the first of which is the average volume density of the cloud, $\overline{\rho}$.  To estimate $\overline{\rho}$, we assume that the ridge is cylindrical with a radius of $R=0.63\pm0.08$~pc.  We defined this radius using the transition between the dense filament and the diffuse emission. Using the column density profile, the transition point appears as a power-break in the profile (see Fig.~\ref{fig:1d_col_log}).  One can then estimate $\overline{\rho}$ via:

\begin{equation}
	\overline{\rho}=\frac{M}{\pi R^2L}
\end{equation} 

\noindent where $L$ is the length, $M$ its mass, and $R$ its radius.  The number density, $\overline{n}=\overline{\rho}/(\mu m_\text{H})$, of \h\ is often preferred to the mass density, where $\mu$ is the molecular weight and is equal to 2.8.  All quantities for both quiescent and active parts are given in Table~\ref{tab:split}. We note that $\overline{n}$ is 2\e$^4$ and 1\e$^4$ for the active and quiescent regions respectively, which are high average values over such a large structure.

	A second important quantity to compute is the mass-per-unit-length (\Mline) of the ridge.  \Mline\ is computed at every pixel along the spine of the ridge by integrating the local perpendicular \h\ column density profile to get the mass {\bf of the corresponding slice}, and dividing it by the physical length of the pixel, providing us with the local mass-per-unit-length. \Mline\ is computed twice, once using the {\it Herschel} \h\ column density map, $M_\text{line-dust}$, and once using a \h\ column density map computed from the \tCO(1--0) integrated intensity, $M_\text{line-13CO}$ (see Fig.~\ref{fig:LTE_col} and Appendix~\ref{sec:LTE} for details on how the \tCO-based \h\ column density map is obtained). The reason behind these two sets of \Mline\ is that, as seen in \tCO, the G316.75 ridge extends beyond the dense part we characterised with {\it Herschel} ($R=0.63$~pc for the dense part, while $R=1.41$~pc when including the diffuse gas traced by \tCO). This implies that the measured \tCO(1--0) velocity dispersions include diffuse gas which is not included in the {\it Herschel}-based \Mline. Therefore, for a fair comparison of the kinetic and gravitational energies of the ridge (see Sec.~\ref{sec:stability}), \tCO-based velocity dispersions have to be compared to \tCO-based mass-per-unit-length measurements. While $M_\text{line-13CO}$ does not require any background subtraction (\tCO(1--0) integrated intensity map used for this only gas at the ridge velocity - see Fig.~\ref{fig:13co_int}), $M_\text{line-dust}$ is obtained by subtracting a constant background column density of 0.23~\tcm (see Sec.~\ref{sec:ppmap} and \ref{subsec:hessian}).

	Figure~\ref{fig:sigma_eff_mline} shows how both \Mline\ measurements vary along the ridge. The uncertainty we show for the dust derived \Mline\ values is calculated from the distance error. Both $M_\text{line-dust}$ and $M_\text{line-13CO}$ are, on average, twice as large in the active part compared to the quiescent part. Interestingly we see that $M_\text{line-13CO}$ is not much different from the $M_\text{line-dust}$ despite encompassing about double the area. This is most likely due to a combination of two factors. First, in order to compute the \tCO-based  \h\ column density map, we used the local thermodynamic equilibrium (LTE) approximation which is known to underestimate column densities by a factor of $\sim2$ \citep{szucs_how_2016}.  Second, cold dense regions such as within infrared dark clouds cause CO to deplete onto dust grains, which also leads to an underestimate of the \h\ mass \citep{hernandez_mapping_2011}. The significance of underestimating the mass in context of the analysis is discussed in the following Sections.

	Finally, the third important quantity for stability analysis is the effective velocity dispersion of the gas, $\sigma_\text{gas}$.  If thermal motions are the only contributors to the kinetic pressure of a cloud, then $\sigma_\text{gas}=\sigma_\text{th}=\sqrt{\frac{k_\text{B}T}{\mu m_\text{H}}}$, where $T$ is the gas temperature and $\sigma_\text{th}$ is thermal sound speed.  As already discussed in Sec.~\ref{sec:nh3},  both the gas and dust temperatures are within error of each other (see Fig.~\ref{fig:temp_col_height}). As a result, we adopt here the PPMAP dust temperature as a proxy for the gas temperature since the angular resolution is $\sim$10 times better, which allows us to investigate smaller scale temperature changes.  If turbulence also contributes to the kinetic pressure then $\sigma_\text{gas}=\sigma_\text{eff}=\sqrt{\sigma_\text{th}^2+\sigma_\text{turb}^2}$ where $\sigma_\text{turb}$ is the turbulent component of the velocity dispersion.  The effective velocity dispersion $\sigma_\text{eff}$ is calculated from the measured velocity dispersion using the three tracers presented in Sec.~\ref{sec:gas_temp} according to the following equation from \citet{fuller_dense_1992}:

\begin{equation}
	\sigma_\text{eff}^2=\sigma_\text{mol}^2+k_\text{B}T\left(\frac{1}{\mu m_\text{H}} - \frac{1}{m_\text{mol}}\right)
\label{eq:sigma_eff}
\end{equation}

\noindent where $\sigma_\text{mol}$ is the observed velocity dispersion for one of the molecular transitions, and $m_\text{mol}$ is the mass of the corresponding molecule.  In Fig.~\ref{fig:sigma_eff_mline} we plot the variation of $\sigma_\text{eff}$ along the ridge for \tCO\ (1--0), \nhhh\ (1,1) and \nhp\ (1--0) and present their mean values for the active and quiescent regions in Table~\ref{tab:analysis}. For a matter of consistency with the \Mline\ measurements, velocity dispersions have been averaged up to $R$=0.63~pc either side of the spine for both dense gas tracers, while up to $R$=1.41~pc for \tCO.  We also note that the observed velocity dispersions can also contain contributions from unresolved systematic gas flows generated by collapse and rotation, and so it is unclear what fraction of the observed velocity dispersion is due to this or due the internal pressure of the cloud \citep{traficante_massive_2018}.  Therefore the effective velocity dispersion represents an upper limit for the kinetic pressure with the lower limit provided by thermal sound speed.

\subsection{Ridge fragmentation} \label{sec:frag}

The first aspect of the analysis concerns the development of gravitational instabilities via linear perturbation.  This theory has been thoroughly investigated in a number of geometries (see \citealt{larson_cloud_1985}).  \cite{jeans_stability_1902} was the first to look at the gravitational instability of an isothermal uniform density infinite medium and found that it exists a length scale above which the amplitude of density perturbation exponentially increases with time. In this particular configuration, the fastest growing perturbation mode is that with the largest length scale, that is to say, clouds that are Jeans unstable should globally collapse. The Jeans length is given by:

\begin{equation}
	\lambda_\text{J}=\sqrt{\frac{\pi\sigma_\text{th}^2}{G\rho}}
	\label{eq:lam_J}
\end{equation}

\noindent where $\rho$ is the density of the medium. If turbulent motions are present, the turbulent Jeans length can be expressed by replacing $\sigma_\text{th}$ with $\sigma_\text{eff}$. The Jeans mass can then be expressed as the mass contained within a sphere of diameter $\lambda_\text{J}$:

\begin{equation}
	M_\text{J}=\frac{4\pi}{3}\rho\left(\frac{\lambda_J}{2}\right)^3=\frac{\pi^{5/2}}{6G^{3/2}}\frac{\sigma_\text{gas}^3}{\rho^{1/2}}
\label{eq:M_J}
\end{equation}

\noindent The situation is very different when considering infinite cylinders in hydrostatic equilibrium. There is still a length scale beyond which the cylinder becomes gravitationally unstable, however, the fastest growing mode is instead one with an intermediate length scale which, depending on the properties of the cylinder, is either a function of the cylinder's radius or its scale height $H$.  The latter is given by \citep{nagasawa_gravitational_1987}:

\begin{equation}
	H=\sqrt{\frac{\sigma_\text{gas}^2}{4\pi G\rho_c}}
\label{eq:H}
\end{equation}

\noindent where $\rho_\text{c}$ is the central density of the cylinder. The scale height is a characteristic of the system.  For instance, for an isothermal and infinitely long cylinder, the density at $r=2H$ drops by a factor 2.25 \citep{ostriker_equilibrium_1964}.  It is interesting to note that the scale height and the Jeans length have very similar forms.  In fact the ratio of the two is $L_\text{J}/H=2\pi$ assuming that $\rho=\rho_\text{c}$.  The most unstable mode for a cylinder in hydrostatic equilibrium depends upon the ratio $R/H$ where $R$ is the radius of the cylinder \citep{nagasawa_gravitational_1987}.  In the two extreme cases where $R/H>>1$ and $R/H <<1$, the most unstable modes have a length scale of $\lambda_\text{cyl}=22.1H$ and $\lambda_\text{cyl}=10.8R$, respectively.  To compute the $R/H$ ratio we estimate the central densities for both parts of the ridge, and the mean $\sigma_\text{eff}$ for \nhhh\ presented in Table~\ref{tab:analysis}. The method used to estimate $\rho_\text{c}$ is shown in Appendix \ref{sec:plummer}.  We chose \nhhh\ over the other two tracers considering that  \nhhh\ is the densest gas tracer with observations that cover the entire ridge.  Using these values we find that the ridge is compatible with $R/H >> 1$, which implies that the most unstable mode for fragmentation is $\lambda_\text{cyl}=22.1H$. 

Following up, we compute the length and mass scales using Jeans fragmentation and cylindrical fragmentation for the active and quiescent regions, and their respective variations from the thermal case to the effective case.  We present this fragmentation analysis in Fig.~\ref{fig:frag}.  On the same figure we over-plot the observed clump masses vs the distance to their nearest neighbour $\lambda_\text{sep}$ (see Table~\ref{tab:core_mass}) where $\lambda_\text{sep}$ is assumed to be a reasonable estimate of the fragmentation length scale, albeit the inclination angle of the ridge with respect to the line of sight is not taken into account.  All quantities presented in Fig.~\ref{fig:frag}a and Fig.~\ref{fig:frag}b are also given in Table~\ref{tab:analysis} and Table~\ref{tab:split}.  From these plots, we see that the fragmentation of the G316.75 ridge into clumps is better explained by the thermal case than by the effective case. This indicates that the large velocity dispersion we measure does not greatly contribute to support the gas against gravity.  It is also interesting to note that all the clumps in the quiescent region are compatible with Jeans fragmentation while it is not necessary the case for the clumps in the active region. Even though we do expect changes in the fragmentation scale of filaments with changing gas properties \citep{kainulainen_high-fidelity_2013}, we do have to keep in mind that the limited angular resolution of our \h\ column density map prevents us from drawing robust conclusions on this fragmentation analysis.

\subsection{Radial stability of the ridge} \label{sec:stability}

\begin{figure*}
	\centering
		\includegraphics[width = \textwidth]{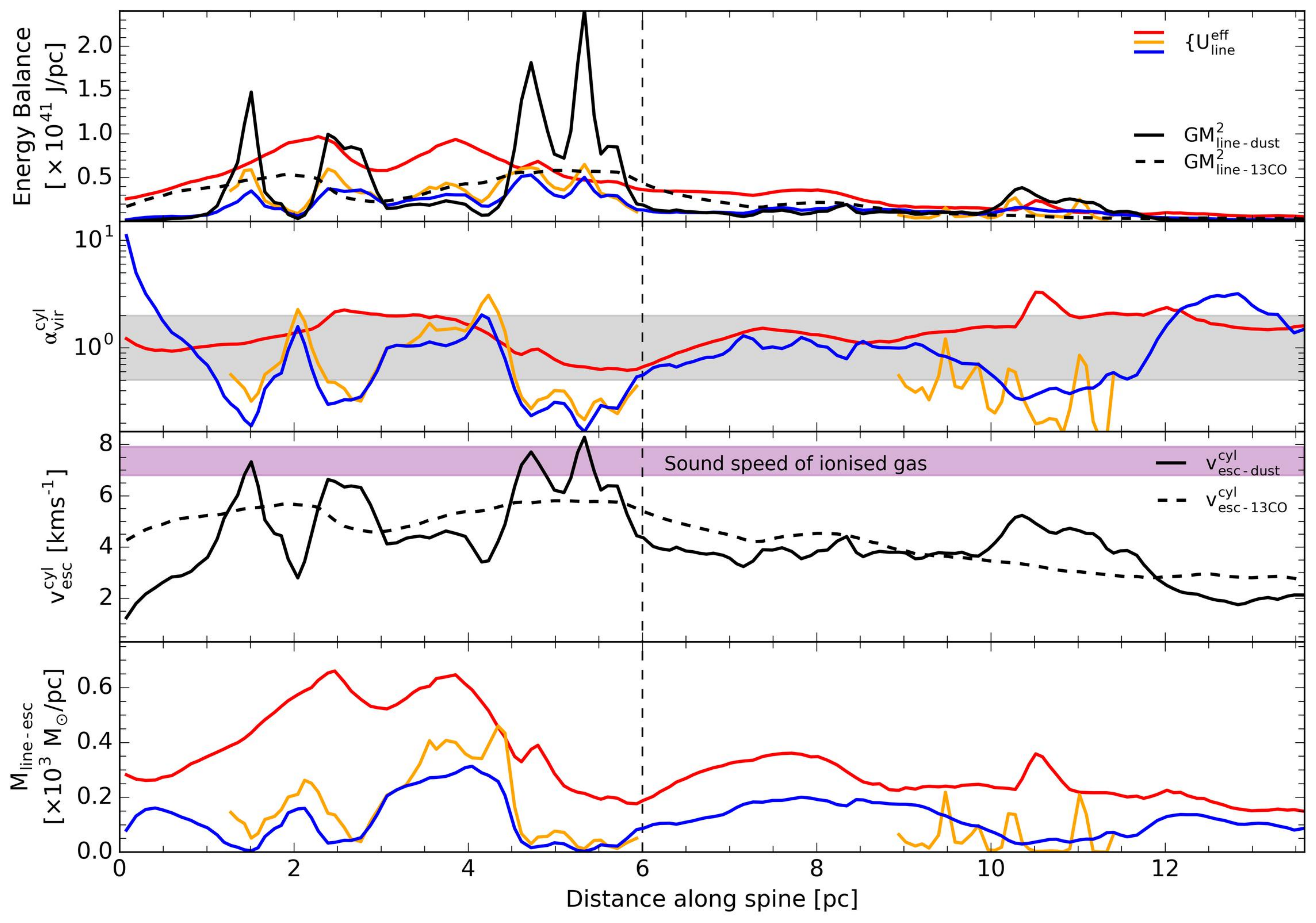}
	\caption{  {\bf First panel:} Kinetic energy per unit line as estimated using the same molecular transitions and line colours as shown in Fig.~\ref{fig:mean kin long}. Also plotted is $GM_\text{line}^2$ (which is proportional to the potential energy per unit line $\Omega_{line}$) calculated using {\it Herschel} (black solid line) and \tCO\ (1-0) (black dashed line).  {\bf Second panel:} Virial ratio as estimated using the same three molecular transitions. The grey shaded region highlights where the gas has a virial ratio between 0.5 to 2.  {\bf Third panel:} Escape velocity as estimated using $M_\text{line-dust}$ (solid line) and $M_\text{line-13CO}$ (dashed line). The purple shaded region indicates the thermal sound speed of ionised gas at $T_\text{e}$=5600-7600~K.  {\bf Fourth panel:} Mass-per-unit-length that can escape the potential calculated using the same three molecular transitions.}
	\label{fig:param_line}
\end{figure*} 

	The G316.75 ridge is a 13.6~pc long filamentary cloud with an aspect ratio of $\sim11$.  When trying to determine how likely it is that the ridge will radially collapse or expand, models that approximate interstellar filaments as cylinders are the most appropriate.  \cite{chandrasekhar_problems_1953} were the first to determine the formalism that describes the virial theorem of a self-gravitating infinite cylinder with internal kinetic pressure and a magnetic field aligned along its axis.  If such a cylinder is in equilibrium, in the absence of magnetic field, it can be shown \citep{chandrasekhar_problems_1953} that the virial theorem is given by:

\begin{equation}
	\frac{4}{3}U_\text{line}-GM_\text{line}^2=0 
\label{eq:vir_eq}
\end{equation}

\noindent where $U_\text{line}$ is the kinetic energy per unit length, and \Mline\ is the mass per unit length of the cylinder.  It is worth noting that the second term on the left side of Eq.~(\ref{eq:vir_eq}) is proportional to $\Omega_\text{line}$, that is to say, the gravitational energy per unit length.  The constant of proportionality is a function of the normalisation of the gravitational potential \citep{ostriker_equilibrium_1964}.  The kinetic energy per unit length can be written as:
 
\begin{equation}
	 U_\text{line}=\frac{3}{2}\frac{k_\text{B}T}{\mu m_\text{H}}M_\text{line}=\frac{3}{2} \sigma^2 M_\text{line}
\label{eq:U_line}
\end{equation}

 \noindent where $T$ is the gas temperature.  Equations~\ref{eq:vir_eq} and \ref{eq:U_line} reveal that in order to characterise the energy balance of the ridge we need $\sigma_\text{th}$, $\sigma_\text{eff}$, and \Mline\ at every position along the spine of the ridge.  These quantities are presented in Fig.~\ref{fig:sigma_eff_mline} (see Sec.~\ref{sec:analysis_param}).  The kinetic energy per unit length $U_\text{line}$ is computed six times in total, thrice for thermal support case $U_\text{line}^\text{th}$ and thrice for the turbulent support case $U_\text{line}^\text{turb}$, using $\sigma_\text{eff}$ from \tCO\ (1--0), \nhhh\ (1,1) and \nhp\ (1--0).  This allows us to track how $U_\text{line}$ varies for different density regimes.  This analysis assumes that the molecular transitions are optically thin, which is most likely true for all tracers in most of the cloud given that the physical resolution of the data of 0.5~pc, 1~pc, and 1.8~pc for \tCO\ (1--0), \nhp\ (1--0) and \nhhh\ (1,1), respectively.  From Eq.~(\ref{eq:vir_eq}) one can derive the expression of the virial ratio for a cylinder $\alpha_\text{vir}^\text{cyl}$,

\begin{equation}
	\alpha_\text{vir}^\text{cyl}=\frac{M_\text{line}^\text{crit}}{M_\text{line}}
\label{eq:vir_cyl}
\end{equation}

\noindent where $M_\text{line}^\text{crit}$ is obtained by combining Eq.~(\ref{eq:vir_eq}) and Eq.~(\ref{eq:U_line}):

\begin{equation}
	M_\text{line}^\text{crit}=\frac{2\sigma^2}{G}
\label{eq:M_line}
\end{equation}
 
 \noindent Equation (\ref{eq:vir_cyl}) can therefore be rewritten as:
 
 \begin{equation}
	\alpha_\text{vir}^\text{cyl}=\frac{2\sigma^2}{GM_\text{line}}.
\label{eq:vir_cyl2}
\end{equation}

\noindent $M_\text{line}^\text{crit}$ corresponds to the mass per unit line needed for a cylinder to remain in virial balance.  By comparing the observed \Mline\ to the effective $M_\text{line}^\text{crit}$ across three density regimes, we can thus assess how bound the ridge is as a function of density and position along the ridge (see Fig.~\ref{fig:sigma_eff_mline}). In order to provide a complete picture of the energy budget of the ridge, Fig.~\ref{fig:param_line} displays the gravitational energy, the kinetic energy, and the $\alpha_\text{vir}^\text{cyl}$ as a function of the position along the ridge spine. We note that the thermal cases for the kinetic energy is a lot smaller than any other quantity represented on the plot and is therefore not added to the plot. This already suggests that the cloud is either collapsing radially, or that another source of support beyond thermal support is preventing radial collapse. When adding the turbulent component to the kinetic energy, $U_\text{eff}$, we see that it does closely match $GM_\text{line}^2$ nearly everywhere along the ridge. This is despite the large differences in average properties of both halves of the ridge. This uniformity of the ridge energy budget between active and quiescent halves of ridge is  emphasised when computing its virial ratio (second panel of Fig.~\ref{fig:param_line}). There, one can see that both halves are very similar, and by looking at this plot it is not obvious which half contains the four O-type stars. The only distinguishing feature between the active and quiescent regions is that $\alpha_\text{vir}^\text{cyl}$ varies slightly more in the active region with a dex of 0.6 compared to 0.3 in the quiescent region (as measured in \nhhh(1,1)), which could be the result of local compression/ejection of matter due to the local injection of momentum/energy by the surrounding young high-mass stars. All this strongly suggests that the physical process that sets the ratio of kinetic to gravitational energy is global to the ridge, and not set by local stellar feedback.  

We also notice that the ridge is gravitationally bound nearly everywhere along ridge, and in all tracers. Even the gas traced by \tCO(1-0), that shows larger velocity dispersion and  seems to be at least partly powered by stellar feedback (see Sec.~\ref{sec:comp_gas_kin} and Fig.~\ref{fig:13co/nh3}), is gravitationally bound. Even more so that the mass estimated from \tCO\ is underestimated.  

Finally, we note that the $\alpha_\text{vir}^\text{cyl}$ traced with \nhhh\ significantly increases  between a longitudinal offset of 0 and 1~pc.  This can be explained by two scenarios. The upturn could be where \h\ column density is currently being pushed away by feedback, however this would be opposite to the observed trend everywhere else in the ridge. Or, it could be a consequence of having subtracted a too large \h\ column density background. This would reduce the ridge mass we measure and would result in an overestimate of the virial ratio. This scenario is supported by the fact that $M_\text{line-13CO}$ in that part of the ridge (see Fig.~\ref{fig:sigma_eff_mline}) is 1.5 to 8 times larger than $M_\text{line-dust}$, the largest difference between the two \Mline\ measurements.

\subsection{Gas expulsion} \label{sec:gasexp}

It is often assumed that stellar feedback from OB stars is powerful enough to destroy the cloud in which they formed.  Two important mechanisms that can cause this disruption are photo-ionisation of the gas and radiation pressure on dust grains.  Depending on the cloud properties, these mechanisms can transfer enough momentum and energy to the gas and dust to counteract the gravitational potential of cloud.  With four O-type stars already formed, we would expect that a significant fraction of mass within the G316.75 ridge is being pushed away by stellar feedback.  In the following, we investigate what fraction of the ridge mass is currently being affected by these mechanisms.

\subsubsection{Molecular gas}

A first evaluation of this can be done by computing what the fraction of the molecular gas mass with velocities beyond the escape velocity of the cloud is.  The escape velocity is defined as the velocity at which the kinetic energy of a test particle equals the gravitational potential energy.  For a spherical cloud of radius $R$ and mass $M$ the escape velocity $v_\text{esc}^\text{sph}$ is given by:

\begin{equation}
	v_\text{esc}^\text{sph}=\sqrt{\frac{2GM}{R}}
	\label{eq:v_esc}
\end{equation} 

However, this equation  is not valid for large aspect ratio filaments as gravitational forces exerted onto a test particle by the mass located on either side of it will tend to cancel each other. In order to derive the escape velocity of a cylinder $v_\text{esc}^\text{cyl}$, one needs to know both $U_\text{line}$ and $\Omega_\text{line}$.  Unfortunately, the latter is only known within a constant (see Sec.~\ref{sec:analysis_param}).  Though, one can reasonably assume that the ratios between the escape velocity and the virial velocity, $v_\text{esc}^\text{sph}/\sigma_\text{vir,3D}^\text{sph}=\sqrt{2}$, in the spherical and cylindrical cases are similar allowing us to write:

 \begin{equation}
	v_\text{esc}^\text{cyl}\simeq\sqrt{2}\sigma_\text{vir,3D}^\text{cyl}
\label{eq:v_esc^cyl}
\end{equation}

\noindent and using equations (\ref{eq:vir_eq}) and (\ref{eq:U_line}) we have:

 \begin{equation}
	\sigma_\text{vir,3D}^\text{cyl}=\sqrt{\frac{3GM_\text{line}}{2}}
\label{eq:v_vir^cyl}
\end{equation}

\noindent leading to:
 \begin{equation}
	v_\text{esc}^\text{cyl}\simeq\sqrt{3GM_\text{line}}
\label{eq:v_esc^cyl}
\end{equation}

\begin{table}
	\centering
	\begin{threeparttable}
	\caption{Total amount of mass that can escape.}
		\begin{tabular}{llll} 
		\hline
		Region & Mass escape  &Mass escape    & Mass escape\tnote{\emph{a}} \\
			   &  \tCO\  (\msun) & \nhhh\ (\msun) & \nhp\ (\msun) \\
		\hline
		Active   & 2120$\pm$730 & 570$\pm$120 & 680$\pm$310 \\
		Quiescent & 1610$\pm$760 & 750$\pm$180 & $120^{+160}_{-120}$ \\
		Total \% &  21\% \tnote{\emph{b}} & 7 \% & 4 \% \\  

\end{tabular} \label{tab:mass_loss}
	\begin{tablenotes}
      \footnotesize
      \item[\emph{a}]{The mass that can escape is a lower estimate since we were unable to calculate the escape fraction where the \nhp\ observations are missing.}
	\item[\emph{b}]{Percentage used for \tCO\ uses \tCO\ derived \h\ column density.}
   \end{tablenotes}
   \end{threeparttable}
	
\end{table}

\noindent In Fig.~\ref{fig:param_line} (third panel) we plot $v_\text{esc}^\text{cyl}$ as a function of the position along the ridge.  With the escape velocity in hand, we can compute what fraction of the ridge mass can escape the gravitational potential for each position.  This mass fraction was calculated by first assuming that $\sqrt{3}\sigma_\text{eff}$ is the standard deviation of the Normal distribution of gas particle velocities and then computing what fraction of that distribution satisfies $|v|> v_\text{esc}^\text{cyl}$. We note that a ridge in virial equilibrium would still have 16\% of its mass that satisfies the escape condition. Overall we find that a maximum of 19\% of the ridge mass as traced by \tCO\ may escape the gravitational potential. We converted this fraction into actual gas mass that may escape the potential (see Fig.~\ref{fig:param_line} fourth panel and Table~\ref{tab:mass_loss}). We see that there is very little difference between the active and quiescent parts of the ridge. If stellar feedback was significantly impacting the gravitational binding of the ridge, we would expect to see much more of the diffuse envelope escaping in the active region compared to the quiescent region.  The fact that the active and quiescent regions can only loose a similar amount of mass suggests that the mass loss has remained unchanged and therefore the O-type stars have little impact in expelling the gas that is already present within the ridge.

\begin{figure}
	\centering
		\includegraphics[width = \columnwidth]{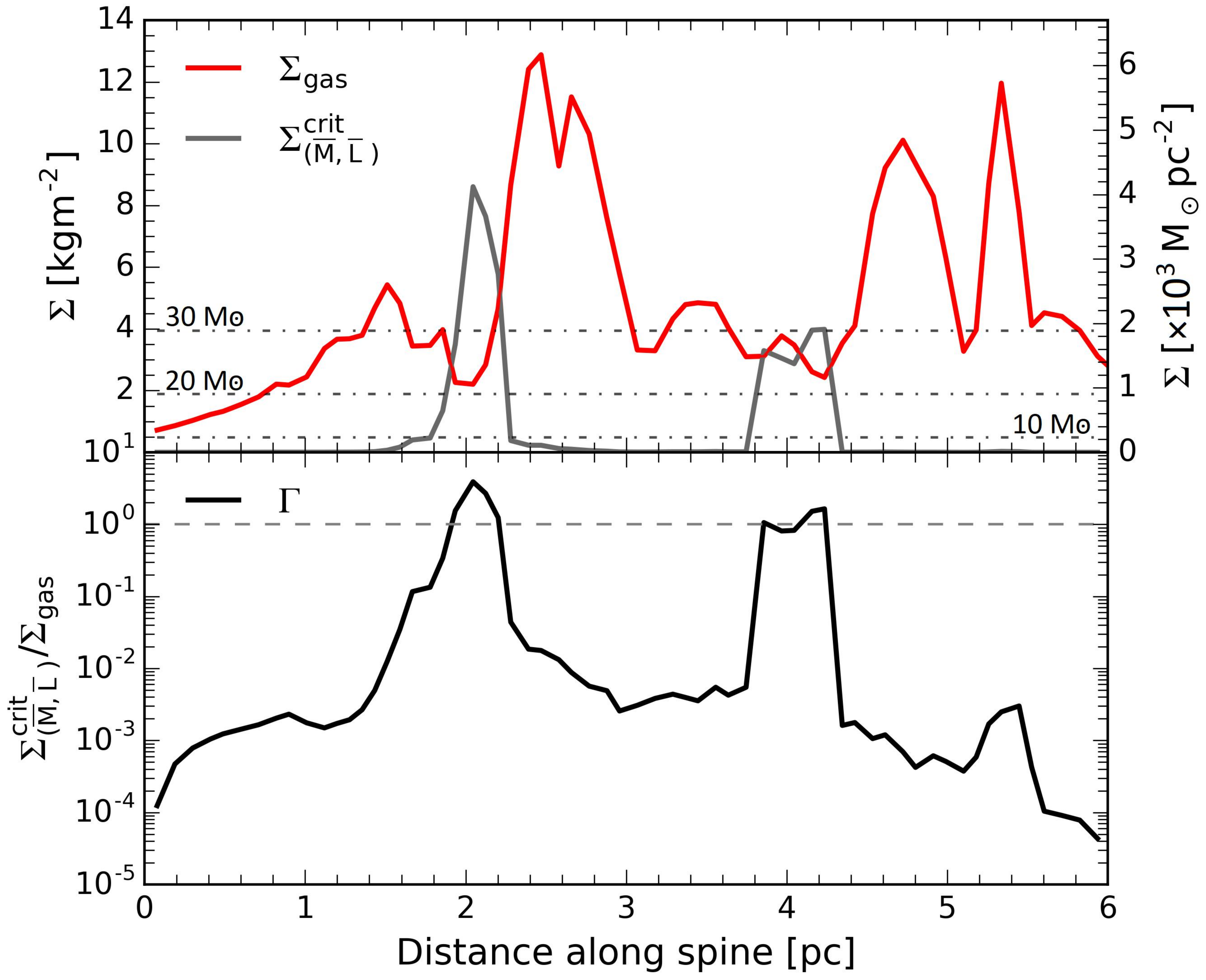}
	\caption{Gas surface density (solid red) and estimated critical gas surface density (solid grey) for the active part of G316.75 {\bf(top)}, and corresponding Eddington ratio {\bf(bottom)}. We note that here, the gas surface density is only half of the observed G316.75 mass surface density as massive stars are assumed to be located at the centre of the ridge. Grey dash-dotted lines in the {\bf top} panel labeled as 30, 20 and 10~\msun\ show the critical gas surface density if a star of that mass were placed every $\sim 0.08$pc (in every pixel) along the ridge spine. The horizontal grey-dashed line in the {\bf bottom} panel shows where the Eddington ratio is equal to 1.}
	\label{fig:Sigma}
\end{figure}

It is interesting to see whether this fraction being expelled is consistent with what we would expect from theoretical arguments.  In the case where dust is optically thick to stellar radiation but optically thin to its own infrared radiation, one can write that the Eddington luminosity, that is to say, the luminosity beyond which radiation pressure wins over gravity, is \citep{thompson_sub-eddington_2016}:

 \begin{equation}
	L_\text{Edd}\simeq4\pi Gc M_* \Sigma_\text{gas}
\label{eq:L_ed}
\end{equation}

\noindent where $c$ is the speed of light, $M_*$ the mass of the star, and $\Sigma_\text{gas}$ is the mass surface density of the gas along any light of sight exposed to the star radiation.  Therefore it exists a critical mass surface density $\Sigma_\text{gas}^\text{crit}$ below which the luminosity of star $L_*$ is enough to eject the gas.  This is given by:

 \begin{equation}
	\Sigma_\text{gas}^\text{crit} = \frac{L_*}{4\pi Gc M_*}
\label{eq:Sigma_gas_crit}
\end{equation}

\noindent From Eq.~(\ref{eq:Sigma_gas_crit}) we can determine the distribution of $\Sigma_\text{gas}^\text{crit}$ along the spine of the ridge by considering the mass and luminosity of the stars already present in G316.75 (see Sec.~\ref{sec:stars}), and evaluate the fraction of its mass that lies below this threshold.

To calculate $\Sigma_\text{gas}^\text{crit}$ we need an estimate of how the cluster mass and luminosity are distributed within the ridge. We do this by injecting a fully sampled IMF of a $\sim$1000~\msun\ cluster along the active region, where each position is weighted by the dust temperature and \h\ column density maps.  By weighting the probability, we assume that high temperature and column density spots are more likely to host more stars than low temperature low column density spots. Finally since we know there are two O stars near the radio emission peaks at 2 and 4~pc, we manually inject these at these locations.  By repeating this analysis many times, we can calculate the average $\Sigma_\text{gas}^\text{crit}$ for each position along the ridge.  We use \cite{reed_luminosity_2001} for the light-to-mass ratio and use the modified Muench IMF from \citet{murray_star_2010} to match our estimate of the cluster mass ($\sim$1000~\msun, see Sec.~\ref{sec:ppmap}). 

Figure~\ref{fig:Sigma} shows the $\Sigma_\text{gas}^\text{crit}$ and $\Sigma_\text{gas}$ at the centre of the ridge as a function of distance along the spine.  We can see that most of the ridge is sub-Eddington which corresponds only to 8--11\% of the mass in the active region.  This is in complete agreement with our previous calculations of the ridge mass fraction that can escape, and therefore confirms that the momentum injected by the stars' radiation is not enough to destroy G316.75. We also show $\Sigma_\text{gas}^\text{crit}$  in the case where a 30~\msun\ star is present in every pixel along the ridge (the same is also displayed for 20 and 10~\msun\ stars). Even though this extreme scenario is unrealistic, it illustrates that dense clumps would still remain bound, and in the case of 20~\msun stars (which is still unrealistic) most of the ridge mass would remain bound. The large mass surface density of G316.75 protects it from radiation pressure and only unrealistic amount of high-mass stars would manage to change this.

	In making these calculations, we made a number of assumptions that can over or underestimate $\Sigma_\text{gas}^\text{crit}$.  Firstly, we assumed that the IMF is fully sampled. If there are less stars, or a few very massive stars, this can over and underestimate $\Sigma_\text{gas}^\text{crit}$ respectively. Secondly, we neglected the potential of the ridge, and assumed that the stars are all located at the centre of the filament where all the luminosity of a star is absorbed by the line of sight column density at the centre. This could potentially underestimate the impact the stars could have. 

\subsubsection{Ionised gas} 

As a result of their large UV luminosities, OB stars efficiently ionise the surrounding gas. The sound speed of this newly ionised gas is often larger than the escape velocity of the parent molecular cloud, leading to its photoevaporation. Here we calculate the ability of the G326.75 O-stars to ionise the ridge, and compare this to observed ionised gas mass. To do this, we first estimate the amount of gas that has already been ionised by assuming that the number density of ionised hydrogen atoms equals the number density of electrons, $n_\text{e}$.  In Sec.~\ref{sec:stars} we calculated that $n_\text{e}=59.5_{-9.0}^{+11.5}$~cm$^{-3}$. Approximating the volume of the HII region as a sphere of radius $R_\text{HII}=6.5$~pc, we find that 1700$\pm$500~\msun\ of gas has been ionised. This only represents $\sim8\%$ of the ridge mass, which is far from the high ionisation mass fractions found in a number of numerical simulations \citep{geen_feedback_2016, kim_modeling_2018}.  Obviously, we can invoke the fact that this cloud is at a very early stage of its evolution, but the dynamical age of the HII region (see Sec.~\ref{sec:stars}) suggests that at least one of the most massive stars has been around for a couple of Myr. Another explanation could be that the ionisation front of the HII region is quickly stalled by large recombination rates.  In order to check whether this scenario is viable, we model G316.75 as a uniform density cylinder of radius $R_\text{cyl}$, and mean density $n_\text{H}$. An O6 star, modelled as a point source that emits ionising photons isotropically $\mathcal{N}_\text{i}$, is illuminating one edge of the cylinder. If no recombinations take place, the rate of ionising photons that reaches the cross section of the cylinder is given by:

 \begin{equation} 
	 \mathcal{N}_\text{i,cyl}=\mathcal{N}_\text{i}\frac{\pi R_\text{cyl}^2}{4\pi l_\text{ion}^2}
 \end{equation}

\noindent where $l_\text{ion}$ is the distance between the star and the edge of the cylinder. Now, a more realistic scenario is one that includes recombination between ionised hydrogen and electrons. The recombination rate per unit volume is given by $\mathcal{R}=\alpha_\text{B}n_\text{e}^2$, where $\alpha_\text{B}$ is the recombination coefficient and $\alpha_\text{B}=$2.59\e$^{-13}$(T$_\text{e}/10^4 \text{K})^{-0.7}$  to excited states of hydrogen \citep{osterbrock_astrophysics_1989}. The total recombination rate is given by the product of $\mathcal{R}$ and the conic volume defined by the star position and the position of the O6 star and the surface area of the cylinder edge. In this case, the rate of ionising photons that reaches the cross section of the cylinder is given by:

\begin{equation} 
	 \mathcal{N}_\text{i,cyl}=\pi R_\text{cyl}^2\left(\frac{\mathcal{N}_\text{i}}{4\pi l_\text{ion}^2}-\frac{l_\text{ion}\mathcal{R}}{3} \right)
 \end{equation}

\noindent Here, we made the implicit assumption that recombinations that occur are instantaneous reionised, an assumption which is valid since the timescale for ionisation is much shorter than for recombination.

\begin{figure}
	\centering
		\includegraphics[width = \columnwidth]{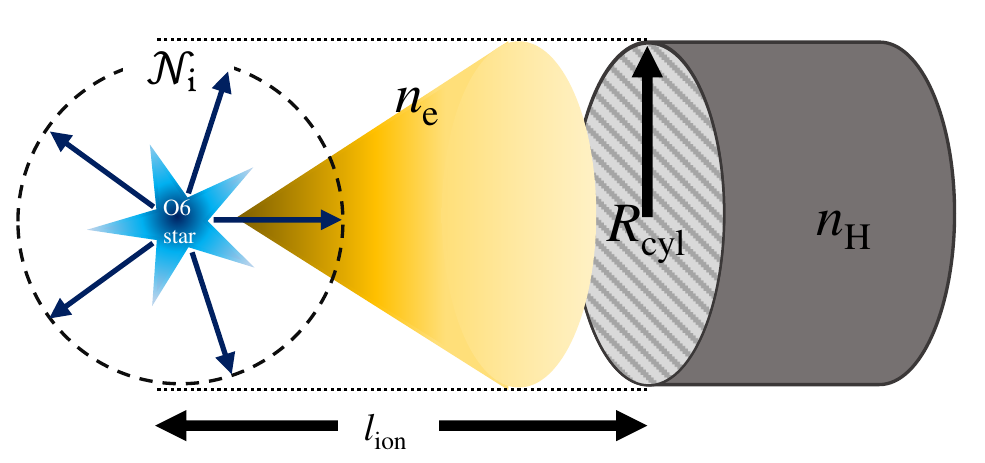}
	\caption{Diagram illustrating our ionisation model of a uniformly dense, cylindrical filament. $\mathcal{N}_\text{i}$ is the total amount of ionising photons emitted by an O6 star, $l_\text{ion}$ is the length ionised by the O6 star, $n_\text{e}$ is the density of electrons,  $R_\text{cyl}$ is the radius of the filament and, $n_\text{H}$ is the mean hydrogen density.} 
	\label{fig:ion}
\end{figure} 

\begin{figure}
	\centering
		\includegraphics[width = \columnwidth]{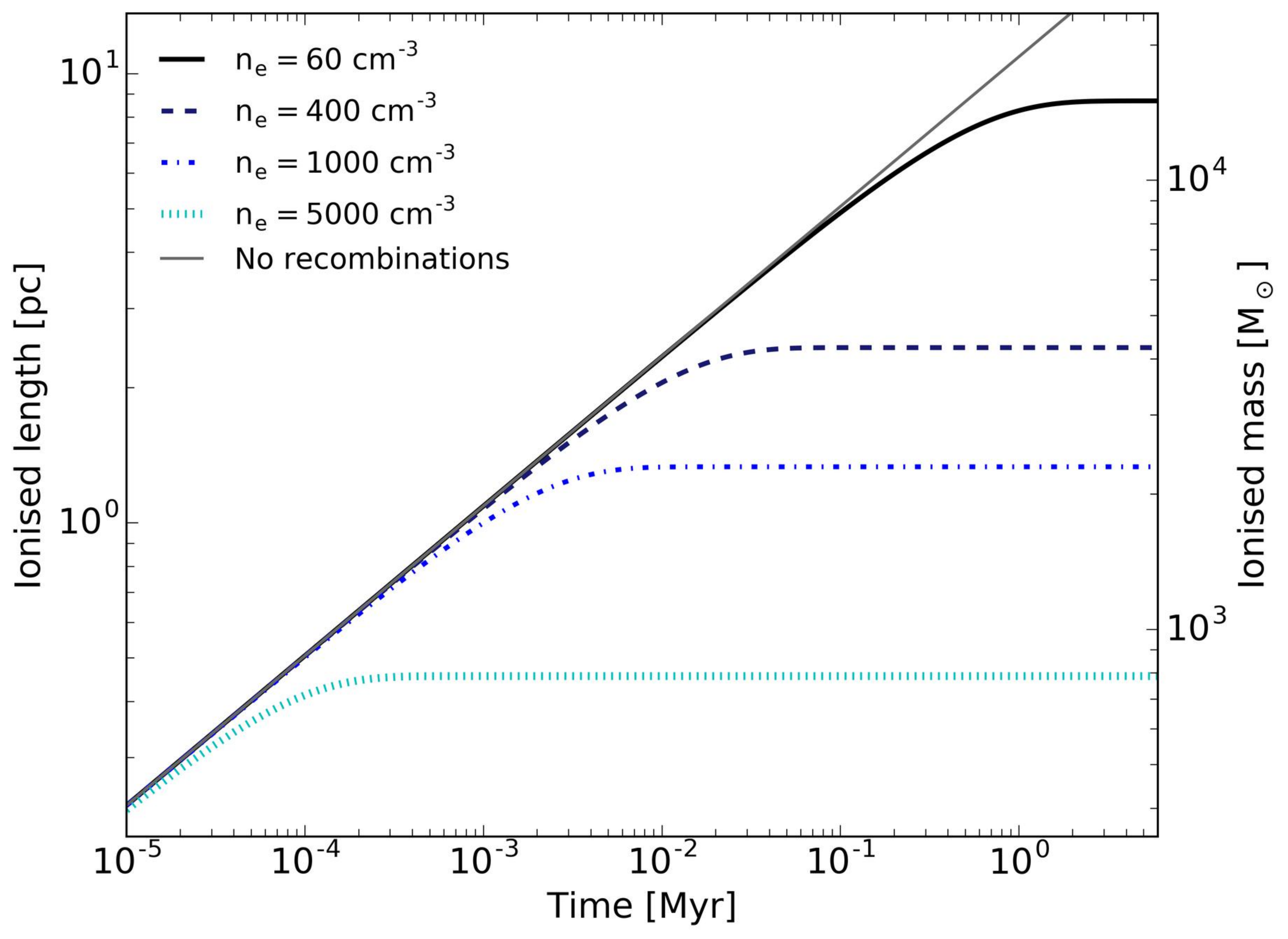}
	\caption{Time evolution of ionised length (2$\times l_\text{ion}$) of a filament of density $n$=2\e$^4$~cm$^{-3}$ exposed to the ionisation rate from an embedded O6 star. This calculation has been done for four different electron density values as indicated on the plot and for the case when there are no recombinations}.
	\label{fig:ion}
\end{figure}

\noindent We can now write the change in $l_\text{ion}$ as a function of the change in time as:

 \begin{equation}
	dl_\text{ion}= \frac{\mathcal{N}_\text{i,cyl}}{\pi R_\text{cyl}^2 n_\text{H}} dt =\frac{1}{n_\text{H}}\left(\frac{\mathcal{N}_\text{i}}{4\pi l_\text{ion}^2}-\frac{l_\text{ion} \alpha_\text{B} n_\text{e}^2}{3} \right)dt.
	\label{eq:dl_ion}
\end{equation}

\noindent  Differential equation~(\ref{eq:dl_ion}) can now be analytically integrated in order to obtain $l_\text{ion}(t)$: 

\begin{equation}
	l_\text{ion} = \bigg(\frac{3\mathcal{N}_\text{i}}{4\pi\alpha_\text{B}n_\text{e}^2}\big[1-\text{exp}(-n_\text{e}^2\alpha_\text{B}t/n_\text{H})\big]\bigg)^\frac{1}{3}
\end{equation}

Using the values we derived for the active part of G316.75 (i.e. $\overline{n}_\text{H}=2\times 10^4$~cm$^{-3}$ -- $\mathcal{N}_\text{i}=1.26\times10^{49}$~s$^{-1}$--$\alpha_\text{B}=$3.47\e$^{-13}$~cm$^3s^{-1}$ -- $n_\text{e}$= [60--5000]~cm$^{-3}$), Fig.~\ref{fig:ion} shows the total length (i.e. 2$\times l_\text{ion}$ when the ionising source is placed within the filament) and mass either side of the star that is ionised against time. On the same figure, we also over-plot the case where recombinations are ignored.

Figure ~\ref{fig:ion} shows that when we ignore recombinations, or when $n_e$ is low, an O6 type star will ionise the entire length of G316.75 by 2~Myr. Since we do not see the level of ionisation suggested by these calculations, and that recombinations do happen, this plot suggests that the relevant electron density is closer to that of the ridge. For instance, for $n_\text{e}$=5000~cm$^{-3}$, 800~\msun\ of gas is predicted to be ionised over a length of 0.5~pc, which better matches the ionised mass and length scales that are observed in G316.75.  We therefore propose that the small ionised gas fraction of G316.75 is the consequence of the large electron and \h\ gas densities in the direct vicinity of the ionising high-mass stars. As a result the ridge itself remains mostly unaffected by ionisation.

\section{Discussion} \label{sec:discuss}

In the following discussion we compare and contrast the properties of the two halves of the G316.75 ridge. By assuming that the gas properties of the quiescent half provide a reasonable proxy for those of the active half before the formation the OB stars, we are able to draw some conclusions on the impact of stellar feedback.  We will also briefly discuss a possible mechanism for the formation of the ridge itself.

\begin{figure}
	\centering
		\includegraphics[width = \columnwidth]{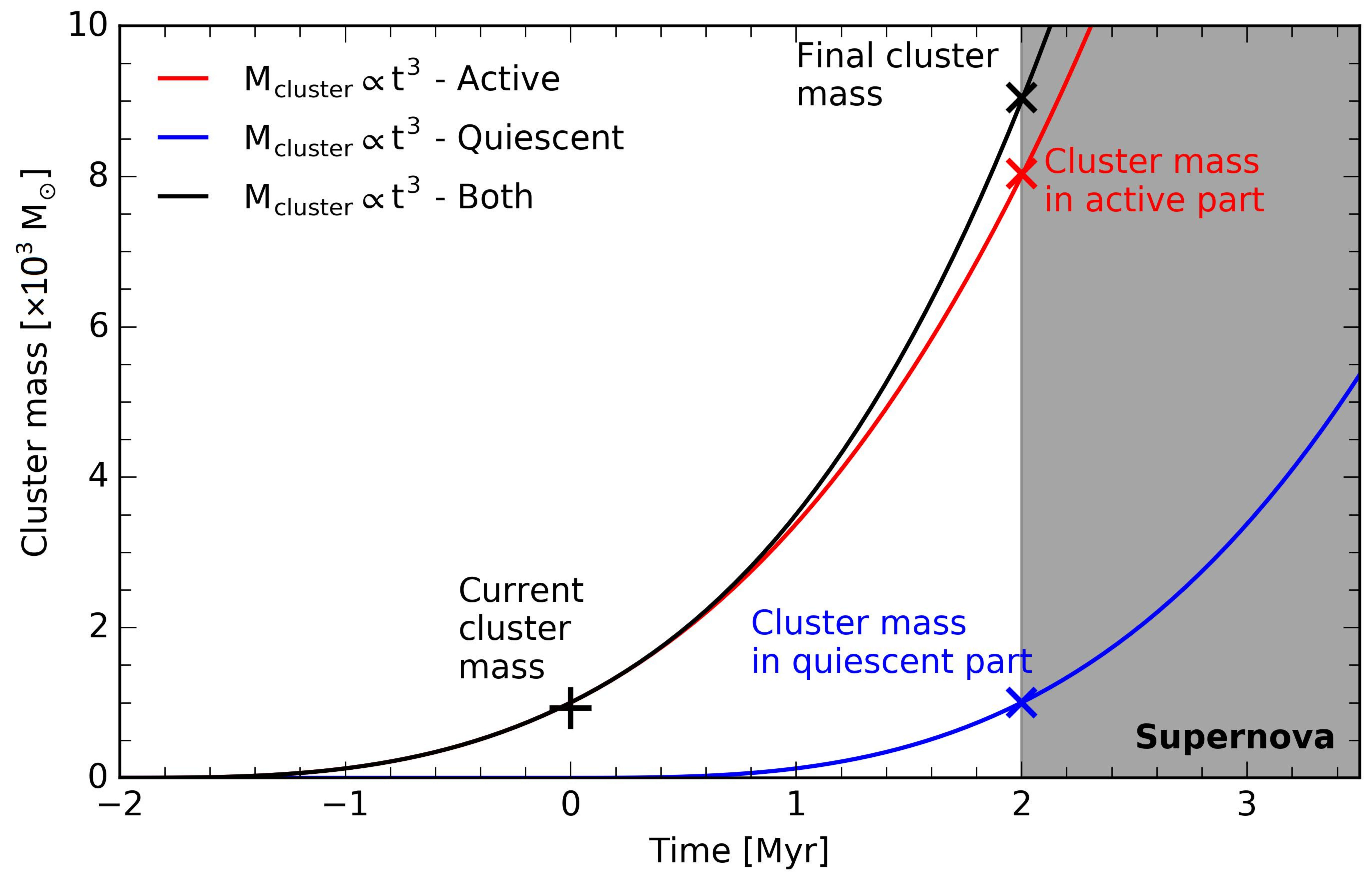}
	\caption{Illustrative figure of how the G316.75 stellar cluster mass might evolve in time.  Time $t=0$~Myr is now. The solid red, blue and  black lines shows the time evolution of the cluster mass in the active region, the cluster mass in the quiescent region and the combined cluster mass respectively.  Grey shaded region labeled as 'supernova' indicates when a supernova explosion is expected.  Labeled markers indicate current and predicted cluster masses as discussed in Sec.~\ref{sec:sfh}.} 
	\label{fig:SFH}
\end{figure}

\subsection{The star formation history of G316.75} \label{sec:sfh}

In this Section, we investigate the past and future star formation history of the G316.75. As discussed in Sec.~\ref{sec:stars}, the radio continuum emission observed in the G316.75 ridge is compatible with the presence of a $\sim1000$~\msun\ embedded cluster in the active part of the ridge. If we use the dynamical age of the HII region as a proxy for the age of the cluster we infer a star formation rate (SFR) averaged over the past 2~Myr of $\sim 5\times 10^{-4}$~\msun~yr$^{-1}$.  If we assume that the present embedded cluster in the active region formed within clumps similar to those currently present in the quiescent region, we can obtain a star formation efficiency per free-fall time, \SFEtff. In the quiescent part of the ridge, we identified a total clump mass of 845~\msun\ ($\sim11\%$ of the mass of the quiescent half) with an average volume density and free-fall time of 0.86$\times 10^5$ cm$^{-3}$ and 0.11~Myr, respectively. This implies a SFR per free-fall time, \SFRtff\ of $\sim8.0\times 10^{-3}$~\msun~yr$^{-1}$, resulting in a \SFEtff $\simeq 5\times 10^{-4}/8.0\times 10^{-3}=6.3\%$ on the clump scale.  If we instead assume that the total clump mass present 2~Myr ago in the active region scales with either the length or the mass of the ridge, we obtain a range of total clump mass between $845\times6.01/7.58=670$~\msun\ and $845\times11200/7700=1229$~\msun\ respectively. Using these scaled masses, we calculate a revised clump-scale star-formation efficiency per free-fall time of \SFEtff$=6.3_{-2.0}^{+1.6}\%$.

\begin{figure}
	\centering
		\includegraphics[width = \columnwidth]{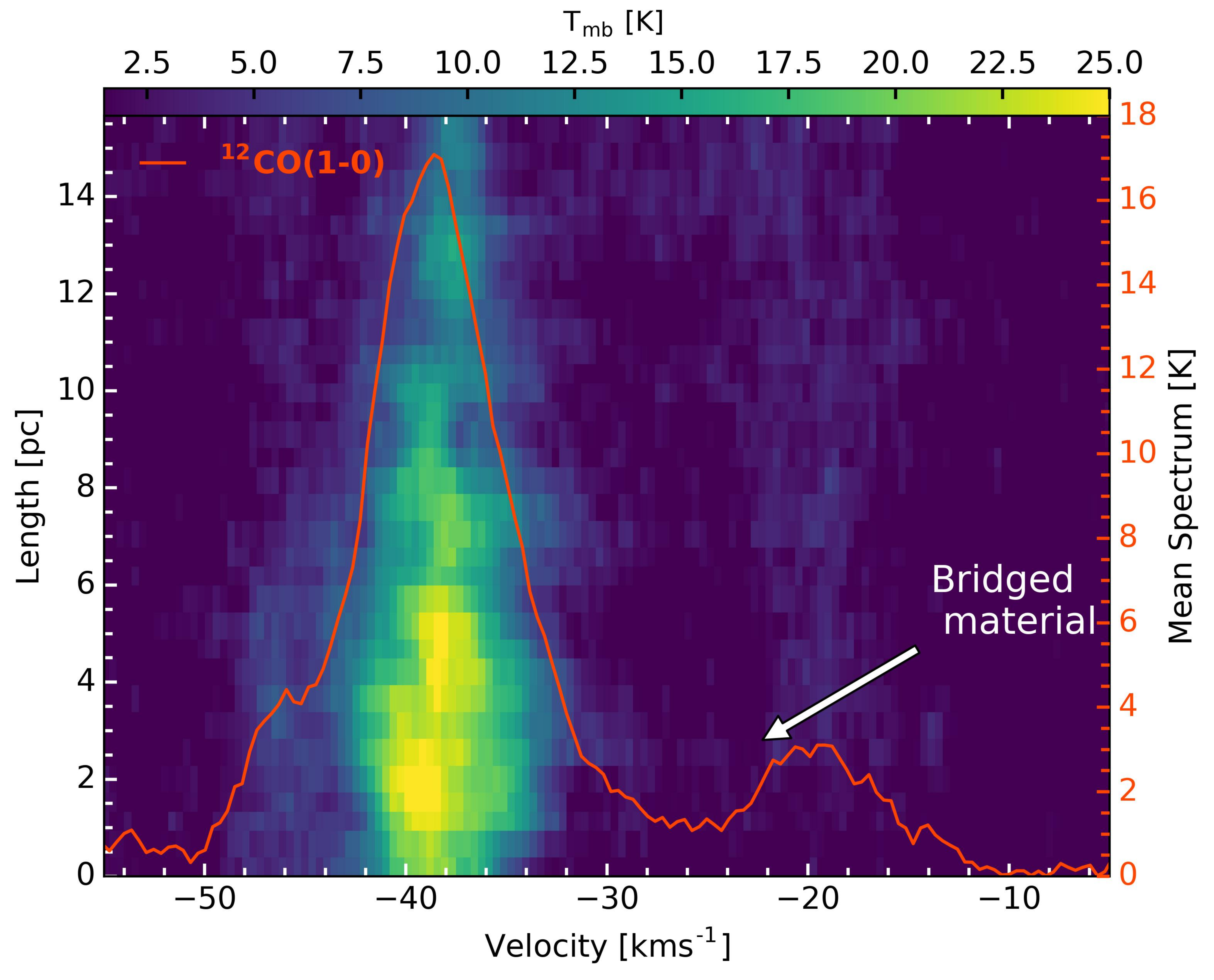}
	\caption{$^{12}$CO (1--0) position-velocity diagram calculated using the same distance as Fig.~\ref{fig:vel_dist}.  Over-plotted in orange is the $^{12}$CO (1--0) spectrum averaged over G316.75. The y axis on the right side corresponds to the mean spectrum.  Arrows indicate a possible bridge feature.}
	\label{fig:12co_VP}
\end{figure}

\begin{figure*}
	\centering
		\includegraphics[width = \textwidth]{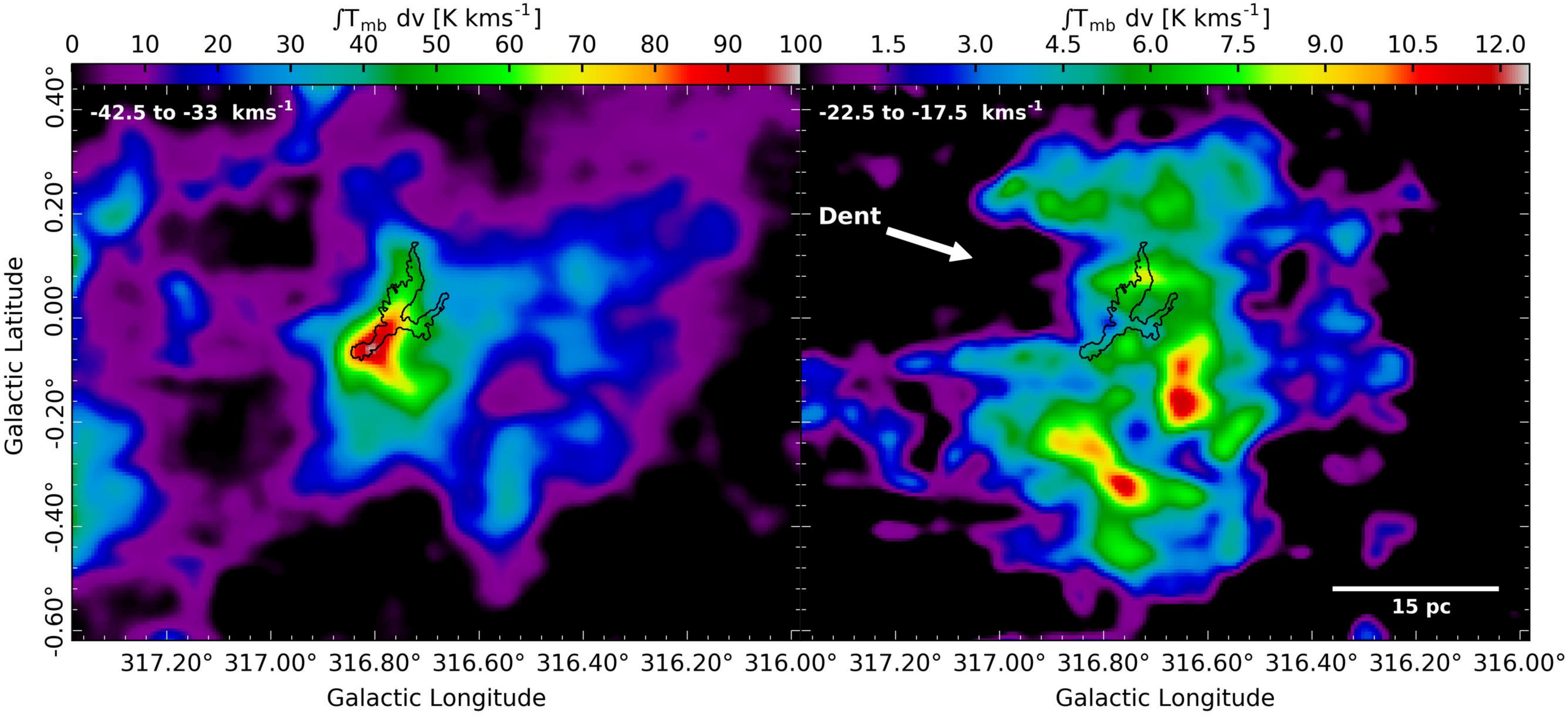}
	\caption{$^{12}$CO (1--0) integrated intensity between -42.5 and -33~\kms ({\bf left}) and -22.5 to -17.5~\kms\ ({\bf right}).   The black contour is identical to Fig~\ref{fig:velo}. Both maps have been convolved by a Gaussian of $\sigma$=1~arcmin to increase the signal to noise ratio.}
	\label{fig:12co low_int}
\end{figure*} 

	What we now observe in the active region is a total clump mass of 3994~\msun.  It is these clumps that will form the next generation of stars therefore, with an average density and free-fall time of 2.46$\times 10^5$ cm$^{-3}$, 0.06~Myr respectively, the \SFRtff\ for this second generation of clumps is $\sim64.3\times 10^{-3}$~\msun~yr$^{-1}$ showing that the total clump mass and \SFRtff\ are 3.4 and 8 times larger than in the quiescent region. As shown by \citet{ginsburg_thermal_2017} in the W51 region, it is very likely that the change in these clump properties is the result of stellar feedback via increased gas temperature and local gas compression (see clumps \#1 and \#3 in Fig.~\ref{fig:col_temp_str}). Given the feedback inefficiency at unbinding the dense gas (see Sec.~\ref{sec:analysis}), it is reasonable to assume that \SFEtff\ will, at the bare minimum, remain constant. In fact, some studies suggest that \SFEtff\ increases with time as \SFEtff $\propto t^2$ \citep{lee_observational_2016}.
The implication of a constant \SFEtff\ would be that in the next 2~Myr, another $\sim 9000$ \msun\ worth of stars could be formed in G316.75. This combined with the already formed cluster would lead to a total stellar mass of $\sim 10,000$~\msun, matching the mass definition limit for young massive clusters of $10^4$~\msun\ \citep{portegies_zwart_young_2010}. These results are summarised in Fig.~\ref{fig:SFH} where the G316.75 cluster mass is plotted as a function of time. In addition to these masses, we added a $M_\text{cluster} \propto t^3$ relationship which, even though admittedly with limited observational constraints, fit the data rather well. Such accelerated SFR has been also derived by \cite{lee_observational_2016} in order to explain the large dispersion of  \SFEtff\ values observed in galactic giant molecular clouds. In the case of G316.75, the accelerated SFR would be the result of the increased mass fraction of the dense star-forming clumps in the active region.

\subsection{The inefficiency of stellar feedback and ridge origin} \label{sec:ridge_origin}

When looking at the mid- to far-infrared composite image of G316.75 presented in Fig.~\ref{fig:rgb_intro}, it is immediately obvious which part of the cloud is currently being affected by the vast amount of energy and momentum released by the embedded OB stars.  However, when looking at the virial ratio profile along the ridge (Fig.~\ref{fig:param_line}), it is not obvious anymore---the active part looks very similar to the quiescent part. This suggests that the process setting the ratio of kinetic to gravitational is global to the ridge and not due to local stellar feedback. An obvious candidate for such global process is radial gravitational collapse. Indeed, collapse is the expression of the conversion of gravitational energy into kinetic energy, and therefore naturally ensures that both quantities remain close to equipartition \citep{lee_formation_2016}

The stellar feedback inefficiency at dispersing the ridge mass is further evidenced through our calculation of the ionised mass in the region. As shown Sec.~\ref{sec:gasexp} only $\sim$8\% of the ridge mass has been ionised so far. This is much lower than reported in simulations of cloud evolution with ionisation feedback, where fractions of 30\% to 95\% are reported at the extreme limits of their parameter space (e.g \citealt{kim_modeling_2018}).  In these simulations, the average cloud is expected to ionise $\sim80\%$ of their parent cloud mass after a couple of initial cloud free-fall times.  The most likely explanation for the difference between our results and such numerical simulations lies in the difference in density structure at the time when the first ionising star was born. In these simulations, the cloud is modeled as a sphere of gas and ionising sources are created when the cloud is still in a porous state. What our study suggests is that the ridge was already a single high-gas-density filamentary structure with a large aspect ratio before the OB stars began to ionise their environment. With such initial conditions, ionising photons would only ionise a very small fraction of the ridge mass (as explained in Sec.~\ref{sec:gasexp}), which would protect the ridge against the damage one typically observes in numerical simulation that use spherical initial conditions and low densities.

	A possible mechanism to rapidly compact a lot of mass into a ridge-like structure could be cloud-cloud collision. In order to test this idea in G316.75, we use the ThrUMMS $^{12}$CO (1--0) data in order to look for signatures in the most diffuse molecular gas observable.  Figure \ref{fig:12co_VP} shows the $^{12}$CO (1--0) spectrum averaged over the ridge.  This spectrum shows two components, the one associated to the ridge in a velocity range -42.5 to -33~\kms, and a weaker one at velocities between -22.5 to -17.5~\kms. In Fig.~\ref{fig:12co low_int} we show the integrated intensity maps of both components, convolved with a Gaussian of $\sigma$=1~arcmin to increase the signal to noise ratio.  We immediately see that the two clouds are spatially coherent in projection suggesting that they could be physically connected to each other.  We also see a drop in intensity morphologically similar to the G316.75 ridge, that may be the 'dent' left by the collision of the two clouds \citep{fukui_molecular_2018}.  By taking the projected distance between the centre of the ridge and the dent, $\sim$20~pc, and the velocity difference between the two clouds, 18~\kms, we can determine that the collision may have happened $t_\text{pcc} \simeq 1$~Myr ago.  This estimate is biased by projection.  The true collision time is given by $t_\text{cc}=t_\text{pcc}\frac{\cos(\alpha)}{\sin(\alpha)}$, where $\alpha$ is the angle between the line of sight and the collision axis.  In order to get a collision time $\ge 2$~Myr, that is to say, the dynamical age of the cluster, we need to have $\alpha \le \pi/7$ (i.e $\sim 27$~deg).

	Observational signatures of cloud-cloud collisions have been investigated in detail (e.g \citealt{haworth_isolating_2015,bisbas_gmc_2017}).  One key signature is the presence of bridges, that is to say, low column density gas at intermediate velocities between the systemic velocities of the two colliding clouds.  These bridges are best seen in position-velocity diagrams (PV).  For this reason, we computed the $^{12}$CO (1--0) PV diagram along the spine of the ridge (see Fig.~\ref{fig:12co_VP}). On this figure, we see the presence of a faint strip of emission connecting the two clouds that are reminiscent of bridging features seen in the aforementioned studies.    
	
	Finally, in their recent work, \cite{inoue_formation_2018} and \cite{whitworth_bipolar_2018} showed how massive filaments, such as ridges, may form via cloud-cloud collision using numerical and analytical methods, respectively. It is interesting to note that both of their studies predict properties that are similar to those observed in G316.75. In \cite{inoue_formation_2018} the velocity and density profiles that develop along their simulated filament has many similarities with that of G316.75. In particular, their modeled filament shows a velocity gradient that flips near the centre of the ridge, which also coincides with large densities, reminiscent to what we observed at 4--5~pc along G316.75. On the other hand, \cite{whitworth_bipolar_2018} predict the formation of bipolar HII regions not unlike what is observed in G316.75.  Even though we cannot be certain that cloud-cloud collision is the mechanism that led to the formation of G316.75, evidence presented here seems to point towards such a mechanism.

\section{Summary and conclusion} \label{sec:conclude}

	We used archive data of the G316.75 massive-star-forming ridge to determine how feedback from high-mass stars impacts the star forming properties. The G316.75 ridge is unique in the sense that half of the ridge is currently actively forming stars, while the other half is quiescent. Comparing and contrasting these two halves gives us the opportunity to quantify the impact of stellar feedback on the surrounding gas. By studying both halves in detail, we showed that despite the presence of four embedded O-types stars, feedback is unable to unbind most of the ridge's mass over a large range of gas densities. Indeed, when looking at the ridge virial ratio we notice very little difference between the active part (i.e. where the massive stars are), and the quiescent part. This is compelling evidence for feedback being inefficient at ionising and pushing away the ridge mass. From theoretical calculations, we showed that such feedback inefficiency is actually expected given the large average density and elongated morphology of the ridge. Due to this inefficiency and the sharp increase of the total clump mass over time, we suggest that G316.75 will continue producing stars to form a massive stellar cluster until a supernova explosion changes the star forming environment in the coming 2~Myr.

	We conclude that the initial morphology of a massive star-forming cloud plays an important role in its future evolution, especially with respect to the impact of the high-mass stars it will form. We suggest that a filamentary configuration strongly reduces the disruptive effects that feedback might have on the gas, and that such morphology could be the result, at least in the case of G316.75, of cloud-cloud collision. As shown in this study, in such configuration, stellar feedback from high-mass stars does very little to the dense gas already present, but is probably able to stop extended lower density gas from accreting onto the ridge. Indeed, as shown in Fig.~\ref{fig:param_line}, the sound speed of the ionised gas is larger than the escape velocity of the ridge. Unlike W49 \citep{rugel_feedback_2019} and 30 Doradus \citep{rahner_forming_2018} where ionised gas can is trapped by the potential of the cloud, the ionised gas in G316.75 is able to freely expand back into the interstellar medium. As shown in Fig.~\ref{fig:rgb_intro}, a large volume around the ridge is directly being exposed to the ionising radiation of the G316.75 O-stars, and therefore the possibility for further gas accretion onto G316.75 is drastically reduced. In other terms, the main impact of feedback in G316.75 is gas exhaustion rather than gas ejection.  These results have strong implications as to how feedback is implemented in cosmological and galaxy-scale simulations, in particular for those reaching parsec-scale resolutions for which star-formation recipes might have to be revised.

\begin{acknowledgements}
	EJW is supported by the Science and Technology Facilities Council (STFC). NP wishes to acknowledge support from STFC under grant number ST/N000706/1. This research has made use of Astropy, a community-developed core Python package for Astronomy (Astropy Collaboration, 2013, 2018). This research  has also made use of the python packages astrodendro, colorcet, matplotlib, multicolorfits, numpy, scipy and skimage. This research also makes use of the software GILDAS and STARLINK.

\end{acknowledgements}



%
%

\bibliographystyle{aa}
\bibliography{G316-75_arxiv} 



\appendix

\section{PPMAP} \label{sec:PPMAP}

\begin{figure*}
	\centering
		\includegraphics[width = \textwidth]{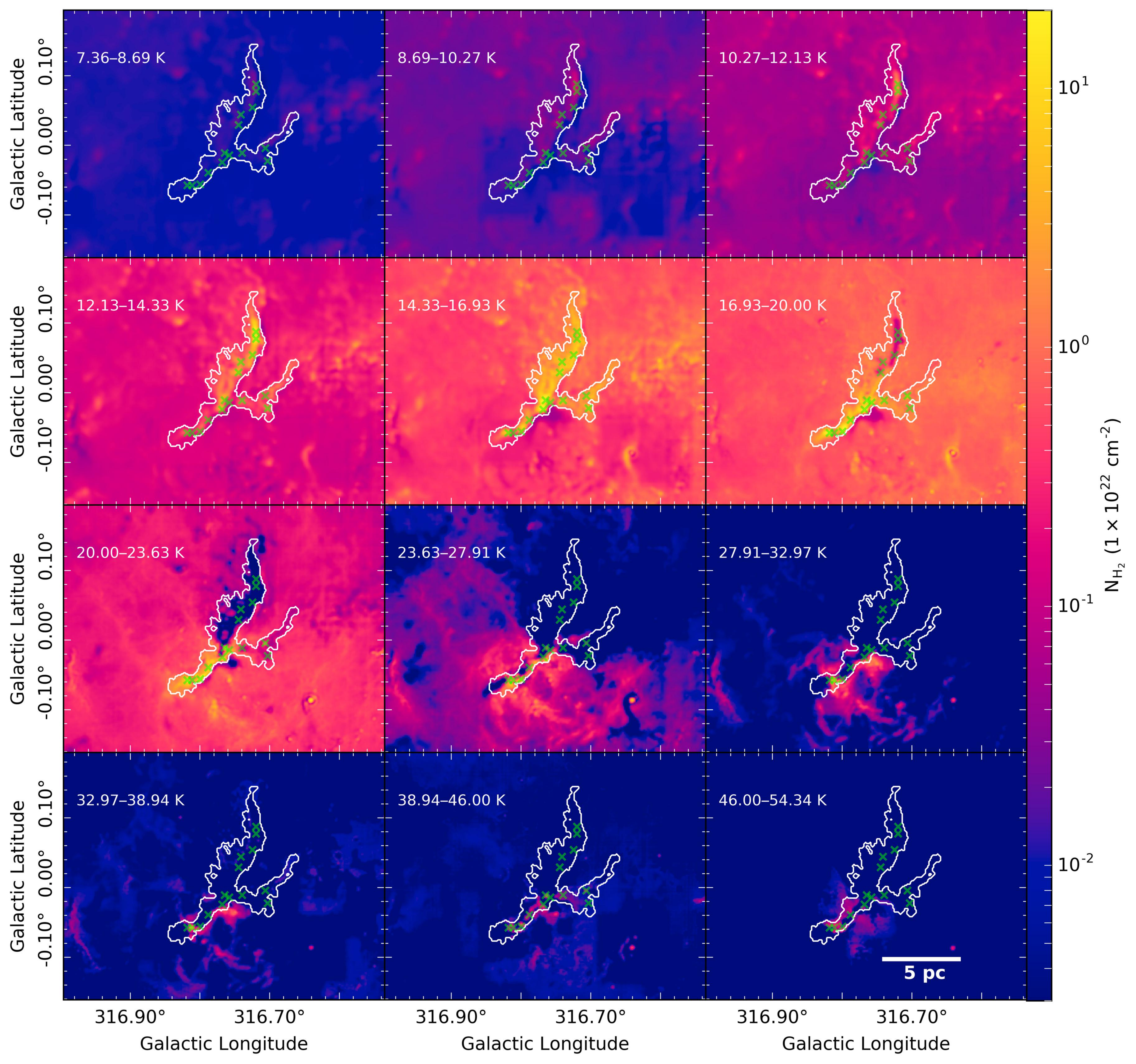}
	\caption{Differential column density of G316.75 over 12 temperature bands.  The temperature bands used are logarithmically distributed from 8 to 50~K.  Each subplot has been labeled with the temperature band the column density represents.  The white contour has the same meaning as Fig~\ref{fig:velo}.  Green crosses mark the location of the clumps present.  Physical scale is 5~pc = 0.106$^\circ$}
	\label{fig:dif_col}
\end{figure*} 

The main advantage of using PPMAP is that it provides, for each target, a set of \h\ column density maps at different temperature bands. Figure~\ref{fig:dif_col} displays such set of maps. As we we can see, these images show a range of features that helps understanding what sources contributes to what temperature band.  For instance, one can nicely see that the quiescent part of the ridge is colder  than the active part, as derived in Sec.~\label{subsec:hessian}. It also shows us exactly how and where the \h\ column density is being affected by feedback. Indeed, we can nicely see that hot dust outlines the shape of multiple bipolar lobes that coincide with the SUMMS radio contours and the spitzer 8~\micron\ seen in Fig.~\ref{fig:rgb_intro}.  Also, we can easily identify where hot embedded sources are located (see clump \#7). This is particularly true in the 38.94--46.00 K temperature band, where $\sim10$ point sources can be seen. 

\section{3D filament modelling} \label{sec:plummer}

To characterise the fragmentation of the ridge (see Sec.~\ref{sec:frag}), we need to derive the central density of ridge $n_\text{c}$. For this purpose, we modelled G316.75 as a cylinder.  The quiescent and active regions were modelled independently. The filament radius was estimated to be $R=0.63\pm0.08$~pc corresponding to the distance from the ridge's spine at which we observe a break in its column density profile (see Fig.~\ref{fig:1d_col_log}).  This break indicates a transition between the dense filament and the diffuse background.  We computed a series of 3D cylindrical models whose density follow a plummer-like profile

\begin{equation}
	\centering
		n(r)=\frac{ n_\text{c}}{( 1+({\frac{r}{ R_\text{flat}})^2 })^{p/2} }
	\label{plummer}
\end{equation}

\noindent where $n_\text{c}$ is the central density, $R_\text{flat}$ is a characteristic length scale, and $p$ is the density power-law index.  The models cover the following parameter space, $n_c$: 4--40\e$^4$~cm$^{-3}$; $R_\text{flat}$: 0.1--0.4~pc; p: 1--4. The resulting models are then projected in 2D. We assume the ridge has zero inclination along the line of sight, which simplifies the 2D projection but  overestimates the derived central density.  Using a $\chi^{2}$ minimisation, each model is then compared to the observed mean profiles calculated in Sec.~\ref{subsec:hessian}.  By doing so, we note that, even though $n_\text{c}$ is well characterised,  $R_\text{flat}$ and $p$ are degenerate (i.e. we do not have enough angular resolution to  properly characterise $R_\text{flat}$). However, since we only need $n_\text{c}$ for the fragmentation analysis, this is not an issue. We use a 1-sigma confidence in the $\chi^2$ residuals to estimate the spread of values in $n_\text{c}$ and obtain $n_\text{c}=17\pm$2\e$^4$~cm$^{-3}$ and $n_\text{c}=7\pm$2\e$^4$~cm$^{-3}$ for the active and quiescent regions, respectively (see Table \ref{tab:analysis}).

\section{Deriving \h\ column density from \tCO}  \label{sec:LTE}

\begin{figure}
	\centering
		\includegraphics[width = \columnwidth]{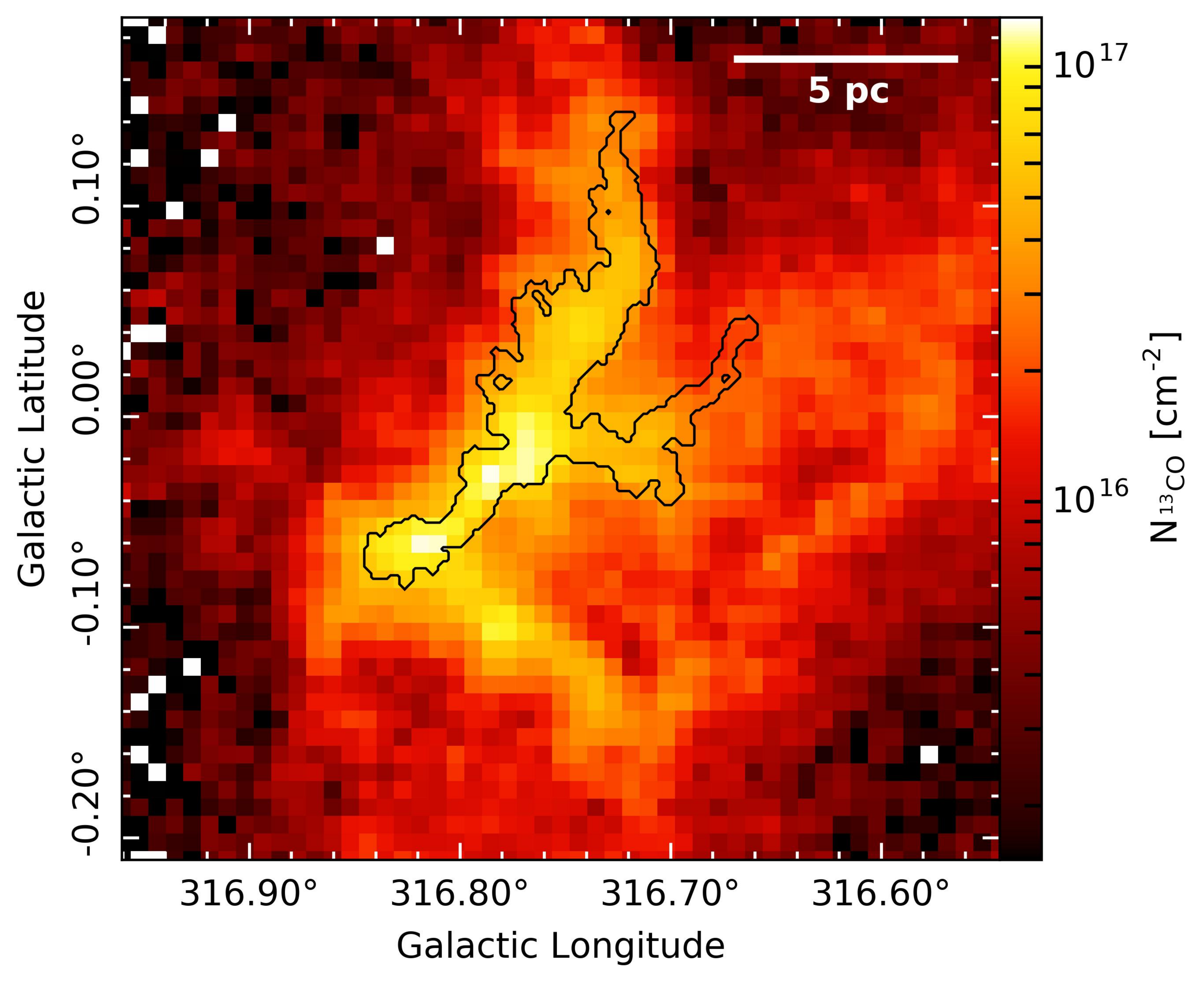}
	\caption{\tCO\ column density map derived using LTE for \tCO.  The black contour is identical to Fig~\ref{fig:velo}.}
	\label{fig:LTE_13co}
\end{figure} 

\begin{figure}
	\centering
		\includegraphics[width = \columnwidth]{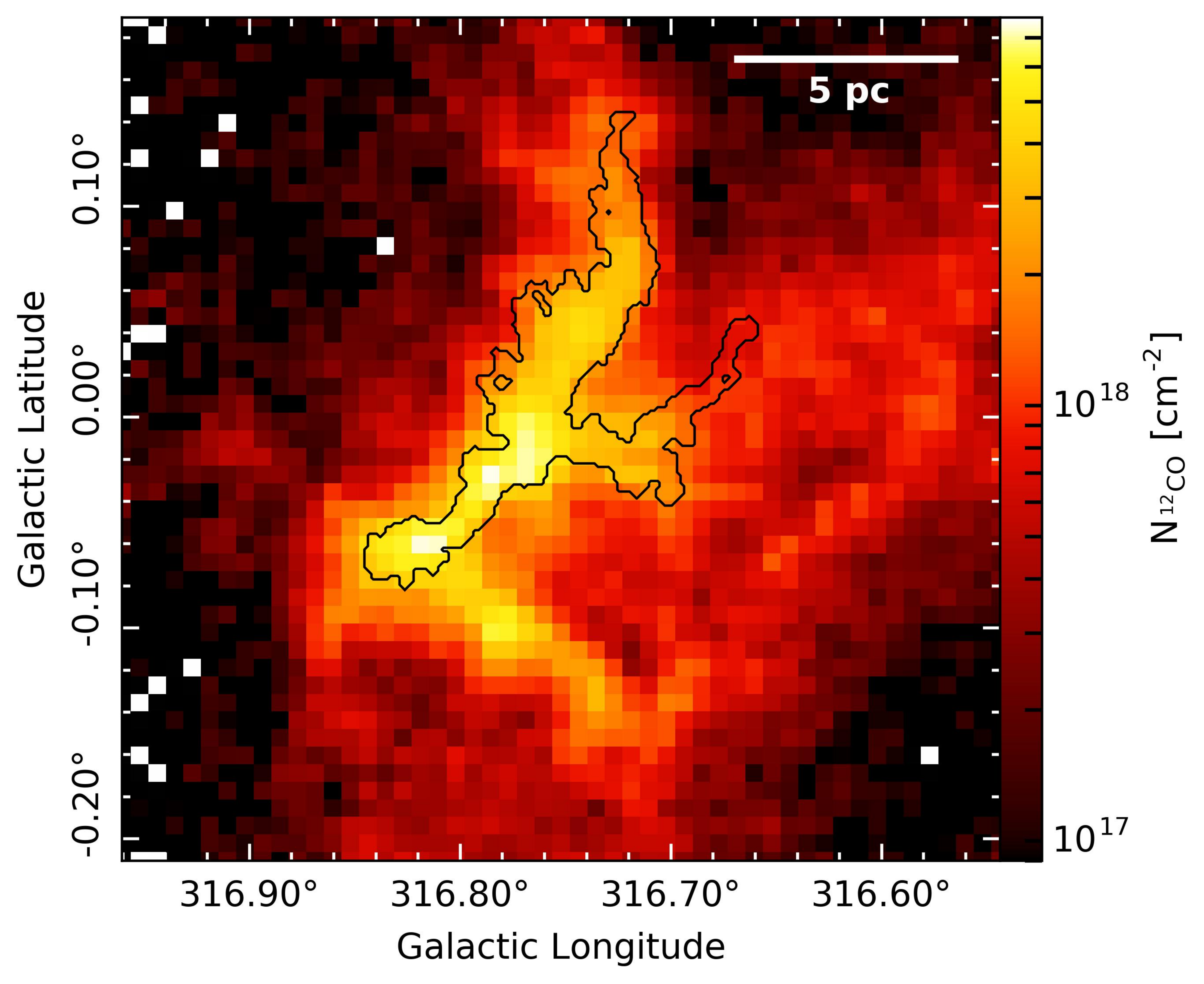}
	\caption{$^{12}$CO column density map derived using LTE for \tCO.  The black contour is identical to Fig~\ref{fig:velo}.}
	\label{fig:LTE_12col}
\end{figure} 

To calculate the \h\ column density from \tCO, we follow the methodology presented in \cite{wilson_tools_2009}.  This method works by assuming that $^{12}$CO and \tCO\ are in LTE, that the excitation temperature, $T_\text{ex}$ is the same for \tCO\ and $^{12}$CO for each line-of-sight, that  $^{12}$CO is optically thick while \tCO\ is optically thin, and \tCO\ and $^{12}$CO emits from the same volume.  These assumptions allows us to assume that the excitation temperature is equal to the kinetic temperature, $T_\text{k}$ which greatly simplifies the expression for the optical depth, $\tau(\nu)$ and for the column density.  Under these assumptions, the \tCO(1-0) excitation temperature and opacity are given by 

\begin{equation}
	\centering
		T_\text{ex} = 5.5\Big/\text{ln}\bigg[1+\frac{5.5}{T_\text{b,peak}^{^{12}\text{CO}}+0.82}\bigg]
	\label{eq:Te}
\end{equation}

\noindent and $\tau_\nu$

\begin{equation}
	\centering
		\tau(\nu) = -\text{ln}\bigg[1-\frac{0.189T_\text{b,peak}^{^{13}\text{CO}}} {(\text{exp}(5.3/T_\text{ex})-1)^{-1} -0.16}\bigg]
	\label{eq:tau}
\end{equation}

\noindent where $T_\text{b,peak}^{^{12}\text{CO}}$ and $T_\text{b,peak}^{^{13}\text{CO}}$ are the peak brightness temperatures of $^{12}$CO and \tCO\ respectively.  Then using the fact that $T_\text{ex}=T_\text{k}$ under the aforementioned assumptions, the \tCO\ column density can be written as

\begin{equation}
	\centering
		N_{^{13}\text{CO}} = 3\times 10^{14}\frac{T_\text{ex}\int \tau(v) \text{d}v}  {1-\text{exp}(-5.3/T_\text{ex})} [\text{cm}^{-2}]
	\label{eq:N_13CO}
\end{equation}

\noindent where $\tau$ is integrated over velocity in units of \kms.  For an optically thin regime we can simplify $T_\text{ex}\int \tau(v) \text{d}v$ leaving us with 

\begin{equation}
	\centering
		N_{^{13}\text{CO}} \cong 3\times 10^{14}\frac{\tau_0\int T_\text{b}^{^{13}\text{CO}}(v) \text{d}v}{[1-\text{exp}(-\tau_0)][1-\text{exp}(-5.3/T_\text{ex})]} [\text{cm}^{-2}]
	\label{eq:N_13COs}
\end{equation}

\noindent where $N_{^{13}\text{CO}}$ is the \tCO\ column density and $\int T_\text{b}^{13\text{CO}}(v) \text{d}v$ is the integrated intensity.  To calculate $N_{^{13}\text{CO}}$, we use the integrated intensity map presented in Fig.~\ref{fig:13co_int}, which has been integrated between -42.5~\kms to -33~\kms.  The result of this has been plotted on Fig.~\ref{fig:LTE_13co} To convert the \tCO\ column density to $^{12}$CO column density, we then follow the method presented in \cite{szucs_12co/13co_2014}.  These authors created synthetic data from numerical simulations of cloud formation/evolution to show that, as a result to chemical fractionation, the \tCO/$^{12}$CO ratio decreases away from the standard ratio of $\sim$60 by up to 50\% when $N_{^{13}\text{CO}}<5$\e$^{16}$~cm$^{-2}$.  However, this variation correlates well with the  \tCO\ column density.  As a result, they could model the \tCO/$^{12}$CO ratio decrease as a function of \tCO\ column density allowing one to correct for this abundance variation. Using their model, we convert the \tCO\ column density to $^{12}$CO column density and plot this on Fig.~\ref{fig:LTE_13co}.   Finally by assuming $^{12}$CO has constant fractional abundance of $\sim$1\e$^{-4}$, we convert the $^{12}$CO column density to \h\ column density and present these maps in Fig.~\ref{fig:LTE_12col} and Fig.~\ref{fig:LTE_col} respectively.

\end{document}